\documentclass[format=acmsmall, review=false, screen=true]{acmart}

\usepackage{color}
\usepackage{graphicx}
\usepackage{balance}
\usepackage{url}
\usepackage{latexsym}
\usepackage{booktabs}
\usepackage{multirow}
\usepackage{tikz}
\usepackage[normalem]{ulem}
\usepackage{soul}

\newtheorem{definition}{Definition}

\begin{document}


\title[Blocking and Filtering]{A Survey of Blocking and Filtering Techniques for Entity Resolution}

\author{George Papadakis}
\affiliation{%
  \institution{National and Kapodistrian University of Athens}
  \country{Greece}}
\email{gpapadis@di.uoa.gr}

\author{Dimitrios Skoutas}
\affiliation{%
  \institution{IMSI, Athena Research Center}
  \country{Greece}
}
\email{dskoutas@imis.athena-innovation.gr}

\author{Emmanouil Thanos}
\affiliation{%
  \institution{KU Leuven}
  \country{Belgium}
}
\email{emmanouil.thanos@kuleuven.be}

\author{Themis Palpanas}
\affiliation{%
  \institution{Paris Descartes University}
  \country{France}
}
\email{themis@mi.parisdescartes.fr}

\renewcommand{\shortauthors}{G. Papadakis et al.}


\vspace{-10pt}
\begin{abstract}
Entity Resolution (ER), a core task of Data Integration, detects different entity profiles that correspond to the same real-world object. Due to its inherently quadratic complexity, a series of techniques accelerate it so that it scales to voluminous data. In this survey, we review a large number of relevant works under two different but related frameworks: Blocking and Filtering. The former restricts comparisons to entity pairs that are more likely to match, while the latter identifies quickly entity pairs that are likely to satisfy predetermined similarity thresholds. We also elaborate on hybrid approaches that combine different characteristics.
For each framework we provide a comprehensive list of the relevant works, discussing them in the greater context. We conclude with the most promising directions for future work in the field.

\end{abstract}
\maketitle

\vspace{-5pt}
\section{Introduction}
\vspace{-5pt}

Entity Resolution (ER) is the task of identifying different entity profiles that describe the same real-world object \cite{DBLP:journals/tkde/ElmagarmidIV07,DBLP:series/synthesis/2015Christophides}. It is a core task for Data Integration, applying to any kind of data, from the structured entities of relational databases \cite{DBLP:books/daglib/0030287} to the semi-structured entities of the Linked Open Data Cloud (\url{https://lod-cloud.net}) \cite{DBLP:series/synthesis/2015Dong,DBLP:series/synthesis/2015Christophides} and the unstructured entities that are automatically extracted from free text \cite{DBLP:journals/tkde/ShenWH15}.
ER consists of two parts: (i) the \textit{candidate selection step}, which determines the entities worth comparing,
and (ii) the \textit{candidate matching step}, or simply \textit{Matching}, which compares the selected entities to determine whether they represent the same real-world object. The latter step 
involves \textit{pairwise comparisons}, i.e., time-consuming operations that typically apply string similarity measures to pairs of entities, dominating the overall cost of ER \cite{DBLP:books/daglib/0030287,DBLP:series/synthesis/2015Christophides,DBLP:series/synthesis/2015Dong}. 

In this survey, we focus on the 
candidate selection step, which is the crucial part of ER with respect to time efficiency and scalability. Without it, ER suffers from a quadratic time complexity, $O(n^2)$, as every entity profile has to be compared with all others. Reducing this computational cost is the goal of numerous techniques from
two dominant frameworks: Blocking and Filtering. The former 
attempts to identify 
entity pairs that are likely to match,
restricting comparisons only between them, while
the latter 
attempts to quickly discard pairs that are guaranteed to not match, 
executing comparisons only between the rest.
The former operates without knowledge of the 
Matching step, while the latter is based on 
it, 
assuming that two entities match if their similarity exceeds a specified threshold.
Hence, Blocking and Filtering 
share the same goal, but are complementary, as they operate under different settings and assumptions. 
So far, though,
they have been developed independently of one another: their combination and, more generally, their relation 
have been overlooked in the literature, with the exception of very few works (e.g., \cite{DBLP:journals/pvldb/KopckeTR10}).

Moreover, the rise of Big Data poses new challenges for both Blocking and Filtering approaches \cite{DBLP:series/synthesis/2015Dong,DBLP:series/synthesis/2015Christophides}:  \textit{Volume} requires techniques to scale to millions of entities, while \textit{Variety} calls for techniques that can cope with an unprecedented schema heterogeneity. Both Blocking and Filtering address Volume primarily through paralellization. Existing techniques were adapted to split their workload into smaller chunks that are distributed across different processing units so that they are executed in parallel. This can be done on a cluster (distributed methods), or through the modern multi-core and multi-socket hardware architectures. Variety, though, is addressed differently in each field. For Blocking, the schema-aware methods are replaced by schema-agnostic techniques, which disregard any schema information, creating blocks of very high recall but low precision. Additionally, a whole new category of methods, called \textit{Block Processing}, intervenes between Blocking and Matching 
to refine the original blocks, 
significantly increasing precision at a negligible (if any) cost in recall. For Filtering, techniques that employ more relaxed matching criteria (e.g., fuzzy set matching or local string similarity join) are proposed, while the case of low similarity thresholds~is~also~considered. 

To the best of our knowledge, this is the first survey to comprehensively cover the aforementioned aspects and to jointly review the two frameworks for efficient ER. We formally define Blocking, Block Processing and Filtering, introducing a common terminology that facilitates their understanding. For each field, we propose a new taxonomy with categories that highlight the distinguishing characteristics of the corresponding methods. Based on these taxonomies, we provide a broad overview of every field, elucidating the functionality of the main techniques as well as the relations among them. As a result, established techniques are now seen in a different light - Canopy Clustering~\cite{DBLP:conf/kdd/McCallumNU00}, for instance, may now be viewed as a Block Processing method.
We also elaborate on the parallelization methods for each field. 
Most importantly, this survey attempts to 
place Blocking and Filtering under a common context, 
taking special care to stress hybrid methods that combine features from both Blocking and Filtering, to analyze works that experimentally compare the two frameworks 
(e.g., \cite{DBLP:conf/semweb/SongH11}) 
and to qualitatively outline their commonalities and differences.
We also investigate the ER tools that incorporate established efficiency techniques and propose a series of open challenges that constitute promising directions for future research.

Parts of the material included in this survey have been presented in tutorials at WWW 2014~\cite{DBLP:conf/www/StefanidisEHC14}, ICDE 2016~\cite{7498364}, ICDE 2017~\cite{DBLP:conf/icde/StefanidisCE17}, and WWW 2018~\cite{PapadakisTutorialWww18}.
A past survey \cite{DBLP:journals/tkde/Christen12} also covers efficiency ER techniques, but is restricted to the schema-aware Blocking methods. 
Other surveys \cite{DBLP:journals/tkde/ElmagarmidIV07} and textbooks \cite{DBLP:books/daglib/0030287,DBLP:series/synthesis/2015Dong,DBLP:series/synthesis/2015Christophides} provide a holistic overview of ER, merely examining the main Blocking and Block Processing techniques. Closer to our work is a recent survey on Blocking \cite{o2019review}, which however offers a more limited coverage and refers neither to parallelization nor to Filtering works. 
Recent surveys on string and set similarity joins also exist, but 
they focus exclusively on 
centralized \cite{DBLP:journals/pvldb/JiangLFL14,DBLP:journals/fcsc/YuLDF16,DBLP:journals/pvldb/MannAB16} or distributed approaches \cite{DBLP:journals/pvldb/FierABLF18}, with the purpose of experimental comparison, and without covering approximate techniques 
that allow for more relaxed matching criteria. Most importantly, none of these surveys considers similarity joins in the broader context of ER.

The rest of the paper is structured as follows: Section \ref{sec:er} provides background knowledge on ER and its efficiency techniques, while Sections \ref{sec:blocking} and \ref{sec:blockProcessing} delve into Blocking and Block Processing, respectively. Section \ref{sec:filtering} is devoted to Filtering, whereas Section \ref{sec:hybrid} elaborates on works that combine Blocking with Filtering. 
Section \ref{sec:tools} enumerates the main ER tools that incorporate efficiency methods, Section \ref{sec:discussion} provides a high-level discussion of the relation between Blocking and Filtering, Section \ref{sec:futureDirections} provides the main directions for future work, and Section \ref{sec:conclusions} concludes the paper.

\section{Preliminaries}
\label{sec:er}

At the core of ER lies the notion of \textit{entity profile}, which constitutes a uniquely identified description of a real-world object in the form of name-value pairs. Assuming infinite sets of attribute names $\mathcal{N}$, attribute values $\mathcal{V}$, and unique identifiers $\mathcal{I}$, an entity profile is formally defined~as~follows~\cite{DBLP:series/synthesis/2015Christophides,DBLP:journals/tkde/PapadakisIPNN13}:

\begin{definition}[Entity Profile]
  An \emph{entity profile} $\mathbf{e_{id}}$ is a tuple $\langle id, A_{id} \rangle$, where $id \in \mathcal{I}$ is a unique identifier, and $A_{id}$ is a set of name-value pairs $\langle n, v \rangle$, with $n \in \mathcal{N}$ and $v \in (\mathcal{V} \cup \mathcal{I})$. A set of entity profiles $\mathbf{\mathcal{E}}$ is called \emph{entity collection}.
\end{definition}

This definition is simple, but flexible enough to accommodate a wide variety of (semi-)structured
representations. 
E.g., nested attributes can be transformed into a flat set of name-value pairs, while links 
may be represented by assigning the id of one entity as the attribute value of the other.

\begin{definition}[Entity Resolution]
    Two entity profiles $e_i$ and $e_j$ \emph{match}, $\mathbf{e_i\equiv e_j}$, if they refer to the same real-world entity.
    Matching entities are also 
    called \emph{duplicates}. The task of Entity Resolution (ER) is to find all matching entities within an entity collection or across two or more entity collections.
\end{definition}

In particular, we distinguish between the following two cases \cite{DBLP:journals/tkde/Christen12,DBLP:books/daglib/0030287}:
\begin{enumerate}
	\item \textit{Deduplication} receives as input an entity collection $\mathcal{E}$ and produces as output the set of all pairs of matching entity profiles within $\mathcal{E}$, i.e., $\mathcal{D}(\mathcal{E}) = \{ (e_i, e_j) : e_i \in \mathcal{E}, \, e_j \in \mathcal{E}, \, e_i\equiv e_j \}$.
	\item \textit{Record Linkage} receives 
	two duplicate-free entity collections, $\mathcal{E}_1$ and $\mathcal{E}_2$, 
	and 
	returns the pairs of matching entity profiles between them, i.e., $\mathcal{D}(\mathcal{E}_1$, $\mathcal{E}_2)$=$\{ (e_i, e_j) : e_i \in \mathcal{E}_1, \, e_j \in \mathcal{E}_2, \, e_i\equiv e_j \}$. 
\end{enumerate}
 
\textit{Multi-source Entity Resolution} involves three or more entity collections and can be performed by applying Deduplication to the union of all collections, or by executing a sequence of pairwise Record Linkage tasks, provided that every input collection is duplicate-free.

ER performance 
is characterized by 
its \textit{effectiveness} and its \textit{efficiency}. The former refers to how many of the actual duplicates are detected, while the latter expresses the computational cost for detecting them -- usually
in terms of the number of performed comparisons,
which is referred to as \textit{cardinality} and 
denoted by $||\mathcal{E}||$. The naive, brute-force approach performs all pairwise comparisons between the input entity profiles, having a quadratic complexity that does not scale to large datasets;
for Record Linkage, $||\mathcal{E}|| = |\mathcal{E}_1| \times |\mathcal{E}_2|$, while for Deduplication $||\mathcal{E}|| = |\mathcal{E}| \cdot (|\mathcal{E}| -1)/2$. 

\vspace{2pt}
\textbf{Blocking.} To tackle ER's inherently quadratic complexity, Blocking 
trades slightly lower effectiveness for significantly higher efficiency. Its goal is to reduce the number of performed comparisons, while missing as few matches as possible. Ideally, one would compare only the pairs of duplicates, whose number grows \textit{linearly} with the number of the input entity profiles~\cite{DBLP:journals/pvldb/GetoorM12,DBLP:conf/icde/StefanidisCE17}. To this end, Blocking clusters potentially matching entities in common blocks 
and exclusively compares entity profiles that co-occur in at least one block.

Internally, a blocking method employs a \textit{blocking scheme}, which applies to one or more entity collections to yield a set of blocks $\mathcal{B}$, 
called
\textit{block collection}. Cardinality $||\mathcal{B}||$ 
denotes the number of comparisons in $\mathcal{B}$, 
given that only entity pairs within the same block are compared, i.e., $||\mathcal{B}||$=$\sum_{b_i \in \mathcal{B}} ||b_i||$, where $||b_i||$ stands for the number of comparisons contained in an individual block $b_i$. We denote the set of \textit{detectable duplicates} in $\mathcal{B}$ as $\mathcal{D}(\mathcal{B})$, while $\mathcal{D}(\mathcal{E})$ stands for all existing duplicates. Since $\mathcal{B}$ reduces the number of performed comparisons,~$\mathcal{D}(\mathcal{B})$$\subseteq$$\mathcal{D}(\mathcal{E})$.

A common assumption in the literature is the \textit{oracle}, i.e., a perfect matching function that, for each pair of entity profiles,
decides correctly whether they match or not \cite{DBLP:conf/icde/StefanidisCE17,DBLP:journals/tkde/Christen12,DBLP:series/synthesis/2015Dong,DBLP:journals/tkde/PapadakisIPNN13,DBLP:journals/tkde/PapadakisKPN14}.
Using an oracle, 
a pair of duplicates is detected as long as they share at least one block. This allows for reasoning about the performance of blocking methods independently of matching methods: there is a clear trade-off between the effectiveness and the efficiency of a blocking scheme \cite{DBLP:conf/icde/StefanidisCE17,DBLP:journals/tkde/Christen12,DBLP:series/synthesis/2015Dong}: the more comparisons are contained in the resulting block
collection $\mathcal{B}$ (i.e., higher $||\mathcal{B}||$), the more duplicates
will be detected (i.e., higher $|\mathcal{D}(\mathcal{B})|$), raising 
effectiveness 
at the cost of lower efficiency.
Thus, a blocking scheme 
should achieve a good balance between these two competing objectives as expressed through
the following 
measures~\cite{DBLP:conf/icdm/BilenkoKM06,DBLP:conf/cikm/VriesKCC09,DBLP:conf/aaai/MichelsonK06,DBLP:conf/wsdm/PapadakisINF11}:

\begin{enumerate}

 \item \textit{Pair Completeness ($PC$)} corresponds to \textit{recall}, estimating the portion of the detectable duplicates in $\mathcal{B}$ with respect to those in $\mathcal{E}$: 
 $PC(\mathcal{B}) = |\mathcal{D}(\mathcal{B})| / |\mathcal{D}(\mathcal{E})| \in [0,1]$.

 \item \textit{Pairs Quality ($PQ$)} corresponds to \textit{precision}, estimating the portion of 
 comparisons in $\mathcal{B}$ that correspond to real duplicates:
 $PQ(\mathcal{B})= |\mathcal{D}(\mathcal{B})|/||\mathcal{B}|| \in [0,1]$.

 \item \textit{Reduction Ratio ($RR$)} measures the reduction in the number of pairwise comparisons 
 in $\mathcal{B}$ with respect to the brute-force approach:
 $RR(\mathcal{B},\mathcal{E}) = 1 - ||\mathcal{B}||/||\mathcal{E}|| \in [0, 1]$..
\end{enumerate}

Higher values for $PC$ indicate higher \textit{effectiveness} of the blocking scheme, while higher values for $PQ$ and $RR$ indicate 
higher \textit{efficiency}.
Note that $PC$ provides an optimistic estimation of recall, presuming
the existence of an oracle, while $PQ$ provides a pessimistic estimation
of precision, treating as false positives the repeated comparisons between duplicates (i.e., only the non-repeated duplicate pairs are considered as true positives).
In this context, 
we can define Blocking as follows:

\begin{definition}[Blocking]
  Given an entity collection $\mathcal{E}$, Blocking clusters similar entities into a block collection $\mathcal{B}$ such that $PC(\mathcal{B})$, $PQ(\mathcal{B})$ and $RR(\mathcal{B}, \mathcal{E})$ are simultaneously maximized.
  \label{def:blocking}
\end{definition}

This definition refers to Deduplication, but can be easily extended to Record Linkage. Simultaneously maximizing $PC$, $PQ$ and $RR$ necessitates that the enhancements in efficiency do not affect the effectiveness of Blocking, 
carefully removing comparisons between non-matching entities. 
Conceptually, Blocking can be viewed as an optimization task, but
this implies that the real duplicate collection $\mathcal{D}(\mathcal{E})$ is known, which is actually what ER tries to compute. Hence, 
Blocking is typically treated as an engineering task that 
provides an approximate solution for the data at hand.

\begin{figure}[t]\centering
	\includegraphics[width=0.75\linewidth]{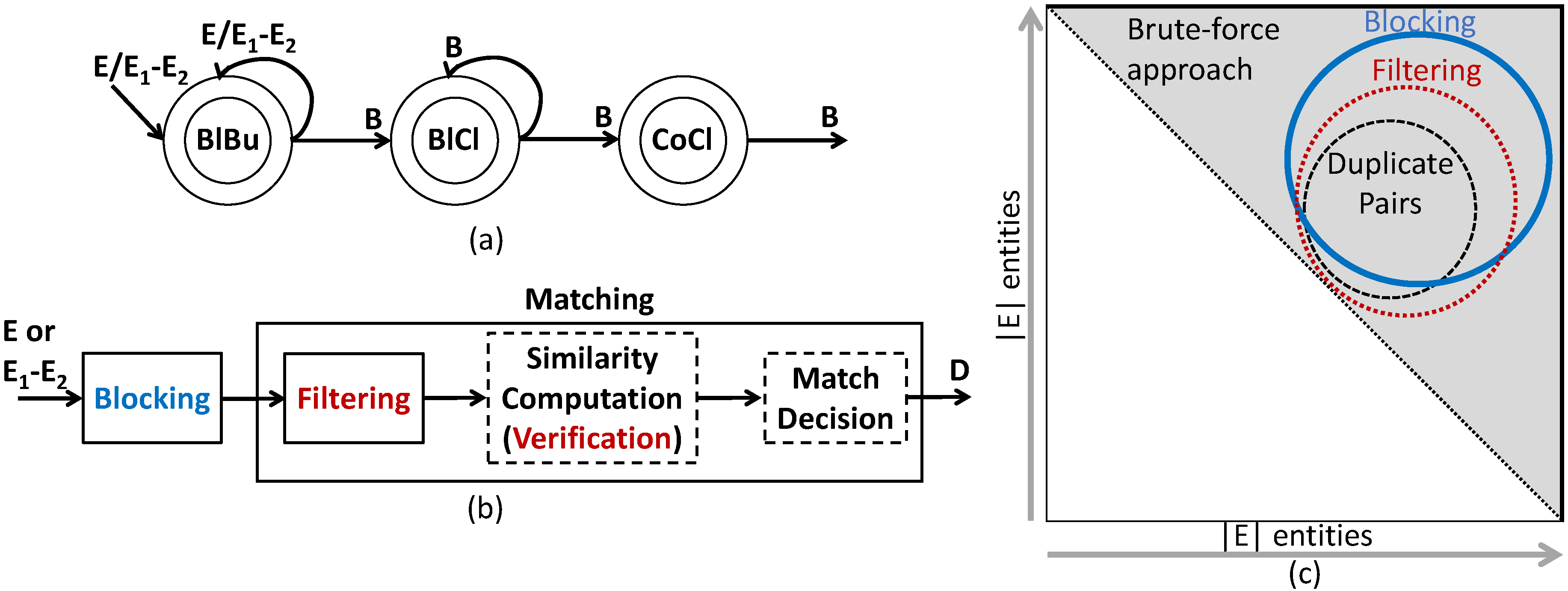}
	\vspace{-9pt}
	\caption{{\small (a) The internal functionality of Blocking modeled as a deterministic finite automaton with three states: Block Building (\textsf{BlBu}), Block Cleaning (\textsf{BlCl}) and Comparison Cleaning (\textsf{CoCl}). (b) The end-to-end workflow for non-learning Entity Resolution \cite{DBLP:journals/pvldb/KopckeTR10}. (c) The relative computational cost for the brute-force approach, Blocking, Filtering and the ideal solution (Duplicate Pairs) over Deduplication.}
	}
	\label{fig:computationalCostPlusWorkflow}
	\vspace{-10pt}
\end{figure}

A blocking-based ER workflow may comprise several stages. 
First, \textit{Block Building} (BlBu)
applies a blocking scheme to produce a block collection $\mathcal{B}$ from the input entity collection(s). 
This step may be repeated several times on the same input,
applying multiple blocking schemes, in order to achieve a more robust performance in the context of highly noisy data. Often, there is a second, optional stage, called \textit{Block Processing}, which
refines $\mathcal{B}$ through additional optimizations that further reduce the number of performed comparisons. This may involve discarding \textit{entire blocks} that primarily contain unnecessary comparisons, 
called \textit{Block Cleaning} (BlCl), and/or discarding \textit{individual comparisons} within certain blocks, 
called \textit{Comparison Cleaning} (CoCl). 
The former may be applied repeatedly, each time enforcing a different, complementary method to discard blocks, but the latter
can be performed only once;
CoCl comprises competitive methods that 
serve exactly the same purpose and, once applied to a block collection, they alter it in such a way that turns all other methods inapplicable.
Figure \ref{fig:computationalCostPlusWorkflow}(a) models this workflow as a deterministic finite automaton with three states, where each state corresponds to one of the blocking sub-tasks. 

\vspace{2pt}
\textbf{Filtering.} 
Given two entity collections $\mathcal{E}_1$ and $\mathcal{E}_2$, a similarity function $f_S : \mathcal{E}_1 \times \mathcal{E}_2 \rightarrow {\rm I\!R}$, and a similarity threshold $\theta$, a \textit{similarity join} identifies all pairs of entity profiles in $\mathcal{E}_1$ and $\mathcal{E}_2$ that have similarity at least $\theta$, i.e., $\mathcal{E}_1 \Join_{\theta} \mathcal{E}_2 = \{ (e_i, e_j) \in \mathcal{E}_1 \times \mathcal{E}_2 : f_S(e_i, e_j) \geq \theta \}$.

\begin{figure}[t]\centering
	\includegraphics[width=0.69\linewidth]{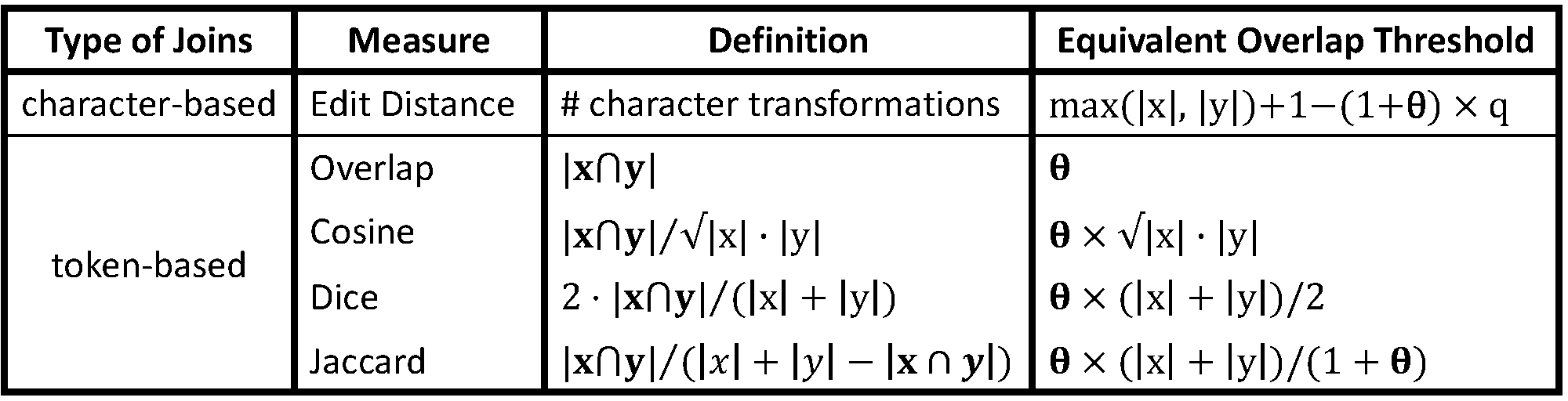}
	\vspace{-10pt}
	\caption{{\small Definition of the main similarity measures used by string similarity join algorithms, and how the input threshold $\theta$ for each measure can be transformed into an equivalent Overlap threshold $\tau$.} 
	}
	\label{fig:measures}
	\vspace{-14pt}
\end{figure}

Similarity joins can be used for defining ER under the intuitive assumption that matching entity profiles are highly similar. In fact, the above formulation corresponds to Record Linkage, while Deduplication can be defined analogously as a self-join operation, where $\mathcal{E}_1 \equiv \mathcal{E}_2$.  

To avoid exhaustive pairwise comparisons, 
similarity joins typically follow the \textit{filter-verification} framework, which involves two~stages~\cite{DBLP:series/synthesis/2013Augsten,DBLP:journals/pvldb/JiangLFL14}:

\begin{enumerate}
    \item \textit{Filtering} computes a set of \textit{candidates} for each entity $e_i$, excluding all those that cannot match with $e_i$. In other words, it prunes all true negatives, but allows some false positives. 
    \item \textit{Verification} computes the actual similarity between candidates (or a sufficient upper bound) to remove the false positives.
\end{enumerate}
 
Due to the relatively straightforward implementation of Verification, in the following we exclusively focus on Filtering. The relevant techniques are defined with respect to three parameters: (i) the representation for each entity, (ii) the similarity function between entity pairs under this representation, and (iii) the similarity threshold above which two entities are considered to match.

The representation typically relies on 
\textit{signatures} extracted from each entity such that two entities match only if their signatures overlap. 
Given that we address ER 
over entities described by one or more textual attributes, we focus on string similarity joins, which 
can be \textit{character-} or \textit{token-based}. The former compare two strings by representing them as sequences of characters and by considering the character transformations required to transform one string into the other. The latter are also called \textit{set similarity joins}, since they transform the strings into sets, typically via tokenization or $q$-gram extraction, and then compare strings using a set-based similarity measure. 

Regarding the similarity function, the most common one for character-based similarity joins is Edit Distance,
which measures the minimum number of edit operations (i.e., insertions, deletions and substitutions) that are required to transform one string to the other \cite{DBLP:series/synthesis/2013Augsten}. For token-based similarity joins, the most commonly used similarity measures include Overlap, Jaccard, Cosine or Dice. The last three are normalized variants of the Overlap \cite{DBLP:series/synthesis/2013Augsten,DBLP:journals/pvldb/MannAB16,DBLP:journals/pvldb/JiangLFL14}. 

Finally, the similarity threshold depends on the data at hand. Note, though, that the join algorithms do not operate directly with thresholds on Jaccard, Cosine or Dice similarity, but 
first translate the given threshold $\theta$ into an equivalent set overlap threshold $\tau$ that depends 
on the size of the sets, as shown in Figure~\ref{fig:measures}. 
A similar transformation is also possible for Edit Distance, which means 
that 
set similarity joins
may be applied to this measure as well \cite{DBLP:series/synthesis/2013Augsten}.

\vspace{2pt}
\textbf{Blocking vs Filtering.} The relation between the two frameworks is illustrated in Figure \ref{fig:computationalCostPlusWorkflow}(b). Blocking, in the sense of the entire process in Figure \ref{fig:computationalCostPlusWorkflow}(a), is applied first, reducing the pairwise comparisons that are considered by Matching. These comparisons are further cut down by Filtering, which is subsequently applied, as the initial part of Matching, given that it requires specifying both a similarity measure and a similarity threshold. Next, Verification is applied to estimate the actual similarity between the compared attribute values. The Entity Resolution process concludes with \textit{Match Decision}, which synthesizes the estimated similarity between multiple attribute values to determine whether the compared entity profiles are indeed duplicates. 

Both Blocking and Filtering are optional steps, but at least one of them should be applied in order to tame the otherwise quadratic computational cost of ER. As shown in Figure~\ref{fig:computationalCostPlusWorkflow}(c), Blocking yields a \textit{super-linear}, but \textit{sub-quadratic} time complexity, lying between the two extremes: the brute-force solution and the ideal one (i.e., Duplicate Pairs). The same applies to the computational cost of Filtering, except that it typically constitutes an \textit{exact} procedure that produces no false negatives, i.e., missed duplicates. It exclusively allows false positives, which are later removed by Verification \cite{DBLP:series/synthesis/2013Augsten}. For this reason, Filtering corresponds to a superset of Duplicate Pairs in Figure~\ref{fig:computationalCostPlusWorkflow}(c). In contrast, Blocking constitutes an inherently \textit{approximate} solution that increases ER efficiency
at the cost of allowing both false positives and false negatives
\cite{DBLP:series/synthesis/2015Christophides}. 
Thus, it intersects Duplicate Pairs, such that the area of their intersection is inversely proportional to the duplicates that are missed by Blocking, while the relative complement of the Duplicate Pairs in Blocking is analogous to the executed comparisons between non-matching entities. 

Note that Figure~\ref{fig:computationalCostPlusWorkflow}(c) corresponds to Deduplication, but can be easily generalized to Record Linkage, as well. Moreover, the relative performance of Blocking and Filtering, i.e., the relative position of their circles, depends on the methods and the data at hand. In most cases, though, the best solution is to use both frameworks, yielding the computational cost that corresponds to their intersection. However, this approach is rarely used in the literature (e.g., \cite{DBLP:journals/pvldb/KopckeTR10}). Most works on Blocking typically omit Filtering (e.g., \cite{DBLP:journals/pvldb/0001APK15,DBLP:journals/pvldb/0001SGP16,DBLP:journals/tkde/Christen12}), whereas most works on Filtering disregard Blocking, applying directly to the input entity collections (e.g., \cite{DBLP:journals/pvldb/MannAB16,DBLP:journals/pvldb/JiangLFL14}). The goal of the present survey is to cover this gap, elucidating the complementarity of the two frameworks.

\section{Block Building}
\label{sec:blocking}

Block Building receives as input one or more entity collections and produces as output a block collection $\mathcal{B}$. The process is guided by a \textit{blocking scheme}, which determines how entity profiles are assigned to blocks. This scheme typically comprises two parts. First, every entity is processed to extract \textit{signatures} (e.g., tokens),
such that the similarity of signatures reflects the similarity of the corresponding 
profiles. Second, every entity is 
mapped to one or more blocks based on these signatures. Let $\mathcal{P(S)}$ denote the power set of a set $S$ and $\mathcal{K}$ denote the universe of signatures appearing in entity profiles. We formally define a blocking scheme as follows:

\begin{definition}[Blocking Scheme]
    Given an entity collection $\mathcal{E}$, a \emph{blocking scheme} is a function $f_B : \mathcal{E} \rightarrow \mathcal{P}(\mathcal{B})$ that maps entity profiles to blocks. It is composed of two functions: (a) a \emph{transformation} function $f_{T} : \mathcal{E} \rightarrow \mathcal{P}(\mathcal{K})$ that maps an entity profile to a set of \emph{signatures} (also called \emph{blocking keys}), and (b) an \emph{assignment} function $f_{A} : \mathcal{K} \rightarrow \mathcal{P}(\mathcal{B})$ that maps each signature to one or more blocks.
\end{definition}

This definition applies to Deduplication, but can be easily extended to Record Linkage. 

\begin{table*}[h]
\centering
\caption{Taxonomy of the Block Building methods discussed in Sections \ref{sec:schemaBasedBB} and \ref{sec:schemaAgnosticBB}.}
\label{tb:bbTaxonomy}
\vspace{-5pt}
{\scriptsize
 \begin{tabular}{| l || c | c | c | c | } 
 \hline
 \multicolumn{1}{|c||}{\textbf{Method}} & \textbf{Key} & \textbf{Redundancy} & \textbf{Constraint} & \textbf{Matching} \\
 & \textbf{type} & \textbf{awareness} & \textbf{awareness} & \textbf{awareness} \\
 \hline
 
 \hline
Standard Blocking (\textsf{SB}) \cite{fellegi1969theory} &  hash-based & redundancy-free & lazy & static \\
Suffix Arrays Blocking (\textsf{SA}) \cite{DBLP:conf/wiri/AizawaO05} &  hash-based & redundancy-positive & proactive & static \\
Extended Suffix Arrays Blocking \cite{DBLP:journals/tkde/Christen12,DBLP:journals/pvldb/0001APK15} &  hash-based & redundancy-positive & proactive & static \\
Improved Suffix Arrays Blocking \cite{DBLP:conf/cikm/VriesKCC09} &  hash-based & redundancy-positive & proactive & static \\  
Q-Grams Blocking \cite{DBLP:journals/tkde/Christen12,DBLP:journals/pvldb/0001APK15} &  hash-based & redundancy-positive & lazy & static \\
Extended Q-Grams Blocking \cite{baxter2003comparison,DBLP:journals/tkde/Christen12,DBLP:journals/pvldb/0001APK15} &  hash-based & redundancy-positive & lazy & static \\
MFIBlocks \cite{DBLP:journals/is/KenigG13} &  hash-based & redundancy-positive & proactive & static \\
\hline
Sorted Neighborhood (\textsf{SN}) \cite{DBLP:conf/sigmod/HernandezS95, DBLP:journals/datamine/HernandezS98,DBLP:conf/edbt/PuhlmannWN06} &  sort-based & redundancy-neutral & proactive & static \\
Extended Sorted Neighborhood \cite{DBLP:journals/tkde/Christen12} &  sort-based & redundancy-neutral & lazy & static \\
Incrementally Adaptive SN \cite{DBLP:conf/jcdl/YanLKG07} &  sort-based & redundancy-neutral & proactive & static \\
Accumulative Adaptive SN \cite{DBLP:conf/jcdl/YanLKG07} &  sort-based & redundancy-neutral & proactive & static \\
Duplicate Count Strategy (\textsf{DCS}) \cite{DBLP:conf/icde/DraisbachNSW12} &  sort-based & redundancy-neutral & proactive & dynamic \\
\textsf{DCS++} \cite{DBLP:conf/icde/DraisbachNSW12} &  sort-based & redundancy-neutral & proactive & dynamic \\
\hline
Sorted Blocks \cite{DBLP:conf/nss/DraisbachN11} &  hybrid & redundancy-neutral & lazy & static \\
Sorted Blocks New Partition \cite{DBLP:conf/nss/DraisbachN11} &  hybrid & redundancy-neutral & proactive & static \\
Sorted Blocks Sliding Window \cite{DBLP:conf/nss/DraisbachN11} &  hybrid & redundancy-neutral & proactive & static \\
\hline
\multicolumn{5}{c}{\textbf{(a) Non-learning, schema-aware methods.}}\\
\hline

ApproxRBSetCover \cite{DBLP:conf/icdm/BilenkoKM06} &  hash-based & redundancy-positive & lazy & static \\
ApproxDNF \cite{DBLP:conf/icdm/BilenkoKM06} &  hash-based & redundancy-positive & lazy & static \\
Blocking Scheme Learner (\textsf{BSL}) \cite{DBLP:conf/aaai/MichelsonK06} &  hash-based & redundancy-positive & lazy & static \\
Conjunction Learner \cite{DBLP:conf/ijcai/CaoCZYLY11} (semi-supervised) &  hash-based & redundancy-positive & lazy & static \\
\textsf{BGP} \cite{DBLP:journals/jidm/EvangelistaCSM10} &  hash-based & redundancy-positive & lazy & static \\
CBlock \cite{DBLP:conf/cikm/SarmaJMB12} &  hash-based & redundancy-positive & proactive & static \\
DNF Learner \cite{giang2015machine} &  hash-based & redundancy-positive & lazy & dynamic \\
\hline
FisherDisjunctive \cite{DBLP:conf/icdm/KejriwalM13} (unsupervised) &  hash-based & redundancy-positive & lazy & static \\
\hline
\multicolumn{5}{c}{\textbf{(b) Learning-based (supervised), schema-aware methods.}}\\
\hline

Token Blocking (\textsf{TB}) \cite{DBLP:conf/wsdm/PapadakisINF11} &  hash-based & redundancy-positive & lazy & static \\
Attribute Clustering Blocking \cite{DBLP:journals/tkde/PapadakisIPNN13} &  hash-based & redundancy-positive & lazy & static  \\
RDFKeyLearner \cite{DBLP:conf/semweb/SongH11} &  hash-based & redundancy-positive & lazy & static \\
Prefix-Infix(-Suffix) Blocking \cite{DBLP:conf/wsdm/PapadakisINPN12} &  hash-based & redundancy-positive & lazy & static \\
TYPiMatch \cite{DBLP:conf/wsdm/MaT13} &  hash-based & redundancy-positive & lazy & static \\
Semantic Graph Blocking \cite{DBLP:conf/ideas/NinMML07} &  - & redundancy-neutral & proactive & static \\
\hline
\multicolumn{5}{c}{\textbf{(c) Non-learning, schema-agnostic methods.}}\\
\hline

Hetero \cite{DBLP:conf/semweb/KejriwalM14a} & hash-based & redundancy-positive & lazy & static \\
Extended DNF BSL \cite{DBLP:journals/corr/KejriwalM15} & hash-based & redundancy-positive & lazy & static \\
\hline
\multicolumn{5}{c}{\textbf{(d) Learning-based (unsupervised), schema-agnostic methods.}}
\end{tabular}
 }
\vspace{-12pt}
\end{table*}

The set of comparisons in the resulting block collection $\mathcal{B}$ is called \textit{comparison collection} and is denoted by $\mathcal{C}(\mathcal{B})$. Every comparison $c_{i,j} \in \mathcal{C}(\mathcal{B})$ belongs to one of the following types~\cite{DBLP:journals/tkde/PapadakisKPN14,DBLP:journals/tkde/PapadakisIPNN13}:

\begin{itemize}
  	\item \textit{Matching comparison}, if $e_i$ and $e_j$ match.
    \item \textit{Superfluous comparison}, if $e_i$ and $e_j$ do not match.
	\item \textit{Redundant comparison}, if $e_i$ and $e_j$ have already been compared in a previous block.
\end{itemize}

We collectively call the last two types \textit{unnecessary comparisons}, as their execution brings no gain.

Note that the resulting block collection $\mathcal{B}$ can be modelled as an inverted index that points from block ids to entity ids.
For this reason, Block Building is also called \textit{Indexing} 
\cite{DBLP:journals/tkde/Christen12,DBLP:books/daglib/0030287}. 

\subsection{Taxonomy}
\label{sec:taxonomy}

To facilitate the understanding of the main methods for Block Building, we organize them into a novel taxonomy 
that consists of the following dimensions:

\begin{itemize}
  \item \textit{Key selection} distinguishes between \textit{non-learning} and \textit{learning-based} methods. The former 
      rely on rules derived from expert knowledge or mere heuristics, while the latter 
      require a training set to learn the best blocking keys using Machine Learning techniques.
 \item \textit{Schema-awareness} distinguishes between \textit{schema-aware} and \textit{schema-agnostic} methods. The former 
 extract blocking keys from specific attributes that are considered to be more appropriate for matching (e.g., more distinctive or less noisy), while the latter 
 disregard schema knowledge, extracting blocking keys from all attributes.
 
 \item \textit{Key type} distinguishes between \textit{hash-} or \textit{equality-based} methods, which map a pair of entities to the same block if they have a common key, and \textit{sort-} or \textit{similarity-based} methods, which map a pair of entities to the same block if they have a similar key. There exist also \textit{hybrid} methods, which combine hash- with sort-based functionality.
 
\item \textit{Redundancy-awareness} classifies methods into three categories based on the relation between their blocks.
\textit{Redundancy-free} methods assign every entity to a single block, thus creating disjoint blocks. \textit{Redundancy-positive} methods place every entity into multiple blocks, yielding overlapping blocks. The more blocks two entities share, the more similar their profiles are. The number of blocks shared by a pair of entities is thus proportional to their matching likelihood. \textit{Redundancy-neutral} methods create overlapping blocks, where most entity pairs share the same number of blocks, or the degree of redundancy is arbitrary, having no implications.
 
 \item \textit{Constraint-awareness} distinguishes blocking methods into \textit{lazy}, which impose no constraints on the blocks they create, and \textit{proactive}, which enforce 
 constraints on their blocks~(e.g., maximum block size), or 
 refine their comparisons by discarding 
 unnecessary ones.
 
 \item \textit{Matching-awareness} distinguishes between \textit{static} methods, which are independent of the subsequent matching process, producing an immutable block collection, and \textit{dynamic} methods, which intertwine Block Building with Matching, updating or processing their blocks dynamically, as more duplicates are detected.
 
\end{itemize}

Table \ref{tb:bbTaxonomy} maps all methods discussed in Sections \ref{sec:schemaBasedBB} and \ref{sec:schemaAgnosticBB} to our taxonomy.

\subsection{Schema-aware Block Building}
\label{sec:schemaBasedBB}

Methods of this type assume that the input entity profiles adhere to a known schema and, based on this schema and respective domain knowledge, one can select the attributes that are most suitable for Blocking.
We distinguish between non-learning methods, reviewed in Section~\ref{sec:nonlearningBlBu}, and learning-based methods, reviewed in Section~\ref{sec:learningBlBu}.

\subsubsection{Non-learning Methods}
\label{sec:nonlearningBlBu}

\begin{figure}[t]\centering
	\includegraphics[width=0.86\linewidth]{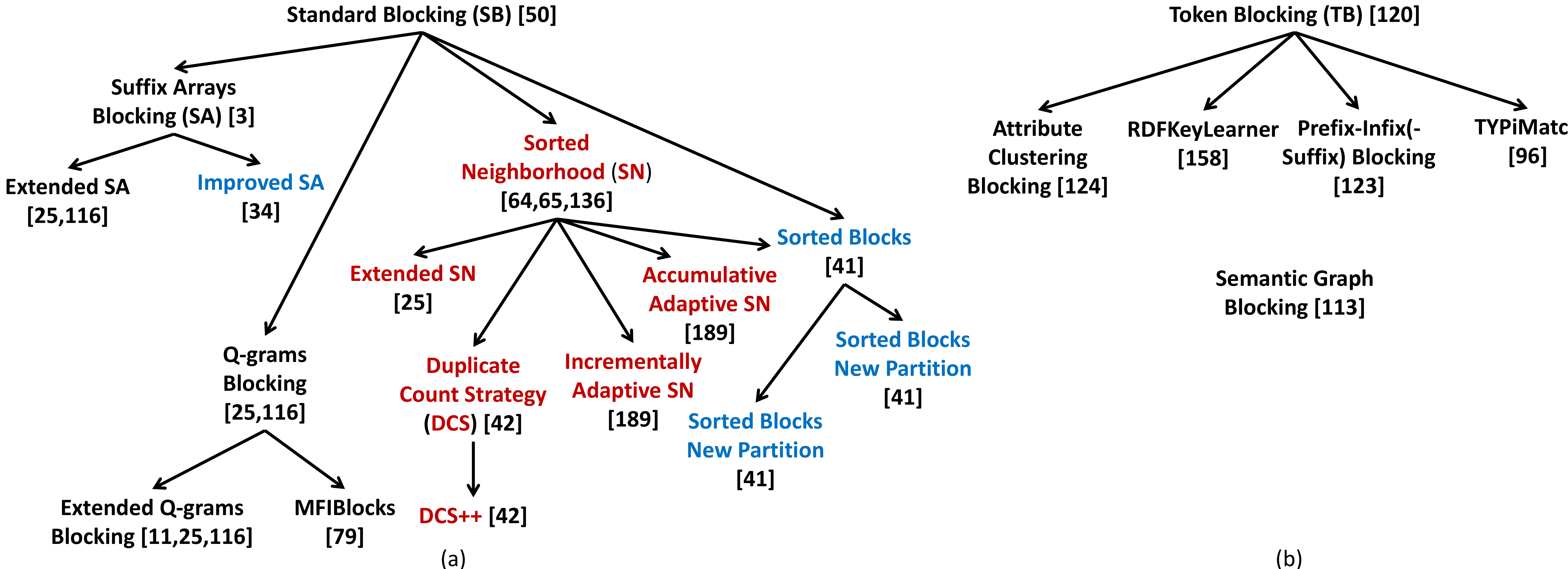}
	\vspace{-8pt}
	\caption{The genealogy trees of non-learning (a) schema-aware and (b) schema-agnostic Block Building techniques. Hybrid, hash- and sort-based methods are marked in {\color{blue}blue}, black and {\color{red}red}, respectively. 
	}
	\label{fig:schemaBasedBlocking}
	\vspace{-10pt}
\end{figure}

The family tree of the methods in this category is shown in Figure \ref{fig:schemaBasedBlocking}(a); a parent-child edge implies that the latter method improves upon the former one. Below, we elaborate on these methods based on their key type.

\textbf{Hash-based Methods.}
\textit{Standard Blocking} (\textsf{SB}) \cite{fellegi1969theory} 
involves the simplest 
functionality:
an expert selects the most suitable attributes, and a transformation function concatenates (parts of) their values to form blocking keys. For every distinct key, a block is created containing all corresponding entities. In short, \textsf{SB} operates as a hash function, conveying two main advantages: (i) it yields redundancy-free blocks, 
and (ii) it has a linear time complexity, $O(|E|)$.
On the flip side, its effectiveness is very sensitive to noise, 
as the slightest difference in the blocking keys of duplicates places them in different blocks.
\textsf{SB} is also a lazy method that imposes no limit on block sizes.

To address these issues, \textit{Suffix Arrays Blocking} (\textsf{SA})~\cite{DBLP:conf/wiri/AizawaO05} converts each blocking key of \textsf{SB} into the list of its suffixes that are longer than a predetermined minimum length $l_{min}$. Then, it defines a block for every suffix that does not exceed a predetermined frequency threshold $b_{max}$, which essentially specifies the maximum block size. This proactive functionality is necessary, as very frequent suffixes (e.g., ``ing") result in large blocks that are dominated by unnecessary comparisons. 

\textsf{SA} 
has two major advantages \cite{DBLP:conf/cikm/VriesKCC09}: (i) it has low time complexity, $O(|E|$$\cdot$$log|E|)$~\cite{DBLP:journals/dke/AllamSK18}, and is very efficient, as it results in a small but relevant set of candidate matches; (ii) it is very effective, due to the robustness to the noise at the beginning of blocking keys and the high levels of redundancy (i.e., it places every entity into multiple blocks). On the downside, \textsf{SA} does not handle noise at the end of \textsf{SB} keys. E.g., two matches with \textsf{SB} keys ``JohnSnith" and ``JohnSmith" have no common suffix if $l_{min}$=4, while for $l_{min}$=3, they co-occur in a block only if the frequency of ``ith" is lower than $b_{max}$.

This problem is addressed by \textit{Extended Suffix Arrays Blocking} \cite{DBLP:journals/tkde/Christen12,DBLP:journals/pvldb/0001APK15}, which uses as keys not just the suffixes
of \textsf{SB} keys,
but all their substrings with more than $l_{min}$ characters. E.g., for $l_{min}$=4, \textsf{SA} extracts from ``JohnSnith" the keys ``JohnSnith", ``ohnSnith", ``hnSnith", ``nSnith", ``Snith" and ``nith", while \textit{Extended SA} additionally extracts the keys ``John", ``ohnS", ``hnSn", ``nSni", ``Snit" as well as all substrings of ``JohnSnith" 
ranging from 5 to 8 characters.

Another extension of \textsf{SB} 
is \textit{Q-grams Blocking} \cite{DBLP:journals/tkde/Christen12,DBLP:journals/pvldb/0001APK15}. Its transformation function converts the blocking keys of \textsf{SB} into sub-sequences of $q$ characters (\textit{$q$-grams}) and defines a block for every distinct $q$-gram. For example, for $q$=3, the key \textit{france} is transformed into the trigrams \textit{fra}, \textit{ran}, \textit{anc}, \textit{nce}. This approach 
differs from \textsf{Extended SA} in that it does not restrict block sizes (lazy method). Also, it is  more resilient to noise than \textsf{SB}, 
but results in more and larger blocks.

To improve it, 
\textit{Extended Q-Grams Blocking} \cite{baxter2003comparison,DBLP:journals/tkde/Christen12,DBLP:journals/pvldb/0001APK15} uses combinations of $q$-grams, instead of individual $q$-grams. Its transformation function concatenates at least $l$ $q$-grams, where $l = max(1,\lfloor k \cdot t \rfloor)$, with $k$ denoting the number of $q$-grams and $t \in [0, 1)$ standing for a user-defined threshold. The larger $t$ is, the larger $l$ gets, yielding less keys from the $k$ $q$-grams. For $T=0.9$ and $q$=3, the key \textit{france} is transformed into the following four signatures ($k$=4 and $l$=3): [\textit{fra}, \textit{ran}, \textit{anc}, \textit{nce}], [\textit{fra}, \textit{ran}, \textit{anc}], [\textit{fra}, \textit{anc}, \textit{nce}], [\textit{ran}, \textit{anc}, \textit{nce}]. In this way, $q$-gram-based blocking keys become more distinctive, decreasing the number and cardinality of blocks. 

A more advanced $q$-gram-based approach is \textit{MFIBlocks} \cite{DBLP:journals/is/KenigG13}. Its transformation function concatenates keys of Q-Grams Blocking into itemsets and uses a maximal frequent itemset algorithm to define as new blocking keys those
exceeding a predetermined support threshold.

\textbf{Sort-based Methods.} 
\textit{Sorted Neighborhood} (\textsf{SN}) \cite{DBLP:conf/sigmod/HernandezS95}
sorts all blocking keys in alphabetical order and arranges the associated entities accordingly. Subsequently, a window of fixed size $w$ slides over the sorted list of entities and compares the entity at the last position with all other entities placed within the same window. The underlying assumption is that the closer the blocking keys of two entities are in the lexicographical order, the more likely they are to be matching. Originally crafted for relational data, \textsf{SN} is extended to hierarchical/XML data based on user-defined keys in \cite{DBLP:conf/edbt/PuhlmannWN06}.

\textsf{SN} has three major advantages \cite{DBLP:journals/tkde/Christen12}: (i) it has low time complexity, $O(|E|\cdot log |E|)$, (ii) it results in linear ER complexity, $O(w \cdot |E|)$, and (iii) it is robust to noise, supporting errors at the end of blocking keys. 
However, it may place two entities in the same block even if their keys are dissimilar (e.g., 
"alphabet" and "apple", 
if no other key intervenes between them).
Its performance also depends heavily on the window size $w$, which is difficult to configure,
especially in Deduplication, where the matching entities form clusters of varying size \cite{DBLP:conf/nss/DraisbachN11,DBLP:journals/tkde/Christen12}.

To ameliorate the effect of $w$,
a common solution is the \textit{Multi-pass SN}
\cite{DBLP:journals/datamine/HernandezS98}, which
applies the core algorithm multiple times, using a different transformation function in each iteration. In this way, more matches can be identified, even if the window is set to low size. Another solution is the \textit{Extended Sorted Neighborhood} \cite{DBLP:journals/tkde/Christen12,DBLP:journals/pvldb/0001APK15}, which slides a window of fixed size over the sorted list of blocking keys rather than the list of entities; this means that each block merges $w$ \textsf{SB} blocks. 

More advanced strategies adapt the window size dynamically 
to optimize the balance between effectiveness and efficiency.
They are grouped into  
three  
categories, depending on the criterion for moving the 
boundaries of the window \cite{DBLP:journals/cj/MaDY15}:

1) \textit{Key similarity strategy.} The window size increases if the similarity of the blocking keys exceeds a predetermined threshold, which indicates that more similar entities should be expected \cite{DBLP:journals/cj/MaDY15}.
    
2) \textit{Entity similarity strategy.} The window size relies on the similarity of the entities within the current window. \textit{Incrementally Adaptive SN} \cite{DBLP:conf/jcdl/YanLKG07} increases the window size if
the distance of the first and the last element in the window is smaller than a predetermined threshold. The actual increase 
depends on the current window size and the selected threshold. \textit{Accumulative Adaptive SN} \cite{DBLP:conf/jcdl/YanLKG07} creates windows with a single overlapping entity and exploits transitivity to group multiple adjacent windows into the same block, as long as the last entity of one window is a potential duplicate of the last entity in the next 
window. After expanding the window, both algorithms apply a retrenchment phase that decreases the window size until all entities 
are potential duplicates.

3) \textit{Dynamic strategy.} 
The core assumption is that the more duplicates are found within a window, the more are expected to be found by increasing its size. 
\textit{Duplicate Count Strategy} (\textsf{DCS}) \cite{DBLP:conf/icde/DraisbachNSW12} defines a window 
$w$ for every entity in \textsf{SN}'s sorted list and executes all its comparisons 
to compute the ratio $d/c$,
where $d$ denotes the newly detected duplicates and $c$ the executed comparisons. The window size is then incremented by one position at a time as long as $d/c \geq \phi$, where $\phi \in (0,1)$ is a threshold that expresses the average number of duplicates per comparison. \textsf{DCS++} \cite{DBLP:conf/icde/DraisbachNSW12} improves \textsf{DCS} by increasing the window size with the next $w-1$ entities, even if the new ratio becomes lower than $\phi$. 
Using transitive closure, 
it skips some windows, saving part of the comparisons. 

\textbf{Hybrid methods.} \textit{Sorted Blocks} \cite{DBLP:conf/nss/DraisbachN11} combines the benefits of \textsf{SB}
and \textsf{SN}.
First, it sorts all blocking keys and the corresponding entities in lexicographical order, like \textsf{SN}. Then, it partitions the sorted entities 
into disjoint blocks, like \textsf{SB}, using a prefix of the blocking keys. Next, all pairwise comparisons are executed within each block. To avoid missing any matches, an overlap parameter $o$ 
defines a 
window of fixed size that includes 
the $o$ last entities in the current block together with the first entity of the next block. The window slides by one position at a time until reaching the $o^{th}$ entity of the next block, executing
all pairwise comparisons between entities from different blocks.

Sorted Blocks is a lazy approach that does not restrict block sizes.
Thus, it may result in large blocks that dominate its processing time. To address this, two proactive variants set a limit on the maximum block size. \textit{Sorted Blocks New Partition} \cite{DBLP:conf/nss/DraisbachN11} 
creates a new block when the maximum 
size is reached for a 
(prefix of) blocking key; the overlap between the blocks ensures that every entity is compared with its predecessors and successors in the sorting order. \textit{Sorted Blocks Sliding Window} \cite{DBLP:conf/nss/DraisbachN11} avoids executing all comparisons within a block that is 
larger than the upper limit; instead, it slides a window 
equal to the maximum block size over the entities of the current block. 

Finally, \textit{Improved Suffix Arrays Blocking} \cite{DBLP:conf/cikm/VriesKCC09}
employs the same blocking keys as \textsf{SA}, but sorts them in alphabetical order, like \textsf{SN}. Then, it compares the consecutive keys with a string similarity measure. If the similarity of two suffixes exceeds a predetermined threshold, the corresponding blocks are merged 
in an effort to detect duplicates even when there is noise at the end of \textsf{SB} keys, or their sole common key is too frequent. For example, \textit{Improved SA} detects the high string similarity of the keys ``JohnSnith" and ``JohnSmith", placing the corresponding entities into the same block. 

\subsubsection{Learning-based Methods}
\label{sec:learningBlBu}

We distinguish these methods into supervised and unsupervised ones. 
Both rely on a labelled dataset that includes pairs of matching and non-matching entities, called \textit{positive} and \textit{negative instances}, respectively. This dataset is used to learn \textit{blocking predicates}, i.e., combinations of an attribute name and a transformation function (e.g., $\{title, First3Characters\}$). Entities sharing the same output for a particular blocking predicate are considered candidate matches (i.e., hash-based functionality). Disjunctions of conjunctions of predicates, i.e., composite blocking schemes, are learned by optimizing an objective function. 

\textbf{Supervised Methods.} 
\textit{ApproxRBSetCover} \cite{DBLP:conf/icdm/BilenkoKM06}
learns disjunctive 
blocking schemes by solving a standard weighted set cover problem. The cover is iteratively constructed by adding in each turn the blocking predicate
that maximizes the ratio of the previously uncovered positive pairs over the covered negative pairs. This is a "soft cover", since some positive instances may remain uncovered.

\textit{ApproxDNF} \cite{DBLP:conf/icdm/BilenkoKM06} alters ApproxRBSetCover so that it learns 
blocking schemes in Disjunctive Normal Form (DNF). Instead of individual predicates, each turn greedily learns a conjunction of up to $k$ predicates that maximizes the ratio of positive and negative covered instances.

A similar approach is \textit{Blocking Scheme Learner} (\textsf{BSL}) \cite{DBLP:conf/aaai/MichelsonK06}. Based on an adaptation of the
Sequential Covering Algorithm, it learns 
blocking schemes that maximize $RR$, while maintaining $PC$ above a predetermined threshold. Its output is a disjunction of conjunctions of blocking predicates.

\textsf{BSL} is improved by \textit{Conjunction Learner} \cite{DBLP:conf/ijcai/CaoCZYLY11}, which minimizes the candidate matches not only in the labelled, but also in the \textit{unlabelled} data, while maintaining high $PC$. The effect of the unlabelled data is determined through a weight $w \in [0,1]$; $w=0$ disregards unlabelled data completely, falling back to \textsf{BSL}, while 
$w=1$ indicates that they are equally important as the labelled ones.

On another line of research, \textit{Blocking based on Genetic Programming} (\textsf{BGP}) \cite{DBLP:journals/jidm/EvangelistaCSM10} employs a tree representation of supervised blocking schemes, where every leaf node corresponds to a blocking predicate. In every turn, a set of genetic programming operators, such as copy, mutation and crossover, are applied to the initial, random set of
blocking schemes. Then, a fitness function 
infers the performance of the new schemes from the harmonic mean of $PC$ and $RR$, and the best ones
are returned as output. 
Yet, \textsf{BGP} involves numerous internal parameters
that are hard to fine-tune.

Another tree-based approach 
is \textit{CBLOCK} \cite{DBLP:conf/cikm/SarmaJMB12}. In this case, every edge is annotated with a hash (i.e., transformation) function and every node $n_i$ comprises the set of entities that result after applying all hash functions from the root to $n_i$. \textit{CBLOCK} is the only proactive learning-based method, restricting the maximum size of its blocks. Every node that exceeds this limit is split into smaller, disjoint blocks through a  greedy algorithm that picks the best hash function based on the resulting $PC$. To minimize the human effort, a drill down approach is proposed for bootstrapping. 

\textbf{Unsupervised Methods.} 
\textit{FisherDisjunctive} \cite{DBLP:conf/icdm/KejriwalM13} 
uses a weak training set generated by 
the TF-IDF similarity between pairs of entities. Pairs with very low (high) values are considered as negative (positive) instances. A boolean feature vector is then associated with every labelled instance. The discovery of DNF 
blocking schemes is finally cast as a Fisher feature selection problem.

Similarly, \textit{DNF Learner}~\cite{giang2015machine} 
applies a matching algorithm 
to a sample of entity pairs to automatically create a labelled dataset. Then, the learning of 
blocking schemes is cast as a DNF learning problem. To scale it to the exponential search space of possible schemes, their complexity is restricted to manageable levels (e.g., they comprise at most $k$ predicates).

\vspace{-8pt}
\subsection{Schema-agnostic Block Building}
\label{sec:schemaAgnosticBB}

Methods of this type make no assumptions about schema knowledge, disregarding completely attribute names; they extract blocks from all attribute values. 
Thus, they inherently support noise in both attribute names and values and are suitable for
highly heterogeneous, loosely structured entity profiles, such as those stemming from the Web of Data~\cite{DBLP:conf/wsdm/PapadakisINF11,DBLP:conf/wsdm/PapadakisINPN12,DBLP:journals/tkde/PapadakisIPNN13}.

\textbf{Non-learning Methods.}
The family tree of this category appears in Figure \ref{fig:schemaBasedBlocking}(b). The cornerstone approach is \textit{Token Blocking} (\textsf{TB}) \cite{DBLP:conf/wsdm/PapadakisINF11}. Assuming that duplicates share at least one common token, its transformation function extracts all tokens from all attribute values of every entity. A block $b_t$ is then defined for every distinct token $t$.
Hence, two entities co-occur in block $b_t$ if they share token $t$ in their values, regardless of the associated attribute names. 

To improve \textsf{TB}, \textit{Attribute Clustering Blocking} \cite{DBLP:journals/tkde/PapadakisIPNN13} requires the common tokens of two entities to appear in \textit{syntactically similar attributes}. These are attribute names that correspond to similar values, but are not necessarily semantically matching (unlike Schema Matching). First, it clusters attributes based on the similarities of their aggregate values. 
Each attribute is connected to its most similar one and the transitive closure of the connected attributes forms disjoint clusters. A block $b_{k,t}$ is then defined for every token $t$ in the values of the attributes belonging to cluster $k$. 

\textit{RDFKeyLearner} \cite{DBLP:conf/semweb/SongH11} applies \textsf{TB} independently to the values of specific attributes, which are selected through the following process: 
each attribute is 
associated with a \textit{discriminability} score, which amounts to the portion of 
distinct values over all values in the given dataset. If this is lower than a predetermined threshold, the attribute is ignored due to limited diversity, i.e., too many entities have the same value(s). For each attribute with high discriminability, its \textit{coverage} is estimated, i.e., the portion of entities that contain it. The harmonic mean of discriminability and coverage is then computed for all valid attributes and the one with the maximum score is selected for defining blocking keys as long as its score exceeds another predetermined threshold. If not, the selected attribute 
is combined with all other attributes and the process is repeated.

\textit{Prefix-Infix(-Suffix) Blocking} \cite{DBLP:conf/wsdm/PapadakisINPN12} 
exploits the naming pattern in entity URIs. The \textit{prefix} describes the domain of the URI, the \textit{infix} is a local identifier, and the optional \textit{suffix} contains details about the format, or a named anchor \cite{DBLP:conf/iiwas/PapadakisDFK10}. E.g., in the URI {\small\texttt{https://en.wikipedia.org/wiki/France\#History}}, the prefix is {\small\texttt{https://en.wikipedia.org/wiki}}, the infix is {\small\texttt{France}} and the suffix is {\small\texttt{History}}. In this context, this method uses as 
keys all (URI) infixes along with all tokens in the literal values. 

\textit{TYPiMatch} \cite{DBLP:conf/wsdm/MaT13} improves \textsf{TB}
by automatically detecting the entity types in the input data.
It creates a co-occurrence graph, where every node corresponds to a token in any attribute value and every edge connects two tokens if both conditional probabilities of co-occurrence exceed a predetermined threshold. The maximal cliques are extracted 
and merged if their overlap exceeds another threshold. The resulting clusters correspond to the entity types,
with every entity participating in all types to which its tokens belong. 
\textsf{TB} is then applied independently to the profiles of each type.

Finally, \textit{Semantic Graph Blocking} \cite{DBLP:conf/ideas/NinMML07} is based exclusively on the relations between entities, be it foreign keys in a database or links in RDF data. It completely disregards attribute values, building a collaborative graph, where every node corresponds to an entity and every edge connects two associated entities. For instance, the collaborative graph for a bibliographic data collection can be formed by mapping every author to a node and adding edges between co-authors.
Then a new block $b_i$ is formed for each node $n_i$, containing all nodes connected with $n_i$ through a path, provided that the path length or the block size do not exceed predetermined limits (proactive functionality). 

\textbf{Learning-based Methods.} 
\textit{Hetero} \cite{DBLP:conf/semweb/KejriwalM14a}
converts the input data into heterogeneous structured datasets using property tables. Then, it maps every entity to a normalized TF vector, and applies an adapted Hungarian algorithm with linear scalability to produce positive and negative feature vectors. Finally, it applies \textit{FisherDisjunctive} \cite{DBLP:conf/icdm/KejriwalM13} with bagging to achieve robust performance.

Similarly, \textit{Extended DNF BSL}
\cite{DBLP:journals/corr/KejriwalM15} combines an established instance-based schema matcher with weighted set covering to learn DNF 
blocking schemes with at most $k$ predicates.

\subsection{Parallelization Approaches}
\label{sec:parallelizationBlBu}

To scale Blocking methods to massive entity collections without altering their functionality, 
the \textit{MapReduce framework} \cite{DeanG04} is typically used, 
as it offers fault-tolerant, optimized execution for applications distributed across a set of independent nodes. 

\textbf{Schema-aware methods.} 
The hash-based, non-learning methods are adapted to MapReduce in a straightforward way. The \texttt{map}
phase implements the transformation function(s), emitting {\small \texttt{(key, entity\_id)}} pairs for each entity. Every reducer acts as an assignment function, placing all entities with blocking key $t$ in block $b_t$.
Dedoop~\cite{DBLP:journals/pvldb/KolbTR12} provides such implementations for various methods.

For sort-based methods, 
the adaptation of \textsf{SN} 
to MapReduce in \cite{DBLP:journals/ife/KolbTR12} can be used as a template.
The \texttt{map} function extracts the blocking key(s) from each input entity, while the ensuing \textit{partitioning} phase 
sorts all entities in alphabetical order of their keys.
The \texttt{reduce} function slides a window of fixed size within every reduce partition. Inevitably, entities close to the partition boundaries need to be compared across different reduce tasks. Thus, the \texttt{map} function is extended to replicate those entities, forwarding them to the respective reduce task and its successor. 

DCS and DCS++ are adapted to the MapReduce framework in \cite{DBLP:conf/sac/MestrePN15}, using three jobs. The first one sorts the originally unordered entities of the data partition assigned to each mapper according to  the selected blocking keys. It also selects the boundary pairs of the sorted partitions. The second job generates the Partition Allocation Matrix, which specifies the sorted partitions to be replicated, while the third job performs DCS(++) locally, to the data assigned to every reducer. 

\textbf{Schema-agnostic methods.} 
A single MapReduce job is required for parallelizing \textsf{TB} \cite{DBLP:series/synthesis/2015Christophides,DBLP:conf/bigdataconf/EfthymiouSC15}. For every input entity $e_i$, the \texttt{map} function emits a ($t$, $e_i$) pair for every token $t$ in the values of $e_i$. Then, all entities sharing a particular token are directed to the same \texttt{reducer} to form a new block. 

For Attribute Clustering Blocking, four MapReduce jobs are required \cite{DBLP:series/synthesis/2015Christophides,DBLP:conf/bigdataconf/EfthymiouSC15}. The first assembles all values 
per attribute.
The second computes the pairwise similarities between all attributes,
even if they are placed in different data partitions. The third connects every attribute 
to its most similar one. The fourth associates every attribute name with a cluster id and adapts 
\textsf{TB}'s \texttt{map} function to emit pairs of the form ($k$.$t$, $i$), where $k$ is the cluster id of $e_i$'s attribute name that contains token $t$.

Finally, the parallelization of Prefix-Infix(-Suffix) Blocking involves three MapReduce jobs \cite{DBLP:series/synthesis/2015Christophides,DBLP:conf/bigdataconf/EfthymiouSC15}. The first parallelizes the algorithm that extracts the prefixes from a set of URIs \cite{DBLP:conf/iiwas/PapadakisDFK10}. The second 
extracts the URI suffixes.
The third  
applies \textsf{TB}'s 
mapper 
to the literal values simultaneously with an infix mapper that emits a pair ($j$, $e_i$) for every infix $j$ that is extracted from $e_i$'s profile.
The final \texttt{reduce} phase ensures that all entities having a common token or infix 
are placed in the same block.

\textbf{Load Balancing.}
For MapReduce, it is crucial to distribute evenly the overall workload among the available nodes, avoiding potential bottlenecks. 
The following methods distribute the execution of comparisons 
in a block collection -
not the cost of building the blocks. 

\textit{BlockSplit} \cite{DBLP:conf/icde/KolbTR12} partitions large blocks into smaller sub-blocks and processes them in parallel. 
Every entity is compared to all entities in its sub-block as well as to all entities of its super-block, even if their sub-block is initially assigned to a different node. This yields an additional network and I/O overhead
and may still lead to unbalanced workload, due to sub-blocks of different size.

To overcome this, \textit{PairRange} \cite{DBLP:conf/icde/KolbTR12} splits evenly the comparisons in a set of blocks into a predefined number of partitions. It involves a single MapReduce job with a mapper that associates every entity $e_i$ in block $b_k$ with the output key $p.k.i$, where $p$ denotes the partition id.
The reducer assembles 
all entities that have the same $p$ and block id, reproducing the comparisons of each partition.

The space requirements of these two algorithms are improved in \cite{DBLP:conf/ipccc/YanXM13}, which minimizes their memory consumption by adapting them so that they work with sketches.

Finally, \textit{Dis-Dedup} \cite{DBLP:journals/pvldb/ChuIK16} is the only method that takes into account both the computational and the communication cost (e.g., network transfer time, local disk I/O time). Dis-Dedup considers all possible cases, from disjoint blocks produced by a single blocking technique to overlapping blocks derived from multiple techniques. It also provides strong theoretical guarantees that the overall maximum cost per reducer is within a small constant factor from the lower bounds. 

\subsection{Discussion \& Experimental Results}

The performance of the above techniques is examined both qualitatively and quantitatively in a series of individual of papers (e.g., \cite{DBLP:journals/tkdd/VriesKCC11,o2018new,DBLP:journals/tkde/PapadakisIPNN13,DBLP:conf/wsdm/PapadakisINF11}) and experimental analyses (e.g., \cite{DBLP:journals/tkde/Christen12,DBLP:journals/pvldb/0001APK15,DBLP:journals/pvldb/0001SGP16}). Below, we summarize the main findings in order to facilitate the use of Block Building techniques. 

Starting with \textit{Standard Blocking} (\textsf{SB}), 
its performance depends heavily on the frequency distribution of attribute values and, thus, of blocking keys. The best case corresponds to a uniform distribution, where $||B||=||E||/|B|$ \cite{DBLP:journals/tkde/Christen12}. Due to its lazy functionality, though, all other key distributions yield a portion of large blocks with many superfluous comparisons, i.e., low $PQ$~and~$RR$.

\textit{Suffix Arrays Blocking} (\textsf{SA}) improves \textsf{SB}'s $PC$, by supporting errors at the beginning of blocking keys. The higher $l_{min}$ is and the lower $b_{max}$ is, the lower $||B||$ and $PC$ get. For the same settings, \textit{Extended SA} raises $PC$ at the cost of higher $||B||$, which inevitably lowers both $PQ$ and $RR$. \textit{Improved SA} is theoretically proven in \cite{DBLP:journals/tkdd/VriesKCC11} to result in a $PC$ greater or equal to that of \textsf{SA}, though at the cost of a higher computational cost and more comparisons, which lower $PQ$ and $RR$.

\textit{Q-grams Blocking} yields higher $PC$ than \textsf{SB}, but decreases both $PQ$ and $RR$. \textit{Extended Q-grams Blocking} raises $PQ$ and $RR$ at a limited, if any, cost in $PC$. \textsf{MFIBlocks} reduces significantly the number of blocks and matching candidates (i.e., very high $PQ$ and $RR$) \cite{DBLP:journals/is/KenigG13}, but it may come at the cost of missed matches (insufficient $PC$) in case the resulting blocking keys are very restrictive for matches with noisy descriptions \cite{DBLP:journals/pvldb/0001SGP16}.

For \textit{Sorted Neighborhood} (\textsf{SN}), a small $w$ leads to high $PQ$ and $RR$ but low $PC$ and vice versa for a large $w$. For \textit{Extended SN}, variations in the window size have a large impact on efficiency ($PQ$ and $RR$), affecting the portion of unnecessary comparisons, but $PC$ is more stable. Among the other \textsf{SN} variants, \textsf{DCS++} is theoretically proven to miss no matches with an appropriate value for $\phi$, while being at least as~efficient~as~\textsf{SN}. \textit{Sorted Blocks New Partition} outperforms most SN-based algorithms, but includes more parameters than \textsf{SN}, involving a more complex configuration.

Most importantly, all these non-learning schema-aware methods 
are quite parameter-sensitive: even small parameter value modifications may yield significantly different performance \cite{DBLP:journals/tkde/Christen12,DBLP:journals/pvldb/0001APK15,DBLP:conf/cikm/VriesKCC09,o2018new}. Their most important parameter is the definition of the blocking keys, which requires fine-tuning by an expert. Otherwise, their $PC$ remains insufficient, placing most duplicates in no common block \cite{DBLP:journals/tkde/Christen12,DBLP:journals/pvldb/0001APK15}. This applies even to methods that employ redundancy for higher recall. 

This shortcoming is ameliorated by schema-agnostic methods, 
which consistently achieve much higher $PC$ than their schema-aware counterparts \cite{DBLP:journals/pvldb/0001APK15}. They also simplify the configuration of Block Building, reducing its sensitivity through the automatic definition of blocking keys \cite{DBLP:journals/pvldb/0001APK15,DBLP:journals/pvldb/0001SGP16}. Rather than human intervention or expert knowledge, their robustness emanates from the high levels of redundancy they employ, placing every entity in a multitude of blocks. On the downside, they yield a considerably higher number of comparisons, resulting in very low $PQ$ and $RR$. Both, however, can be significantly improved by Block Processing \cite{DBLP:conf/icde/PapadakisN11,DBLP:journals/pvldb/0001SGP16}.

Regarding the relative performance of schema-agnostic methods, \textsf{TB} yields very high $PC$, at the cost of very low $PQ$ and $RR$. It constitutes a very efficient approach, iterating only once over the input entities, and it is the sole parameter-free Block Building technique in the literature as well as the most generic one, applying to any entity collection with textual values. Its performance is improved by 
\textit{Attribute Clustering} and \textit{Prefix-Infix(-Suffix) Blocking} for specific type of datasets: highly heterogeneous ones, with a large variety of attribute names \cite{DBLP:journals/tkde/PapadakisIPNN13,DBLP:journals/pvldb/0001SGP16}, and semi-structured (RDF) ones \cite{DBLP:conf/wsdm/PapadakisINF11}, respectively. In these cases, both methods yield a much larger number of smaller blocks, significantly raising $PQ$ at a minor cost in $PC$. Both methods, though, involve a much higher computational cost than \textsf{TB}. The same applies to 
\textit{TYPiMatch}, where the detection of entity types is a rather time-consuming process. Yet, its $PC$ is consistently insufficient, because it falsely divides duplicate entities different entity types, due to the sensitivity to its parameter~configuration~\cite{DBLP:journals/pvldb/0001SGP16}.

Finally, the learning-based Block Building techniques 
typically suffer from the scarcity of labelled datasets; even if a training set is available for a particular dataset, it cannot be directly used for learning supervised blocking schemes for another dataset. Instead, a complex transfer learning procedure is typically required \cite{DBLP:conf/cikm/NegahbanRG12,DBLP:journals/corr/abs-1809-11084}. Regarding their efficiency, \textsf{BSL} is typically faster than \textit{ApproxRBSetCover} and \textit{ApproxDNF}, as it exclusively considers positive instances, thus requiring a smaller training set. \textit{Conjunction Learner} 
requires every supervised blocking scheme to be applied to the large set of unlabelled data, which is impractical. To accelerate it, a random sample of the unlabelled data is used in practice. \textit{CBLOCK} is also the only learning-based method that is suitable for the MapReduce framework: every entity runs through the learned tree and is directed to the machine corresponding to its leaf node. In terms of effectiveness, there is no clear winner. \textsf{BSL} and \textit{FisherDisjunctive} achieve the top performance in \cite{o2018new}. The latter addresses the scarcity of labelled data, but is not scalable to large datasets.

\section{Block Processing}
\label{sec:blockProcessing}

Block Processing receives as input an existing block collection $\mathcal{B}$ and produces as output a new block collection $\mathcal{B'}$ that improves the balance between effectiveness and efficiency,  i.e., $PQ(B) \ll PQ(B')$, $RR(B',B)\gg 0$, while $PC(B) \sim PC(B')$. We distinguish Block Processing methods into \textit{Block Cleaning} ones, which decide whether 
entire blocks
should be retained or modified, and \textit{Comparison Cleaning} ones, which 
decide whether 
individual comparisons
are unnecessary. 

\vspace{-10pt}
\subsection{Block Cleaning}
\label{sec:blcl}

We classify Block Cleaning methods into two categories: (i) \textit{static}, which are independent of matching results, and (ii) \textit{dynamic}, which are interwoven with the matching process. 

\textbf{Static Methods.} A core idea is the assumption that the larger a block is, the less likely it is to contain unique duplicates, i.e., matches that share no other block. Such large blocks are typically produced by lazy schema-agnostic techniques and correspond to stop words.
In this context, \textit{Block Purging}  
discards 
blocks that exceed an upper limit on block cardinality \cite{DBLP:journals/tkde/PapadakisIPNN13} or size \cite{DBLP:conf/wsdm/PapadakisINPN12}.
\textit{Block Filtering} \cite{DBLP:conf/edbt/0001PPK16} applies this assumption to individual entities, removing every entity from the largest blocks that contain it. In other words, it
retains every entity in $r\%$ of its smallest blocks.

On a different line of research, \textit{Size-based Block Clustering} \cite{DBLP:conf/kdd/FisherCWR15} applies hierarchical clustering to transform a set of blocks into a new one where all block sizes lie within a specified size range. It merges recursively small blocks that correspond to similar blocking keys, while splitting large blocks into smaller ones. A penalty function controls the trade-off between block quality and block size. 
A similar approach is the MapReduce-based dynamic blocking algorithm 
in~\cite{mcneill2012dynamic}, which splits large blocks into sub-blocks.
\textit{MaxIntersectionMerge} \cite{nascimento2019exploiting} ensures that all blocks involve at least $|b|_{min}$ entities.
To this end, it merges each block smaller than $|b|_{min}$ entities with the block that has the most entities in common and is larger than $|b|_{min}$.
Similarly,
\textit{Rollup Canopies} \cite{DBLP:conf/cikm/SarmaJMB12} receives as input a training set with positive examples, a limit on the maximum block size and a set of disjoint blocks; using a learning-based greedy algorithm, it merges pairs of small blocks to increase $PC$. 

Finally, \cite{DBLP:conf/icdm/RanbadugeVC16} generalizes Meta-blocking (see Section \ref{sec:cocl}) to Multi-source ER: it constructs a graph, where the nodes correspond to blocks and the edges connect blocks whose blocking keys are more similar than a predetermined threshold. The edges are weighted using various functions and all pairs of blocks are then processed in descending edge weights in an effort to maximize the redundant and superfluous comparisons that are skipped.

\textbf{Dynamic Methods.}
\textit{Iterative Blocking} \cite{DBLP:conf/sigmod/WhangMKTG09} merges 
any new pair of detected duplicates, $e_i$ and $e_j$, into a new entity, $e_{i,j}$,
and replaces both $e_i$ and $e_j$ with $e_{i,j}$ in all blocks that contain them, even if they have already been processed.
The new entity $e_{i,j}$ is compared with all co-occurring entities, as 
the new content in $e_{i,j}$ might 
identify previously missed matches. 
The ER process terminates when all blocks have been processed without finding new duplicates. 

Iterative Blocking applies exclusively to Deduplication. In Record Linkage, there is no need for merging two matching entities, 
due to the 1-1 restriction. Still,
the detected duplicates should be propagated in order to save the superfluous comparisons with 
their co-occurring entities 
in the subsequently processed blocks. 
The earlier the matches are detected, the more superfluous comparisons are saved.
To this end, \textit{Block Scheduling} 
optimizes the processing order 
of blocks in a non-iterative way, sorting them in decreasing order of
the probability $p_i(d)$ that a block $b_i$ contains a pair of duplicates. This is set inversely proportional to block cardinality, i.e., $p_i(d)=1/||b_i||$ \cite{simonini2018schema}, or to the minimum size of the inner block, i.e., $p_i(d)=1/min{|b_{i,1}|,|b_{i,2}|}$, where $|b_{i,k}| \subset \mathcal{E}_k$ \cite{DBLP:conf/wsdm/PapadakisINF11}. 
The former definition also applies to Iterative Blocking, which does not specify the exact block processing order, even though this affects significantly the resulting performance \cite{DBLP:journals/pvldb/0001SGP16}.

\textit{Block Pruning} \cite{DBLP:conf/wsdm/PapadakisINF11} extends Block Scheduling by exploiting the decreasing density of detected matches in its processing order (i.e., the later a block is processed, the less 
unique duplicates it contains). 
After processing the latest block, it estimates the average number of executed comparisons per new duplicate. 
If this ratio falls below a specific threshold, it terminates the ER process.

\subsection{Comparison Cleaning}
\label{sec:cocl}

\begin{figure}[t]\centering
	\includegraphics[width=0.59\linewidth]{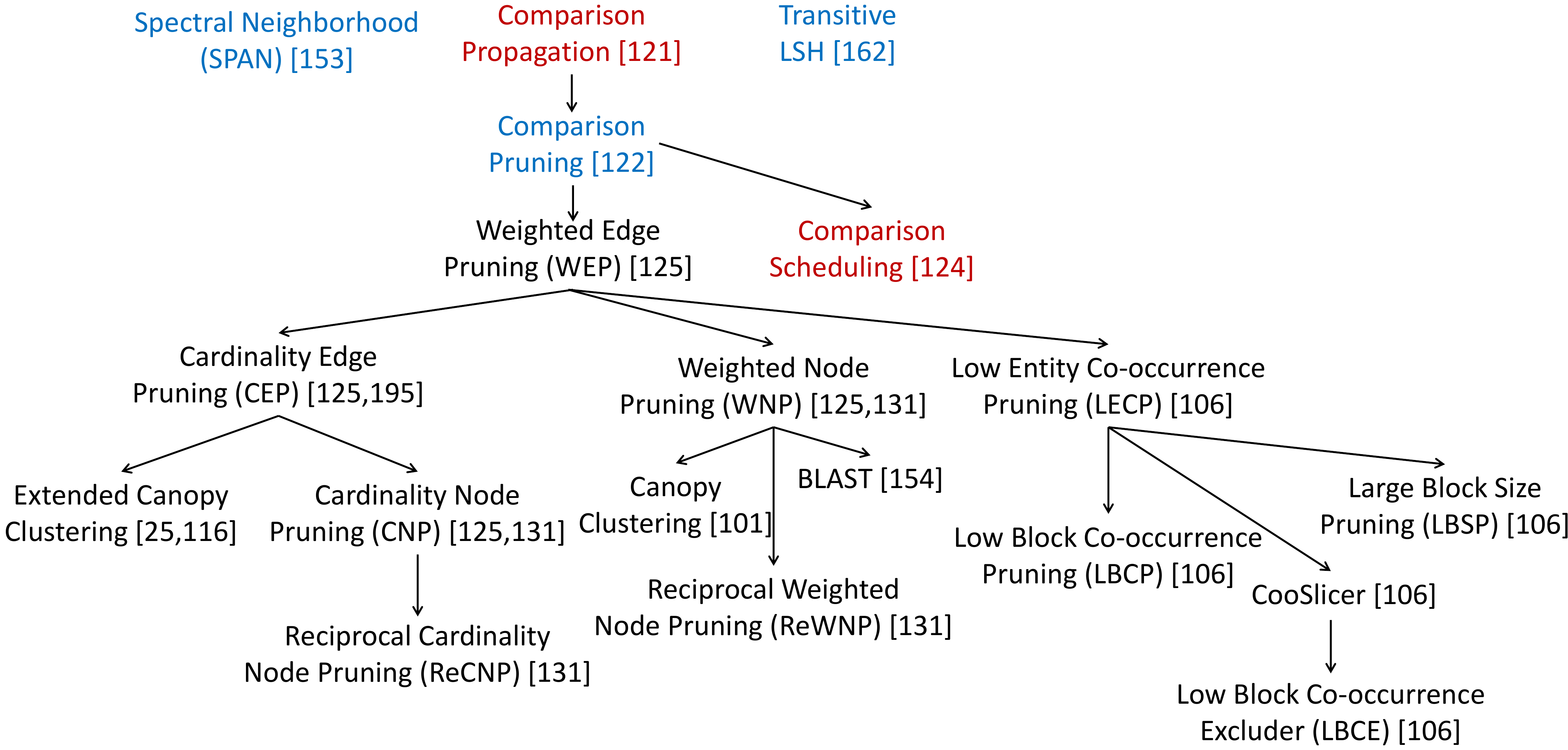}
	\vspace{-8pt}
	\caption{The genealogy tree of non-learning Comparison Cleaning methods. Methods in black conform to the Meta-blocking framework in Figure \ref{fig:coclExample}, methods in {\color{blue}blue} are Meta-blocking techniques following a (partially) different approach and methods in {\color{red}red} are not part of the Meta-blocking framework.
	}
	\label{fig:taxonomyCC}
	\vspace{-14pt}
\end{figure}

\begin{figure}[t!]\centering
	\includegraphics[width=0.99\linewidth]{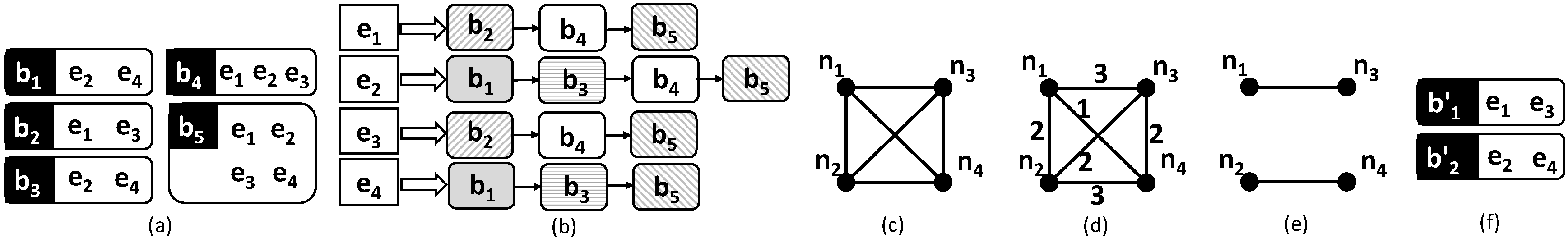}
	\vspace{-8pt}
	\caption{(a) A block collection $B$ with $e_1$$\equiv$$e_3$ and $e_2$$\equiv$$e_4$, (b) the corresponding Entity Index, (c) the corresponding blocking graph $G_B$, (d) the weighted $G_B$, (e) the pruned $G_B$, and (f) the new block collection $B'$.}
	\vspace{-19pt}
\label{fig:coclExample}
\end{figure}

\textbf{Non-learning Methods.}
Figure \ref{fig:taxonomyCC} illustrates the family tree of the methods belonging to this category.
The cornerstone method is \textit{Comparison Propagation} \cite{DBLP:conf/jcdl/PapadakisINPN11}, which propagates all executed comparisons to the subsequently processed blocks. In this manner, it eliminates all redundant comparisons in a given block collection without losing any pair of duplicates, thus raising $PQ$ and $RR$ at no cost in $PC$.  
It builds an inverted index that points from entity ids to block ids, called \textit{Entity Index}, and with its help, it compares two entities $e_i$ and $e_j$ in block $b_k$ only if $k$ is their least common block id.
For example, consider the blocks in Figure \ref{fig:coclExample}(a) and their Entity Index in Figure \ref{fig:coclExample}(b). The least common block id of $e_1$ and $e_3$ is 2. Thus, they are compared in $b_2$, but neither in $b_4$~nor~in~$b_5$.

Given a redundancy-positive block collection, the Entity Index allows for identifying the blocks shared by a pair of co-occurring entities. This allows for weighting all pairwise comparisons in proportion to the matching likelihood of the corresponding entities, based on the principle that the more blocks two entities share, the more likely they are to be matching. This gives rises to a family of \textit{Meta-blocking} techniques \cite{DBLP:journals/tkde/PapadakisKPN14,DBLP:conf/edbt/0001PPK16,DBLP:journals/pvldb/SimoniniBJ16} that
go beyond Comparison Propagation by discarding not only all redundant comparisons, but also the vast majority of the superfluous ones.

The first relevant method is \textit{Comparison Pruning} \cite{DBLP:conf/sigmod/PapadakisINPN11}, which
computes the Jaccard co-efficient of the block lists of two entities. If it does not exceed a conservative threshold that depends on the average number of blocks per entity, the 
comparison is pruned, as it designates an unlikely match.

Meta-blocking was formalized into a more principled approach in \cite{DBLP:journals/tkde/PapadakisKPN14}. The given redundancy-positive block collection $\mathcal{B}$ is converted into a blocking graph $G_B$, where the nodes correspond to entities and the edges connect every pair of co-occurring entities - see Figure \ref{fig:coclExample}(c). Given that no parallel edges are allowed, all redundant comparisons are discarded by definition. The edges are then weighted proportionately to the likelihood that the adjacent entities are matching. In Figure \ref{fig:coclExample}(d), the edge weights indicate the number of common blocks. Edges with low weights are pruned, because they correspond to superfluous comparisons. In Figure \ref{fig:coclExample}(e), all edges with a weight lower than the average one are discarded. The resulting pruned blocking graph $G_{B'}$ is transformed into a restructured block collection $\mathcal{B}'$ by forming one block for every retained edge - see Figure \ref{fig:coclExample}(f). As a result, $\mathcal{B}'$ exhibits a much higher efficiency, $PQ(B')$$\gg$$PQ(B)$ and $RR(B',B)$$\gg$$0$, for similar effectiveness, $PC(B')$$\sim$$PC(B)$; in our example, the 12 comparisons in the input blocks of Figure \ref{fig:coclExample}(a) are reduced to 2 matching comparisons in the output blocks in Figure \ref{fig:coclExample}(f).

Four main pruning algorithms exist: (i) \textit{Weighted Edge Pruning} (\textsf{WEP}) removes all edges that do not exceed a specific threshold, e.g., the average edge weight \cite{DBLP:journals/tkde/PapadakisKPN14}; (ii) \textit{Cardinality Edge Pruning} (\textsf{CEP}) retains the globally $K$ top weighted edges, where $K$ is static \cite{DBLP:journals/tkde/PapadakisKPN14} or dynamic \cite{zhang2017pruning}; 
(iii)~\textit{Weighted Node Pruning} (\textsf{WNP}) retains in each node neighborhood the entities that exceed a local threshold, which may be the average edge weight of each neighborhood \cite{DBLP:journals/tkde/PapadakisKPN14}, or the average of the maximum weights in the two adjacent node neighborhoods, as in \textit{BLAST} 
\cite{DBLP:journals/pvldb/SimoniniBJ16};
(iv) \textit{Cardinality Node Pruning} (\textsf{CNP}) retains the top-$k$ weighted edges in each node neighborhood \cite{DBLP:journals/tkde/PapadakisKPN14}.
\textit{Reciprocal WNP} and \textit{CNP} \cite{DBLP:conf/edbt/0001PPK16} apply an aggressive pruning that retains edges satisfying the pruning criteria
in both adjacent node neighborhoods. \textsf{WNP} and \textsf{WEP} are combined through the weighted sum of their thresholds in \cite{DBLP:conf/iscc/AraujoPN17}.

Another family of pruning algorithms is presented in \cite{nascimento2019exploiting}, focusing on the 
edge weights between the entities in each block. \textit{Low Entity Co-occurrence Pruning} (\textsf{LECP}) cleans every block from a specific portion of the entities with the lowest average edge weights. \textit{Large Block Size Pruning} (\textsf{LBSP}) applies \textsf{LECP} only to the blocks whose size exceeds the average block size in the input block collection. \textit{Low Block Co-occurrence Pruning} (\textsf{LBCP}) removes every entity from the blocks, where it is connected with the lowest weights, on average, with the rest of the entities. \textit{CooSlicer} enforces a maximum block size constraint, $|b|_{max}$, to all input blocks. In blocks larger than $|b|_{max}$ all entities are sorted in decreasing order of average edge weight, and the $|b|_{max}$ top-ranked entities are iteratively placed into a new block. \textit{Low Block Co-occurrence Excluder} (\textsf{LBCE}) discards a specific portion of the blocks with the lowest average edge weight among their entities.

All these pruning algorithms can be coupled with any 
\textit{edge weighting scheme} \cite{DBLP:journals/tkde/PapadakisKPN14}. \textsf{ARCS} sums the inverse cardinalities of the common blocks, giving higher weights to 
entity pairs that co-occur in smaller blocks. \textsf{CBS} counts the number of blocks shared by two entities, as in Figure \ref{fig:coclExample}(c), with \textsf{ECBS} extending it to discount the contribution from entities placed in many blocks.
\textsf{JS} corresponds to the Jaccard coefficient of two block lists, while
\textsf{EJS} extends it to discount the contribution from entities appearing in many non-redundant comparisons. 
Finally, Pearson's $\chi^2$
test assesses whether two adjacent entities appear independently in blocks and can be combined with
the aggregate attribute entropy associated with the tokens forming their common blocks~\cite{DBLP:journals/pvldb/SimoniniBJ16}.

Note that 
Meta-blocking covers established methods that are
considered as Block Building methods in the literature:
given that Block Building is equivalent to indexing \cite{DBLP:journals/tkde/Christen12}, any method based on indexes is in fact a Meta-blocking technique. 
For example, \textit{Transitive LSH} \cite{DBLP:conf/psd/SteortsVSF14}
converts the blocks extracted from \textsf{LSH} into an unweighted blocking graph and applies a community detection algorithm (e.g., \cite{clauset2004finding})
to partition the graph nodes into disjoint clusters, which will become the new blocks. The process finishes when the size of the largest cluster is lower than a predetermined threshold. This approach can be applied on top of any Block Building method, not just \textsf{LSH}.

The generalization principle also applies to \textit{Canopy Clustering} \cite{DBLP:conf/kdd/McCallumNU00}, which 
places all entities in a pool and, in every iteration, it removes a random entity $e_i$ from the pool to create a new block. Using a cheap similarity measure, all entities still in the pool are compared with $e_i$. Those exceeding a threshold $t_{ex}$ are removed from the pool and placed into the new block. Entities exceeding another threshold $t_{in}$ ($< t_{ex}$) are also placed in the new block, without being removed  from the pool. 
As the cheap similarity measure, we can use any of the above weighting schemes on top of any Block Building method,
thus turning Canopy Clustering into a pruning algorithm for Meta-blocking. 

The generalization applies to \textit{Extended Canopy Clustering} \cite{DBLP:journals/tkde/Christen12,DBLP:journals/pvldb/0001APK15}, too,  
which replaces the sensitive 
weight thresholds
with cardinality ones:
for each randomly selected entity, the $n_1$ nearest entities are placed in its block, while the $n_2 (\leq n_1)$ nearest entities are removed from the pool.

On another line of research, \textit{SPAN} \cite{DBLP:conf/icde/ShuCXM11} converts a block collection into a matrix $M$, where the rows correspond to entities and the columns to the tf-idf of blocking keys (tokens or $q$-grams). Then, the entity-entity matrix is defined as $A=MM^T$. A spectral clustering algorithm converts $A$ into a binary tree, where the root node contains all entities and every leaf node is a disjoint subset of entities. The Newman-Girvan modularity is used as the stopping criterion for the bipartition of the tree. Blocks are then derived from a search procedure that carries out pairwise comparisons based on the blocking keys, inside the leaf nodes and across the neighboring ones.

Finally, the sole dynamic non-learning method 
is \textit{Comparison Scheduling} \cite{DBLP:journals/tkde/PapadakisIPNN13}. Its goal is 
to detect most matches upfront so as to maximize the superfluous comparisons that are skipped, due to the 1-1 restriction. It orders all comparisons in decreasing matching likelihood (edge weight) and executes a comparison only if none of the involved entities has already been matched. 

\textbf{Learning-based Approaches.}
\textit{Supervised Meta-blocking} \cite{DBLP:journals/pvldb/0001PK14} 
treats edge pruning 
as a binary classification problem, where every edge is labelled "\texttt{likely match}" or "\texttt{unlikely match}". 
Every edge is represented by a feature vector that comprises five features: \textsf{ARCS}, \textsf{ECBS}, \textsf{JS} and the Node Degrees of the adjacent entities. Undersampling is employed to tackle the class imbalance problem: 
the training set comprises just 5\% of the minority class ("\texttt{likely match}") and an equal number of majority class instances. Several established classification algorithms are used for \textsf{WEP}, \textsf{CEP} and \textsf{CNP}, with all of them exhibiting robust performance with respect to their internal configuration.

\textit{BLOSS} \cite{DBLP:journals/is/BiancoGD18}  
restricts the labelling cost of Supervised Meta-blocking by 
carefully selecting a training set that is up to 40 times smaller, but retains the original performance. Using $ECBS$ weights, it partitions the unlabelled instances into similarity levels and applies rule-based active sampling inside every level. Then, it cleans the sample from non-matching outliers with high $JS$ weights. 

\textbf{Parallelization Approaches.}
Meta-blocking is adapted to the MapReduce framework in three ways \cite{DBLP:journals/is/Efthymiou0PSP17}:
(i) The \textit{edge-based strategy} stores the blocking graph 
on the disk, bearing a significant I/O cost.
(ii) The \textit{comparison-based strategy} builds the blocking graph \textit{implicitly}. A pre-processing job enriches every block with the list of block ids associated with every entity. The Map phase of the second job computes the edge weights and discards all redundant comparisons, while the ensuing Reduce phase prunes superfluous comparisons.
This strategy  maximizes the efficiency of \textsf{WEP} and \textsf{CEP} 
and is adapted to Apache Spark 
in \cite{DBLP:conf/iscc/AraujoPN17}. (iii) The \textit{entity-based strategy} 
aggregates for every entity the bag of all entities that co-occur with it in at least one block. Then, it estimates the edge weight that corresponds to each neighbor based on its frequency in the co-occurrence bag. This approach offers the best implementation for \textsf{WNP} and \textsf{CNP} and their variations (e.g., \textsf{BLAST}). It is adapted to Apache Spark in \cite{DBLP:journals/is/SimoniniGBJ19}, leveraging the broadcast join for higher efficiency.

To avoid the underutilization of the available resources, these strategies employ \textit{MaxBlock} \cite{DBLP:journals/is/Efthymiou0PSP17} for load balancing. Based on the highly skewed distribution of block sizes in redundancy-positive block collections, it splits the input blocks into partitions of equivalent computational cost, which is equal to the total number of comparisons in the largest input block. 

The \textit{multi-core parallelization} of Meta-blocking is examined in \cite{DBLP:conf/i-semantics/0001BPK17}. 
The input is transformed into 
an array of chunks, with an index indicating the next chunk to be processed. Following the established fork-join model, every thread retrieves the current value of the index and is assigned to process the corresponding chunk. Depending on the definition of chunks, three alternative strategies are proposed: (i) \textit{Naive Parallelization} treats every entity as a separate chunk, ordering all entities in decreasing computational cost (i.e., the aggregate number of comparisons in the associated blocks). (ii) \textit{Partition Parallelization} uses MaxBlock to group the input entities into an arbitrary number of disjoint clusters with identical computational cost. (iii) \textit{Segment Parallelization} 
sets the number of clusters equal to the number of available cores. 

\subsection{Discussion \& Experimental Results}

\begin{figure}[t]\centering
	\includegraphics[width=0.40\linewidth]{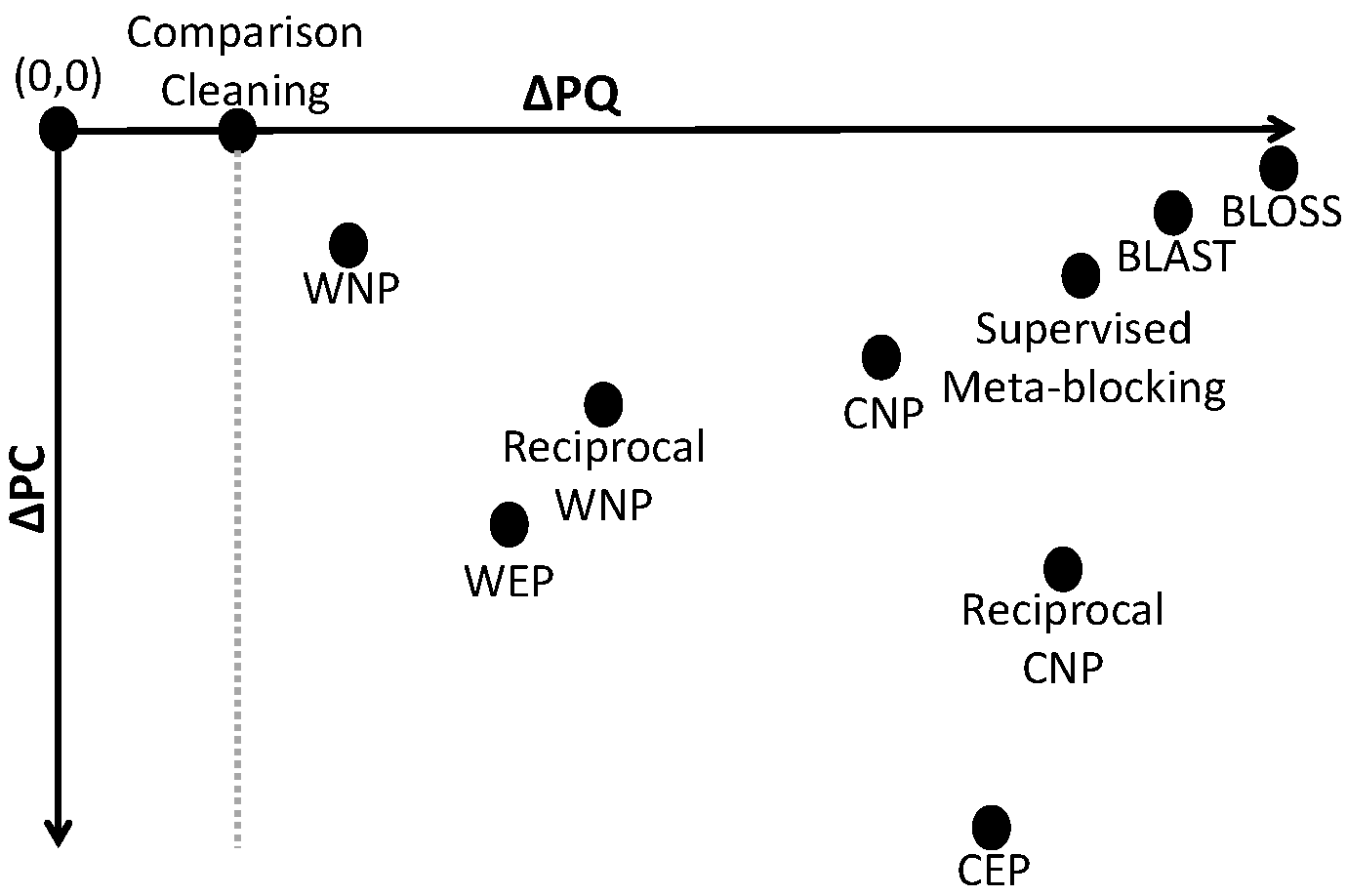}
	\vspace{-8pt}
	\caption{The relative performance of the main Comparison Cleaning methods.
	}
	\label{fig:relativePerformance}
	\vspace{-14pt}
\end{figure}

The core characteristic of Block Processing methods is their 
schema-agnostic functionality, which typically
relies 
on block features, such as size, cardinality and overlap. This is no surprise, as they are primarily crafted for boosting the performance of schema-agnostic Block Building methods. In fact, extensive experiments demonstrate that Block Processing is indispensable for these methods, raising precision
by whole orders of magnitude, at a minor cost in 
recall
\cite{DBLP:journals/pvldb/0001SGP16,DBLP:journals/bdr/PapadakisPPK16,DBLP:journals/tkde/PapadakisIPNN13}. 

Regarding their relative performance, there is no clear winner among the Block Cleaning methods. For example,
both Block Filtering and Block Purging boost $PQ$ and $RR$ by orders of magnitude, while exhibiting a low computational cost and a negligible impact on $PC$ \cite{DBLP:journals/tkde/PapadakisIPNN13,DBLP:conf/edbt/0001PPK16}. However, the top performer among them depends not only on their parameter configuration, but also on the data at hand \cite{DBLP:journals/pvldb/0001SGP16}.
Most importantly, though, Block Cleaning techniques are usually complementary in the sense that multiple ones can be applied consecutively in a single blocking workflow, as depicted in Figure \ref{fig:computationalCostPlusWorkflow}(b). For example, Block Filtering is typically applied after Block Purging by lowering $r$ to $50\%$ instead of $80\%$, which is the best configuration when applied independently \cite{DBLP:journals/bdr/PapadakisPPK16,DBLP:conf/edbt/0001PPK16,DBLP:journals/pvldb/0001SGP16}.

In contrast, Comparison Cleaning methods are incompatible with each other in the sense that at most one of them can be part of a blocking workflow. The reason is that applying any Comparison Cleaning technique to a redundancy-positive block collection deprives it from its co-occurrence patterns and renders all other techniques inapplicable. 
These techniques also involve a much higher computational cost than Block Cleaning methods, due to their finer level of granularity. Their relative performance is summarized in Figure \ref{fig:relativePerformance}, based on empirical evidence from experimental studies \cite{DBLP:journals/pvldb/0001SGP16} and individual publications \cite{DBLP:journals/is/BiancoGD18,DBLP:journals/pvldb/0001PK14,DBLP:journals/pvldb/SimoniniBJ16,DBLP:journals/bdr/PapadakisPPK16,DBLP:conf/edbt/0001PPK16,DBLP:journals/is/SimoniniGBJ19}. Note that we exclude methods not compared to other Comparison Cleaning techniques (e.g., the techniques presented in \cite{nascimento2019exploiting}).

In more detail, Figure \ref{fig:relativePerformance} maps the performance of the main Comparison Cleaning methods to a two dimensional space defined by $\Delta PC$=$PC(\mathcal{B}')-PC(\mathcal{B})$ on the vertical axis and $\Delta PQ$=$PQ(\mathcal{B}')-PQ(\mathcal{B})$ on the horizontal axis, where $B$ and $B'$ stand for the input and the output block collections, respectively. Given that Comparison Cleaning techniques trade lower recall ($PC$) for higher precision ($PQ$), $\Delta PC$ and $\Delta PQ$ take exclusively negative and positive values, respectively. Therefore, the higher a method is placed, the better recall it achieves, whereas the further to the right it lies, the better is its precision. This means that the ideal overall performance corresponds to the upper~right~corner. 

We observe that $\Delta PC$ is delimited by two extremes: Comparison Cleaning on the top left corner and \textsf{CEP} on the bottom right corner. The former has no impact on recall, as it 
increases precision only by removing redundant comparisons. All other Comparison Cleaning techniques discard superfluous comparisons, too, thus achieving larger $\Delta PQ$ at the cost of a negative $\Delta PC$. On the other extreme, \textsf{CEP} prunes a large portion of superfluous comparisons, yielding very high precision, but the lowest recall. \textsf{WEP} replaces \textsf{CEP}'s cardinality constraint with a weight threshold, dropping precision to a large extent for a significantly higher recall. Still, \textsf{WEP}'s performance is a major improvement over the input block collection, while being rather robust across numerous datasets. \textsf{WNP} moves further towards this direction, shrinking the decrease in recall and the increase in precision. This is further improved by \textit{Reciprocal WNP}, which significantly raises \textsf{WNP}'s precision for slightly lower recall. Thus, it dominates \textsf{WEP}, albeit being sensitive to the characteristics of the data at hand. Compared to \textsf{CEP}, \textsf{CNP} confines its pruning inside individual node neighborhoods. In this way, it achieves a much higher recall for a limited decrease in precision. This is further improved by \textit{Reciprocal CNP}, which reduces \textsf{CNP}'s recall slightly for much higher precision and, thus, it often dominates \textsf{CEP}.
\textsf{WNP}, \textsf{CNP} and their variants are improved by \textit{Supervised Meta-blocking} and \textsf{BLAST}, which achieve comparable recall for significantly higher precision. \textsf{BLAST} takes a lead in precision, partially because it employs the most effective weighting scheme, namely Pearson's $\chi^2$ test. Another advantage is that \textsf{BLAST} requires no labeling effort, due to its unsupervised functionality. \textsf{BLOSS}, however, achieves almost perfect recall ($\Delta PC \approx 0$) for the highest precision among all Comparison Cleaning techniques, while requiring merely $\sim$50 labeled instances. Note that exceptions to these general patterns of performance are possible for a particular dataset. 

\section{Filtering}
\label{sec:filtering}

\begin{table*}[t]
\centering
\setlength{\tabcolsep}{3.5pt} 
\caption{Overview of string and set similarity join methods.}
\label{tab:filtering_table}
\vspace{-5pt}
{\scriptsize
\begin{tabular}{| l || l | l | l | l |}
\hline
\textbf{Method} & \textbf{Operation} & \textbf{Similarity} & \textbf{Filters} & \textbf{Index} \\
\hline
\hline
GramCount~\cite{DBLP:conf/vldb/GravanoIJKMS01} & string join & Edit Distance & length, count, position & $q$-grams table \\
MergeOpt~\cite{DBLP:conf/sigmod/Sarawagi04} & set join & Overlap & count & inverted index \\
FastSS~\cite{BoHuSt07} & string join & Edit Distance & deletion neighborhood & dictionary \\
\hline
SSJoin~\cite{DBLP:conf/icde/ChaudhuriGK06} & set join & Overlap & prefix & DBMS \\
All-Pairs~\cite{DBLP:conf/www/BayardoMS07} & vector join & Cosine & prefix & inverted index \\
DivideSkip~\cite{DBLP:conf/icde/LiLL08} & string search & Edit Distance, Overlap & length, position, prefix & inverted index \\
Ed-Join~\cite{DBLP:journals/pvldb/XiaoWL08} & string join & Edit Distance & prefix+mismatching $q$-grams & inverted index \\
QChunk~\cite{DBLP:conf/sigmod/QinWLXL11} & string join & Edit Distance & prefix+$q$-chunks & inverted index \\
VChunkJoin~\cite{DBLP:journals/tkde/WangQXLS13} & string join & Edit Distance & prefix+chunks & inverted index \\
PPJoin~\cite{DBLP:conf/www/XiaoWLY08,DBLP:journals/tods/XiaoWLYW11} & set join & Overlap & prefix, positional & inverted index \\
PPJoin+~\cite{DBLP:conf/www/XiaoWLY08,DBLP:journals/tods/XiaoWLYW11} & set join & Overlap & prefix, positional, suffix & inverted index \\
MPJoin~\cite{DBLP:journals/is/RibeiroH11} & set join & Overlap & min-prefix & inverted index \\
GroupJoin~\cite{DBLP:journals/pvldb/BourosGM12} & set join & Overlap & prefix+grouping & inverted index \\
AdaptJoin~\cite{DBLP:conf/sigmod/WangLF12} & set join & Overlap & adaptive prefix & inverted index \\
SKJ~\cite{DBLP:journals/pvldb/WangQLZC17} & set join & Overlap & prefix-based+set relations & inverted index \\
TopkJoin~\cite{DBLP:conf/icde/XiaoWLS09} & top-$k$ set join & Overlap & prefix-based & inverted index \\
JOSIE~\cite{DBLP:conf/sigmod/ZhuDNM19} & top-$k$ set search & Overlap & prefix, position & inverted index \\
\hline
PartEnum~\cite{DBLP:conf/vldb/ArasuGK06} & set join & Hamming, Jaccard & partition-based & clustered index \\
PassJoin~\cite{DBLP:journals/pvldb/LiDWF11} & string join & Edit Distance & partition-based & inverted index \\
PTJ~\cite{DBLP:journals/pvldb/DengLWF15} & set join & Overlap & partition-based & inverted index \\
\hline
B$^{ed}-$Tree~\cite{DBLP:conf/sigmod/ZhangHOS10} & string search/join & Edit Distance & string orders & B$^+$-tree \\
PBI~\cite{DBLP:journals/tkde/LuDHO14} & string search & Edit Distance & reference strings & B$^+$-tree \\
MultiTree~\cite{DBLP:conf/icde/ZhangLWZXY17} & set search & Jaccard & tree traversal & B$^+$-tree \\
Trie-Join~\cite{DBLP:journals/pvldb/WangLF10} & string join & Edit Distance & subtrie pruning & trie \\
HSTree~\cite{DBLP:journals/vldb/YuWLZDF17} & string search & Edit Distance & partition-based & segment tree \\
Trans~\cite{zhang2018transformation} & top-$k$ set search & Jaccard & transformation distance & R-tree \\
\hline
\multicolumn{5}{c}{\textbf{(a) Exact, centralized, single predicate algorithms}}\\
\hline
FuzzyJoin~\cite{DBLP:conf/icde/AfratiSMPU12} & set/string join & Hamming, ED, Jaccard & ball-hashing, splitting, anchor points & lookup tables \\
VernicaJoin~\cite{DBLP:conf/sigmod/VernicaCL10} & set join & Overlap & prefix, positional, suffix & inverted index \\
MGJoin~\cite{DBLP:journals/tkde/RongLWDCT13} & set join & Overlap & multiple prefix & inverted index \\
MRGroupJoin~\cite{DBLP:journals/pvldb/DengLWF15} & set join & Overlap & partition-based & inverted index \\
FS-Join~\cite{DBLP:conf/icde/RongLSWLD17} & set join & Overlap & segment-based & inverted index \\
Dima~\cite{DBLP:journals/pvldb/SunSLDB17,DBLP:journals/pvldb/SunS0BD19} & search, join, top-$k$ & Jaccard, ED & segment-based & global \& local \\
\hline
\multicolumn{5}{c}{\textbf{(b) Parallel \& distributed algorithms}}\\
\hline
ATLAS~\cite{DBLP:conf/sigmod/ZhaiLG11} & vector join & Jaccard, Cosine & random permutations & inverted index \\
BayesLSH~\cite{DBLP:journals/pvldb/SatuluriP12} & set join & Jaccard, Cosine & All-Pairs / LSH & All-Pairs / LSH \\
CPSJoin~\cite{DBLP:conf/icde/ChristianiPS18} & set join & Jaccard & LSH-based & sketches \\
\hline
\multicolumn{5}{c}{\textbf{(c) Approximate algorithms}}\\
\hline
LS-Join~\cite{DBLP:journals/tkde/WangYWL17} & local string join & Edit Distance & length, count & inverted index \\
pkwise~\cite{DBLP:conf/sigmod/WangXQWZI16} & local set join & Overlap & $k$-wise signatures & inverted index \\
pkduck~\cite{DBLP:journals/pvldb/TaoDS17} & abbreviation matching & Custom & extension of prefix filter & trie \\
\hline
Fast-Join~\cite{DBLP:journals/tods/WangLF14} & fuzzy set join & Bipart. graph matching & token sensitive signatures & inverted index \\
SilkMoth~\cite{DBLP:journals/pvldb/DengKMS17} & fuzzy set join & Bipart. graph matching & weighted token signatures & inverted index \\
MF-Join~\cite{DBLP:conf/icde/WangLZ19} & fuzzy set join & Bipart. graph matching & partion-based & inverted index \\
\hline
MultiAttr~\cite{DBLP:conf/sigmod/LiHDL15} & set search/join & Overlap & tree traversal & prefix tree \\
Smurf~\cite{DBLP:journals/pvldb/CADA18} & string matching & Jaccard, Edit Distance & random forest & inverted indexes \\
AU-Join~\cite{DBLP:journals/pvldb/XuL19} & string join & Syntactic, Synonym, Taxonomy & pebbles & inverted indexes \\
\hline
\multicolumn{5}{c}{\textbf{(d) Algorithms for complex matching}}\\
\end{tabular}
\vspace{-15pt}
}
\end{table*}

Given specific similarity predicates, comprising a similarity measure and a corresponding threshold, Filtering techniques receive as input an entity or a block collection and produce as output pairs of entities satisfying these predicates.
Next, we present the main filtering methods in the literature, organized in four groups: basic filters proposed by earlier works; prefix filtering and its extensions; partition-based filtering; and methods using tree indexes.
An overview of the discussed methods is presented in Table~\ref{tab:filtering_table}, characterized by the type of operation they perform (e.g., search or join), the similarity measure they assume (e.g., token- or character-based), the type of filters they use (e.g., prefix- or partition-based) and the index structure they employ (e.g., inverted index or tree).

\textbf{Basic filtering.}
\texttt{GramCount}~\cite{DBLP:conf/vldb/GravanoIJKMS01} focuses on incorporating string similarity joins inside a 
DBMS based on $q$-grams and edit distance. It is the first work to propose 
the following techniques:

\textit{Length filtering} states that if two strings $r$ and $s$ are within edit distance $\theta$, their lengths cannot differ by more than $\theta$. In the case of set similarity joins, the length filter has been adapted to deal with set sizes~\cite{DBLP:conf/vldb/ArasuGK06}; e.g., for Jaccard similarity threshold $\theta$, the condition becomes $\theta \cdot |s| \leq |r| \leq |s| / \theta$. Length filtering is a simple but effective criterion that is employed by many other works alongside more advanced filters. 
A \textit{position-enhanced} length filter 
offers a tighter upper bound
\cite{DBLP:conf/gvd/MannA14}.

\textit{Count filtering} states that if two strings $r$ and $s$ are within edit distance $\theta$, they must have at least $max(|r|, |s|) - 1 - (\theta - 1) \cdot q$ common $q$-grams. This filter has also been adapted to sets, in particular in \texttt{MergeOpt}~\cite{DBLP:conf/sigmod/Sarawagi04}, which proposed various optimizations for applying count filtering with both character-based and token-based similarity measures, and in \texttt{DivideSkip}~\cite{DBLP:conf/icde/LiLL08}, which proposed efficient techniques for merging the inverted lists of signatures. 

\textit{Position filtering} also considers the positions of $q$-grams in the strings. It states that if two strings $r$ and $s$ are within edit distance $\theta$, a positional $q$-gram in one cannot correspond to a positional $q$-gram in the other that differs from it by more than $\theta$ positions.

On another line of research, \texttt{FastSS}~\cite{BoHuSt07}
introduces the concept of \textit{deletion neighborhood}, a filtering criterion specifically tailored to edit distance. For a string $s$, its deletion neighborhood contains substrings of $s$ derived by deleting a certain number of characters. These are then used as signatures for filtering. However, this method is practical only for very short strings.

\textbf{Prefix-based filtering.}
\textit{Prefix filtering} has been proposed by \texttt{SSJoin}~\cite{DBLP:conf/icde/ChaudhuriGK06}, which focuses on 
similarity joins inside a DBMS, 
and \texttt{All-Pairs}~\cite{DBLP:conf/www/BayardoMS07},
which is a main memory algorithm. Prefix filter applies to sets and can also be used for strings represented as sets of $q$-grams. The elements of each set are first sorted in a global order, typically in increasing order of frequency. Then, the $\pi$-prefix of each set is formed by selecting its $\pi$ first elements in that order. Prefix filter states that for two sets to be similar, their prefixes must contain at least one common element. The prefix size $\pi$ of a set $r$ is determined based on the similarity measure and threshold being used; e.g., for edit distance threshold $\theta$, $\pi = q \cdot \theta + 1$, while for Jaccard similarity threshold $\theta$, $\pi = \lfloor (1 - \theta) \cdot |r| \rfloor + 1$. As described next, numerous subsequent algorithms have adopted prefix filtering and proposed various optimizations and extensions over it, both for edit distance and set-based similarity joins.

For edit distance, \texttt{DivideSkip}~\cite{DBLP:conf/icde/LiLL08} uses prefix filtering in combination with length and position filtering, 
taking special care to efficiently merge the inverted lists of signatures. \texttt{Ed-Join}~\cite{DBLP:journals/pvldb/XiaoWL08} proposes two optimizations
based on analyzing the locations and contents of mismatching $q$-grams to further reduce the prefix length by removing unnecessary elements. \texttt{QChunk}~\cite{DBLP:conf/sigmod/QinWLXL11} introduces the concept of \textit{$q$-chunks}, which are substrings of length $q$ that start at 1+$i$$\cdot$$q$ positions in the string, for $i \in [0, (|r|-1)/q]$. Given two strings $r$ and $s$, QChunk extracts $q$-grams from the one and $q$-chunks from the other; 
if $r$ and $s$ are within edit distance $\theta$, the size of the intersection between the $q$-grams of $r$ and the $q$-chunks of $s$ should be at least $\lceil|s|/q\rceil$-$\theta$.
\texttt{VChunkJoin}~\cite{DBLP:journals/tkde/WangQXLS13} uses non-overlapping substrings called \textit{chunks}, ensuring that each edit operation destroys at most two chunks. This yields a tight lower bound on the number of common chunks that two strings must share if they match.

For set similarity joins,
\texttt{PPJoin}~\cite{DBLP:conf/www/XiaoWLY08,DBLP:journals/tods/XiaoWLYW11} extends prefix filtering with \textit{positional filtering}. This takes also into consideration the positions where the common tokens in the prefix occur, thus deriving a tighter upper bound for the overlap between the two sets. In addition, \texttt{PPJoin}+~\cite{DBLP:conf/www/XiaoWLY08,DBLP:journals/tods/XiaoWLYW11} further uses \textit{suffix filtering}. Following a divide-and-conquer strategy, this partitions the suffix of the one set into two subsets of similar sizes. The token separating the two partitions is called \textit{pivot} and is used to split the suffix of the other set. This allows to calculate the maximum number of tokens in each pair of corresponding partitions between the two sets that can match.

\texttt{MPJoin}~\cite{DBLP:journals/is/RibeiroH11} adds a further optimization over \texttt{PPJoin} that allows for dynamically pruning the length of the inverted lists. This reduces the computational cost of candidate generation, rather than the number of candidates. \texttt{GroupJoin}~\cite{DBLP:journals/pvldb/BourosGM12} extended \texttt{PPJoin} with \textit{group filtering}, whose candidate generation 
treats all sets with identical prefixes as a single set.
Multiple candidates may thus be pruned in batches. \texttt{AdaptJoin}~\cite{DBLP:conf/sigmod/WangLF12} proposed \textit{adaptive prefix filtering}, which generalizes prefix filtering by adaptively selecting an appropriate prefix length for each set. It supports longer prefixes dynamically, extending their length by $n-1$, and then prunes a pair of sets if they contain less than $n$ common tokens in their extended prefixes. Prefix filtering is a special case where $n=1$.

A different perspective for speeding up set similarity joins is proposed by \texttt{SKJ}~\cite{DBLP:journals/pvldb/WangQLZC17}. The idea is based on the following observation: existing approaches examine each set individually when computing the join; however, it is possible to improve efficiency through computational cost sharing between \textit{related sets}. To this end, the \texttt{SKJ} algorithm introduces \textit{index-level skipping}, which groups related sets in the index into blocks, and \textit{answer-level skipping}, which incrementally generates the answer of one set from an already computed answer of another related set.

Finally, there are Filtering techniques 
for computing top-$k$ results progressively, instead of requiring the user to select a similarity threshold.
\texttt{TopkJoin}~\cite{DBLP:conf/icde/XiaoWLS09} retrieves the top-$k$ pairs of sets ranked by their similarity score, 
based on prefix filtering and 
on the monotonicity of maximum possible scores of unseen pairs.
\texttt{JOSIE}~\cite{DBLP:conf/sigmod/ZhuDNM19} 
presents a method for top-$k$ set similarity search. It exploits prefix and position filtering but,
instead of dealing with sets of relatively small size (e.g., $\sim$100 tokens), it 
is crafted for
finding joinable tables in data lakes, where sets represent the distinct values of a table column,
comprising millions of tokens. 
This introduces new challenges, which are tackled by proposing an algorithm that minimizes the cost of set reads and inverted index probes.

\textbf{Partition-based filtering.} The algorithms in this category partition each string or set into multiple disjoint segments in such a way that matching pairs have at least one common segment.
\texttt{PartEnum}~\cite{DBLP:conf/vldb/ArasuGK06} generates a signature scheme based on the principles of \textit{partitioning} and \textit{enumeration}. The former states that if two vectors with Hamming distance not higher than $k$ are partitioned into $k$ + 1 equi-sized partitions, then they must have at least one common partition. The latter states that if these vectors are partitioned instead into $n > k$ equi-sized partitions, then they must have in common at least $n - k$ partitions. \texttt{PassJoin}~\cite{DBLP:journals/pvldb/LiDWF11} partitions a string into a set of segments and creates inverted indices for the segments. Then, for each string, it selects some of its substrings and uses them to retrieve candidates from the index. A method is proposed to minimize the number of segments required to find the candidates pairs. \texttt{PTJ}~\cite{DBLP:journals/pvldb/DengLWF15} proposes an approach to increase the pruning power of partition-based filtering by using a mixture of the subsets and their 1-deletion neighborhoods, which are subsets derived from a set after eliminating one element.

Essentially, these methods are based on the \textit{pigeonhole principle}, which states that if $n$ items are contained in $m$ boxes, 
at least one box has no more than $\lfloor n / m \rfloor$ items. This is extended by
the \textit{pigeonring principle} 
\cite{DBLP:journals/pvldb/QinX18}, which organizes the boxes in a ring and constrains the number of items in multiple boxes rather than a single one, thus offering tighter bounds. 
Applying it to various similarity search problems
shows that pigeonring always produces less or equal number of candidates than the pigeonhole principle does and that pigeonring-based algorithms can be implemented on top of existing pigeonhole-based ones with minor modifications \cite{DBLP:journals/pvldb/QinX18}.

\textbf{Tree-based filtering.} Most methods presented so far build inverted indexes on the signatures extracted from the strings or sets. Next, we present algorithms employing tree-based indexes.

Most approaches are based on the B$^+$-tree. \texttt{B}$^{ed}$-\texttt{Tree}~\cite{DBLP:conf/sigmod/ZhangHOS10} proposes a B$^+$-tree based index for range and top-$k$ similarity queries as well as similarity joins, using edit distance. It is based on a mapping from the string space to the integer space to support efficient searching and pruning. \texttt{PBI}~\cite{DBLP:journals/tkde/LuDHO14} uses a B$^+$-tree index and exploits the fact that edit distance is a metric. The string collection is partitioned according to a set of selected \textit{reference strings}. Then, the strings in each partition are indexed based on their distances to their corresponding reference strings. The proposed approach supports both range and $k$-NN queries and can be integrated inside a DBMS. In \texttt{MultiTree}~\cite{DBLP:conf/icde/ZhangLWZXY17}, each element in a set is represented as a vector and is mapped to an integer according to a defined global ordering, which is then used to insert the element in the B$^+$-tree index. Searching for similar elements is then done via a range query on the index.

On another line of research, \texttt{Trie-Join}~\cite{DBLP:journals/pvldb/WangLF10} 
proposes a trie-based technique for string similarity joins with edit distance. Each trie node represents a character in the string. Thus, strings with a common prefix share the same ancestors. 
A trie node is called an \textit{active node} of a string $s$ if their edit distance is not larger than the given threshold. This leads to a technique called \textit{subtrie pruning}: 
given a trie $T$ and a string $s$, if node $n$ is not an active node for every prefix of $s$, then $n$'s descendants cannot be similar to $s$. \texttt{HSTree}~\cite{DBLP:journals/vldb/YuWLZDF17} recursively partitions strings into disjoint segments and builds a hierarchical segment tree index. This is then used to support both threshold-based and top-$k$ string similarity
search based on edit distance. Finally, a transformation-based framework for top-$k$ set similarity search is presented in~\cite{zhang2018transformation}. It transforms sets of various lengths into fixed-length vectors in such a way that similar sets are mapped closer to each other. An R-tree is then used to index these records and prune the space during search.

\subsection{Parallel \& Distributed Algorithms}
\label{subsec:filtering_distributed}

MapReduce-based approaches have been proposed to tackle scalability issues when dealing with very large collections of sets or strings. A theoretical analysis of different methods for performing similarity joins on MapReduce is presented in~\cite{DBLP:conf/icde/AfratiSMPU12}. 
It considers algorithms that 
operate in a single MapReduce job, avoiding the overhead associated with initiating multiple ones. 
It shows that different algorithms provide different tradeoffs with respect to map, reduce and 
communication~cost.

\texttt{VernicaJoin}~\cite{DBLP:conf/sigmod/VernicaCL10} is based on prefix filtering. It computes prefix tokens and builds an inverted index on them. Then, it generates candidate pairs from the inverted lists, using additionally the length, positional and suffix filters to prune candidates. A deduplication step is finally employed to remove duplicate result pairs generated from different reducers. \texttt{MGJoin}~\cite{DBLP:journals/tkde/RongLWDCT13} follows a similar approach to \texttt{VernicaJoin}, but introduces multiple prefix orders and a load balancing technique that partitions sets based on their length. \texttt{MRGroupJoin}~\cite{DBLP:journals/pvldb/DengLWF15} is a MapReduce extension of \texttt{PTJ}~\cite{DBLP:journals/pvldb/DengLWF15}. It applies a partition-based technique, where records are grouped by length and are partitioned in subrecords, such that matching records share at least one subrecord. The process is performed in a single MapReduce job. \texttt{FS-Join}~\cite{DBLP:conf/icde/RongLSWLD17} sorts the tokens in each set in increasing order of frequency, and then splits each set into disjoint subsets using appropriate pivot tokens. These subsets are then grouped together so that subsets from different groups are non-overlapping.

Finally, \texttt{Dima}~\cite{DBLP:journals/pvldb/SunSLDB17,DBLP:journals/pvldb/SunS0BD19} is a distributed in-memory system built on top of Spark that supports threshold and top-$k$ similarity search and join with both token-based and character-based similarities. It relies on signature-based global and local indexes for efficiency. The proposed signatures are adaptively selectable based on the workload, which allows to balance the workload among partitions. \texttt{Dima} extends the Catalyst optimizer of Spark SQL to introduce cost-based optimizations.

\subsection{Approximate Algorithms}
\label{subsec:filtering_approx}

Approximate algorithms for similarity search and join
increase the efficiency of Filtering step at the cost of allowing both false positives and false negatives, thus missing some matches
\cite{DBLP:journals/corr/WangSSJ14,DBLP:series/synthesis/2013Augsten}. They typically rely on \textit{locality sensitive hashing} (\textsf{LSH})~\cite{DBLP:conf/vldb/GionisIM99}, which transforms an item to a low-dimensional representation such that similar items have much higher probability to be mapped to the same hash code than dissimilar ones. This property allows \textsf{LSH} to be exploited in the filtering phase to generate candidates~\cite{DBLP:conf/vldb/LvJWCL07,DBLP:conf/sigmod/TaoYSK09,DBLP:conf/compgeom/DatarIIM04}. The basic idea is that each object is hashed several times using randomly chosen hash functions. Then, candidates are those pairs of objects that have been hashed to the same code by at least one hash function. 

\texttt{ATLAS}~\cite{DBLP:conf/sigmod/ZhaiLG11} is a probabilistic algorithm that is based on random permutations both to generate candidates and to estimate the similarity between candidate pairs. It also proposes a method to efficiently detect cluster structures within the data, which are then exploited to search for similar pairs only within each cluster.
\texttt{BayesLSH}~\cite{DBLP:journals/pvldb/SatuluriP12} combines Bayesian inference with \textsf{LSH} to estimate similarities to a user-specified level of accuracy. It uses \textsf{LSH} 
for both Filtering 
and Verification,
providing probabilistic guarantees on the resulting
accuracy and recall. \texttt{CPSJoin}~\cite{DBLP:conf/icde/ChristianiPS18} is a randomized algorithm for set similarity joins. It uses a recursive filtering technique, building upon a previously proposed index for set similarity search~\cite{DBLP:conf/stoc/ChristianiP17}, as well as sketches for estimating set similarity. The algorithm has 100\% precision and provides a probabilistic guarantee on recall.

\subsection{Algorithms for Complex Matching}
\label{subsec:filtering_advanced}

The works discussed so far assume a single similarity predicate, i.e, they apply to the values of a specific attribute. Moreover, when comparing sets, they assume binary matching between their 
elements,
while in the case of strings, 
they compare strings in their entirety. In the following, we present methods that employ \textit{multiple} similarity predicates or more complex ones.

\textbf{Local matching.}
A local string similarity join
finds pairs of strings that contain similar \textit{substrings}. Under edit distance constraints, it can be defined as matching any $l$-length substring with up to $k$ errors. \texttt{LS-Join}~\cite{DBLP:journals/tkde/WangYWL17} is based on the observation that if two strings are locally similar, they must share at least one common $q$-gram, for a suitably calculated gram length $q$. An inverted index is constructed incrementally during the search. For every examined string, its $q$-grams are generated and the candidates are retrieved from the index by finding those strings that have matching $q$-grams.

\texttt{pkwise}~\cite{DBLP:conf/sigmod/WangXQWZI16} detects pieces of text in a given collection that share similar \emph{sliding windows}, i.e., multisets containing $w$ consecutive tokens of a given document. The similarity of two sliding windows is defined as the overlap of those sets. Prefix filtering is used but instead of relying on single tokens to build the signatures, it proposes \emph{$k$-wise signatures}, which comprise combinations of $k$ tokens. Larger values of $k$ increase the signatures' selectivity but also the cost of signature generation.
An additional optimization is to share common signatures across adjacent windows.

Finally, \texttt{pkduck}~\cite{DBLP:journals/pvldb/TaoDS17} 
matches strings with \textit{abbreviations},
based on a new similarity measure 
that accounts for
abbreviations.
It also proposes an appropriate signature scheme that extends prefix filtering and generates signatures without iterating over all strings derived from an
abbreviation.

\textbf{Fuzzy matching.} Rather than assuming a binary match, in this setting, the similarity between the elements of two sets may take any value between 0 and 1. In fact, it is defined 
as the maximum matching score in the bipartite graph representing the matches between their elements.

In \texttt{Fast}-\texttt{Join}~\cite{DBLP:journals/tods/WangLF14}, edge weights in this bipartite graph denote the edit similarities between matching elements. The proposed method follows the filter-verification framework, creating a signature for each set such that matching sets have overlapping signatures. The signature of a set comprises an appropriately selected subset of its tokens. \texttt{SilkMoth}~\cite{DBLP:journals/pvldb/DengKMS17} generalizes and improves upon this work, providing a formal characterization of the space of valid signatures. 
Given
that finding the optimal signature is NP-complete, it proposes heuristics to select signatures. To further reduce candidates, a refinement step is added: it compares each set with its candidates and rejects those for which certain bounds do not hold. 
Both edit distance and Jaccard coefficient are supported 
for measuring the similarity between elements. \texttt{MF-Join}~\cite{DBLP:conf/icde/WangLZ19} performs element-~and~record-level filtering. The former utilizes a partition-based signature scheme with a frequency-aware partition strategy, while the latter exploits count filtering and an upper bound on record-level similarity.

\textbf{Multiple predicates.} A method for similarity search and join on \textit{multi-attribute} data is presented in~\cite{DBLP:conf/sigmod/LiHDL15}. For instance, given an entity collection 
where each entity is described
by its name and address, this work identifies pairs of entities having \textit{both} similar names and similar addresses. To enable simultaneous filtering on multiple attributes, a combined prefix tree index is built on these attributes. The construction of the index is guided by a cost model and a greedy algorithm. In another direction, \texttt{Smurf}~\cite{DBLP:journals/pvldb/CADA18} performs string matching between two collections of strings based on multiple-predicate matching conditions in the form of a \textit{random forest} classifier that is learned via active learning. Filtering techniques for string similarity joins are exploited to speed up the execution of the random forest. The focus and novelty of this work is on how to reuse computations across the trees in the forest to further increase efficiency. Finally, \texttt{AU}-\texttt{Join}~\cite{DBLP:journals/pvldb/XuL19} presents a new framework for string similarity joins that supports not only syntactic similarity measures, such as Jaccard similarity on $q$-grams, but also \textit{semantic} similarities, including \textit{synonym-based} and \textit{taxonomy-based} matching. It 
partitions strings into segments and applies
different types of similarity measures on different pairs of segments. A new signature scheme, called \textit{pebble}, 
handles multiple similarity measures:
pebbles are  $q$-grams for gram-based similarity,
the left-hand side of a synonym rule for synonym similarity, and
ancestor nodes in the taxonomy for taxonomy similarity.

\vspace{-10pt}
\subsection{Discussion \& Experimental Results}
\label{subsec:filtering_discussion}

Filtering techniques for string and set similarity joins have attracted a lot of research interest over the past two decades. Early works 
view this operation as an extension of the standard join operator in relational databases, where the join condition is based on similarity rather than equality~\cite{DBLP:conf/vldb/GravanoIJKMS01,DBLP:conf/icde/ChaudhuriGK06}. The same perspective is shared by more recent works, like those 
proposing B$^+$-tree based indexes, which 
can be easily integrated into an existing DBMS
\cite{DBLP:conf/sigmod/ZhangHOS10,DBLP:journals/tkde/LuDHO14,DBLP:conf/icde/ZhangLWZXY17}. Another characteristic example is Dima~\cite{DBLP:journals/pvldb/SunSLDB17,DBLP:journals/pvldb/SunS0BD19}, which extends the Catalyst optimizer of Spark SQL to 
support 
similarity-based queries. In this sense, similarity joins are sometimes referred to as \textit{approximate} or \textit{fuzzy} joins, although this should not be confused with the approximate algorithms in Sec.~\ref{subsec:filtering_approx}, or the fuzzy set joins in Sec.~\ref{subsec:filtering_advanced}. Numerous Filtering techniques have been proposed by more recent works, which focus on main memory execution. 
\textit{Prefix-based} filtering is the most popular
approach~\cite{DBLP:conf/icde/ChaudhuriGK06,DBLP:conf/www/BayardoMS07,DBLP:conf/icde/LiLL08,DBLP:journals/pvldb/XiaoWL08,DBLP:conf/sigmod/QinWLXL11,DBLP:journals/tkde/WangQXLS13,DBLP:conf/www/XiaoWLY08,DBLP:journals/tods/XiaoWLYW11,DBLP:conf/www/XiaoWLY08,DBLP:journals/tods/XiaoWLYW11,DBLP:journals/is/RibeiroH11,DBLP:journals/pvldb/BourosGM12,DBLP:conf/sigmod/WangLF12}, 
followed by 
\textit{partition-based} filtering
\cite{DBLP:conf/vldb/ArasuGK06,DBLP:journals/pvldb/LiDWF11,DBLP:journals/pvldb/DengLWF15}. Furthermore, to scale similarity joins to large collections, distributed~\cite{DBLP:conf/icde/AfratiSMPU12,DBLP:conf/sigmod/VernicaCL10,DBLP:journals/tkde/RongLWDCT13,DBLP:journals/pvldb/DengLWF15,DBLP:conf/icde/RongLSWLD17,DBLP:journals/pvldb/SunSLDB17,DBLP:journals/pvldb/SunS0BD19} and approximate~\cite{DBLP:conf/sigmod/ZhaiLG11,DBLP:journals/pvldb/SatuluriP12,DBLP:conf/icde/ChristianiPS18} algorithms have been proposed.

More recently, there has been an increasing focus and interest on works that deal with more complex similarity predicates. These include the matching of strings based on \textit{substrings} or \textit{abbreviations}~\cite{DBLP:journals/tkde/WangYWL17,DBLP:conf/sigmod/WangXQWZI16,DBLP:journals/pvldb/TaoDS17}, matching of sets based on \textit{fuzzy matching} of their elements~\cite{DBLP:journals/tods/WangLF14,DBLP:journals/pvldb/DengKMS17,DBLP:conf/icde/WangLZ19}, and the combination of \textit{multiple} similarity predicates~\cite{DBLP:conf/sigmod/LiHDL15,DBLP:journals/pvldb/CADA18}. These works can be considered as more closely relevant to matching entity profiles in Entity Resolution.

Regarding performance,
a series of experimental analyses provides interesting insights \cite{DBLP:journals/pvldb/JiangLFL14,DBLP:journals/pvldb/MannAB16,DBLP:journals/pvldb/FierABLF18}.
However, each study focuses on a certain subset of the aforementioned methods. 
Below, we 
briefly summarize their findings, 
including 
additional results from individual papers to fill the gaps.

\textit{Similarity joins using Edit Distance}. A comparison between \texttt{FastSS}, \texttt{All-Pairs}, \texttt{DivideSkip}, \texttt{Ed-Join}, \texttt{QChunk}, \texttt{VChunkJoin}, \texttt{PPJoin}, \texttt{PPJoin}+, \texttt{AdaptJoin}, \texttt{PartEnum}, \texttt{PassJoin} and \texttt{Trie}-\texttt{Join} is conducted in~\cite{DBLP:journals/pvldb/JiangLFL14}. The results demonstrate that \texttt{PassJoin} is the most efficient 
algorithm, with \texttt{FastSS} providing a reliable alternative 
in the case of very short strings.
    
\textit{Similarity joins using set-based measures}. \texttt{AdaptJoin} and \texttt{PPJoin}+ are reported as the best algorithms in the aforementioned study~\cite{DBLP:journals/pvldb/JiangLFL14}. Different results, though, are reported in 
a subsequent 
study
that compares \texttt{All-Pairs}, \texttt{PPJoin}, \texttt{PPJoin}+, \texttt{MPJoin}, \texttt{AdaptJoin} and \texttt{GroupJoin}.
It indicates that the plain prefix filtering, i.e., \texttt{All-Pairs}, is still quite competitive, 
winning in the majority of cases. 
\texttt{PPJoin} and \texttt{GroupJoin} 
exhibit
the best median and average performance, respectively, while
more sophisticated filters are found to provide only moderate improvements in some cases or even to negatively affect performance.
The difference with the 
results 
in \cite{DBLP:journals/pvldb/JiangLFL14} is attributed to the 
more efficient  
verification step; 
reducing the cost of Verification means that 
complex and, thus, time-consuming filters 
often do not pay off, despite 
reducing the number of candidate pairs.

\textit{Prefix vs. partition filtering}. \texttt{PTJ} is compared against \texttt{PPJoin+} and \texttt{AdaptJoin} in \cite{DBLP:journals/pvldb/DengLWF15}, showing that it outperforms both methods. 
The same comparison is performed in \cite{DBLP:journals/pvldb/WangQLZC17}, showing that \texttt{PTJ} does not outperform those methods in most cases. As noted in~\cite{DBLP:journals/pvldb/WangQLZC17}, this discrepancy seems to be 
caused by 
differences in implementation; specifically, the comparison in~\cite{DBLP:journals/pvldb/DengLWF15} 
uses the original implementations of \texttt{PPJoin+} and \texttt{AdaptJoin}, while the one in~\cite{DBLP:journals/pvldb/WangQLZC17} uses the optimized implementations provided by~\cite{DBLP:journals/pvldb/MannAB16}. Overall, \texttt{PTJ} may generate fewer candidates, but uses complex index structures, thus spending much more time on the filtering phase compared to prefix-based
algorithms. Another factor that affects the performance of prefix filtering is the frequency distribution of the tokens in the dataset. The core idea of prefix filtering is to select rare tokens as signatures
so as to reduce the number of candidates. In~\cite{DBLP:journals/pvldb/JiangLFL14}, an experiment 
involving different dataset distributions shows that 
\texttt{PPJoin(+)} and \texttt{AdaptJoin} perform better in datasets with Zipfian distribution than uniform one.

\textit{Set relations}. Another interesting finding is that set relations can be effectively exploited to speed up the computation of similarity joins~\cite{DBLP:journals/pvldb/WangQLZC17}. In the presented experiments, the proposed algorithm, \texttt{SKJ}, 
consistently outperforms \texttt{PPJoin}, \texttt{PPJoin+}, \texttt{AdaptJoin} and \texttt{PTJ} across all datasets.

\textit{Tree-based algorithms}. These algorithms typically focus on similarity search rather than join. \texttt{HSTree} and \texttt{PBI} are compared against \texttt{B}$^{ed}$-\texttt{Tree} in~\cite{DBLP:journals/vldb/YuWLZDF17} and~\cite{DBLP:journals/tkde/LuDHO14}, respectively, reporting better performance. Also, \texttt{Trans} shows better performance than \texttt{MultiTree} in~\cite{zhang2018transformation}. \texttt{BiTrieJoin}, an improved variant of \texttt{TrieJoin}, 
is reported in~\cite{DBLP:journals/pvldb/JiangLFL14} to have comparable performance to \texttt{PassJoin} for short strings, but it underperforms
for medium and long strings.

\textit{Distributed algorithms}. \texttt{VernicaJoin}, \texttt{MGJoin}, \texttt{MRGroupJoin} and \texttt{FS}-\texttt{Join} are experimentally compared in~\cite{DBLP:journals/pvldb/FierABLF18}. 
\texttt{VernicaJoin} exhibits the best performance
in most cases, but  
all algorithms are often outperformed by non-distributed ones. This should be attributed 
to the overhead introduced by the MapReduce framework 
as well as 
to high or skewed data replication between map and reduce tasks. The latter constitutes 
an inherent limitation of the distributed algorithms that cannot be overcome by simply increasing the number of nodes in the cluster. In~\cite{DBLP:journals/pvldb/SunS0BD19}, \texttt{Dima} is shown to outperform
the adaptation of \texttt{VernicaJoin}
to Apache Spark.

\textit{Approximate algorithms}. The experimental survey in~\cite{DBLP:journals/pvldb/JiangLFL14} included a comparison between \texttt{BayesLSH}-\texttt{lite} and exact algorithms. Moreover, \texttt{ATLAS}, \texttt{BayesLSH} and \texttt{CPSJoin} have been compared against \texttt{All-Pairs} in~\cite{DBLP:conf/sigmod/ZhaiLG11}, \cite{DBLP:journals/pvldb/SatuluriP12} and \cite{DBLP:conf/icde/ChristianiPS18}, respectively. Overall, the experiments indicate that approximate algorithms are preferable for low similarity thresholds, e.g., for Jaccard similarity below 0.5, while exact algorithms perform better for high thresholds.

\section{Join-based Blocking Methods}
\label{sec:hybrid}

\begin{figure}[t]\centering
    \includegraphics[width=0.46\linewidth]{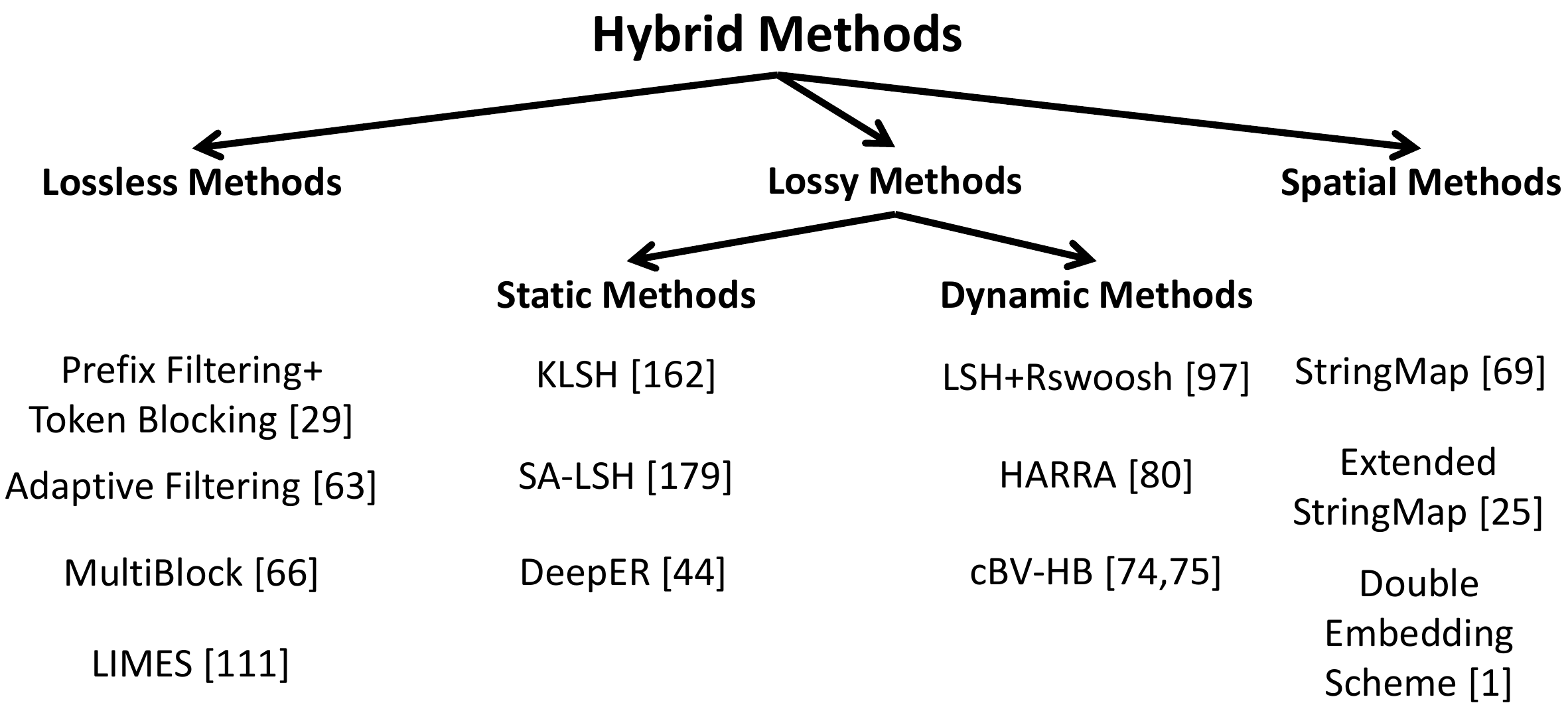}
	\includegraphics[width=0.46\linewidth]{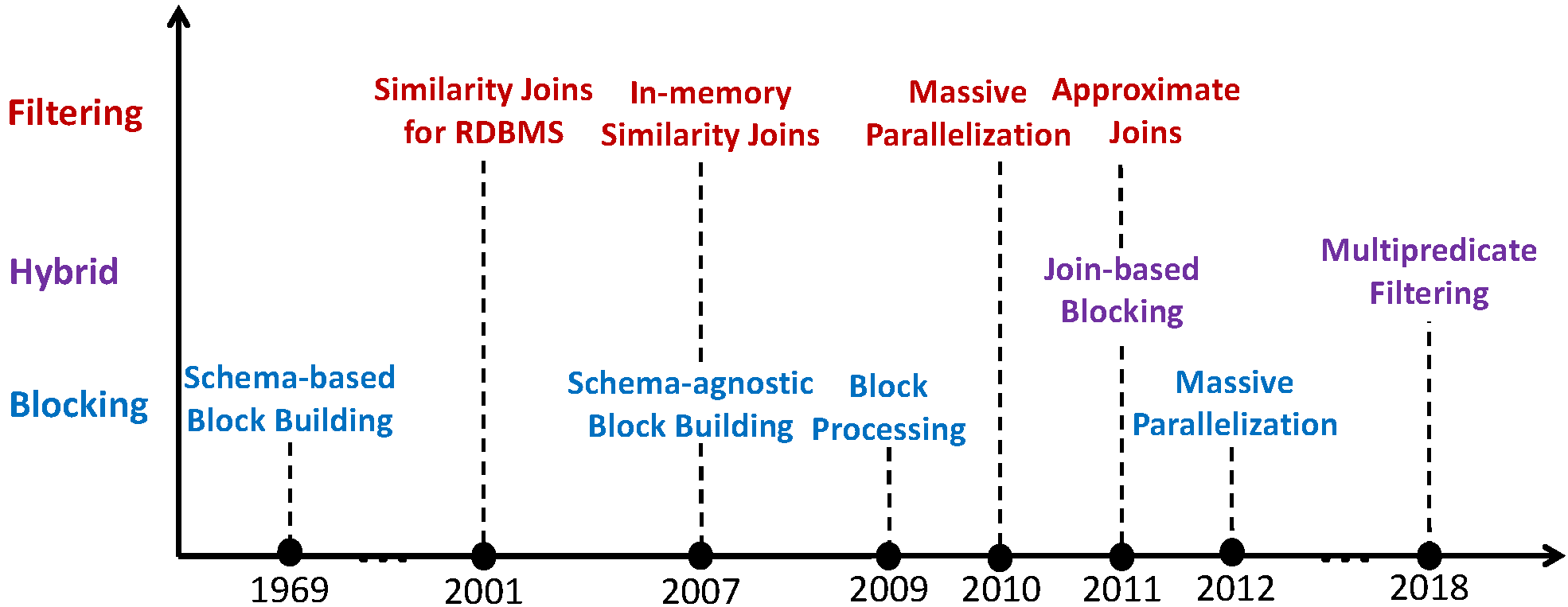}
	\vspace{-8pt}
	\caption{(a) The taxonomy of the hybrid, join-based blocking methods. (b) Timeline of the landmarks in the evolution of {\color{blue}Blocking}, {\color{red}Filtering} and {\color{purple}their convergence}. 
	}
	\label{fig:taxonomyTimeline}
	\vspace{-14pt}
\end{figure}

We now elaborate on Block Building 
methods that incorporate Filtering techniques,
converting Blocking into a nearest neighbor search. As illustrated in Figure \ref{fig:taxonomyTimeline}(a), we categorize these hybrid techniques into three major categories according to the filtering techniques they employ:  the \textit{lossless} ones rely on exact, single predicate filtering techniques (cf. Table \ref{tab:filtering_table}(a)), the \textit{lossy} ones
rely on approximate filtering (cf. Section \ref{subsec:filtering_approx}), while the \textit{spatial} ones
leverage spatial join techniques for filtering. Note that the lossy hybrid methods are further distinguished into \textit{static} and \textit{dynamic} ones, depending on whether they are independent or interwoven with Matching, respectively.

Starting with the lossless hybrid methods, the simplest approach is to combine Prefix Filtering with Token Blocking, creating one block for every token that appears in the prefix of at least two entities \cite{DBLP:series/synthesis/2015Christophides}. Another approach is \textit{Adaptive Filtering} \cite{DBLP:conf/sdm/GuB04}, which couples schema-aware, non-learning Block Building techniques with two filtering methods. First, blocks are created by extracting keys from specific attributes. In every block with a size exceeding a predetermined threshold, Length and Count Filtering are applied for Comparison Cleaning, using an edit distance threshold on an attribute that is not considered by the initial transformation function. 

Another lossless hybrid method is \textit{LIMES}, which 
operates only on metric spaces \cite{DBLP:conf/ijcai/NgomoA11}. Its core idea is to 
leverage the triangle inequality to approximate the distance between entities based on previous comparisons. Utilizing sets of entities as reference points, called \textit{exemplars}, this method computes lower and upper bounds to filter out superfluous comparisons before their execution.

In another direction, \textit{MultiBlock} \cite{isele2011efficient}
optimizes the execution of complex matching rules that comprise special similarity functions for textual, geographic and numeric values. A block collection is created for every similarity function such that similar entities 
share multiple blocks. E.g., edit distance is supported for textual values and blocks are created for character $q$-grams such that entity pairs satisfying the distance threshold co-occur in a sufficient number of blocks. Then, all block collections are aggregated into a multidimensional index 
that respects the co-occurrence patterns of similar entities and guarantees no false dismissals, i.e., $PC$=1.

Regarding the lossy approaches, they are dominated by techniques based 
on \textsf{LSH} \cite{DBLP:conf/vldb/GionisIM99}, which
efficiently estimates the similarity between two attribute values $v_i$ and $v_j$ by randomly sampling hash functions $f$ from a \textit{sim-sensitive} family $F$ such that the probability $Pr(f(v_i) = f(v_j))$ equals to $sim(v_i, v_j)$ for any pair of attribute values and any function $f \in F$. This means that \textsf{LSH} derives $sim(v_i, v_j)$ from the proportion of hash functions $f$ such that $f(v_i)$ = $f(v_j$). Typically, the required number of these functions is relatively small for a sufficiently small sampling error; e.g., for 500 functions, the maximum sampling error is about $\pm$4.5\% with 95\% confidence interval \cite{DBLP:conf/semweb/DuanFHKSW12}.

In the context of ER, LSH is typically combined with MinHash signatures \cite{DBLP:conf/sequences/Broder97}, which efficiently estimate the Jaccard similarity as follows \cite{DBLP:journals/corr/abs-1907-08667,DBLP:conf/psd/SteortsVSF14}. Given an entity collection $\mathcal{E}$, the values of selected attribute names are converted into a bag of $k$-\textit{shingles}, i.e., $k$ consecutive words or characters. Then, a matrix $M$ of size $K \times |\mathcal{E}|$ is formed, with the rows corresponding to the $K$ distinct shingles that appear in all attribute values and the columns to the input entities. The value of every cell $M(i,j)$ indicates whether the entity $e_j$ contains the shingle $s_j$, $M(i,j)$=1, or not, $M(i,j)$=0. Given that $M$ is a sparse matrix, $p$ random minhash functions are used to reduce its dimensionality: they are applied to each column, deriving a new matrix $M'$ of size $p \times |\mathcal{E}|$. 
The $p$ rows are then partitioned into $b$ non-overlapping bands and a hash function is applied to every band of each column. The resulting buckets are treated as blocks that provide probabilistic guarantees that the pairs of similar entities co-occur in at least one block. In fact, the desired probabilistic guarantees can be used for configuring the parameters of \textsf{LSH}, i.e., the number of hash functions, rows and bands \cite{DBLP:journals/corr/abs-1907-08667}.

In this context, LSH is combined with K-Means in \textsf{KLSH} \cite{DBLP:conf/psd/SteortsVSF14}. 
KMeans is applied to the low-dimensional columns of $M'$, which represent the input entities. The resulting clusters form a disjoint block collection $\mathcal{B}$, with $|\mathcal{B}|$ determined by the desired average number of entities per block.

\textsf{LSH} is applied to the distributed representations (i.e., embeddings) of the input entities in \textit{DeepER} \cite{DBLP:journals/corr/abs-1710-00597}. Every entity is transformed into a dense, real-valued vector by aggregating the embeddings of all attribute value tokens, which are pre-trained by word2vec \cite{DBLP:conf/nips/MikolovSCCD13}, Glove \cite{DBLP:conf/emnlp/PenningtonSM14} etc. This vector is then hashed into multiple buckets with \textsf{LSH}. A block is then created for every entity containing its top-$N$ most likely matches, which are detected using Multiprobe-LSH \cite{DBLP:conf/vldb/LvJWCL07}.

\textsf{LSH} is also combined with a semantic similarity in \textsf{SA-LSH} (i.e., semantic-aware LSH) \cite{DBLP:journals/tkde/WangCL16}. 
A taxonomy tree is used to model the concepts that describe the input entity collection.
The semantic similarity of two entities is inversely proportional to the length of the paths that connect the corresponding concepts and their children: the longer the paths, the lower the semantic similarity.
The concepts of every entity are converted into a hash signature through a semantic hashing algorithm. The resulting low-dimensional signatures are directly combined with the 
signatures that are extracted from the n-grams of selected attribute values, capturing the textual similarity of entities. However, the construction of the taxonomy tree requires heavy human intervention.

Regarding the dynamic lossy methods, \textsf{LSH} is combined with \textsf{R-Swoosh} \cite{DBLP:journals/vldb/BenjellounGMSWW09} 
in \cite{DBLP:conf/edbt/MalhotraAS14} through a MapReduce parallelization. Initially, a job is used for defining blocks using \textsf{LSH}. Then, a graph-parallel Pregel-based platform 
applies R-Swoosh,
iteratively executing the non-redundant comparisons in the blocks and computing the transitive closure of the detected duplicates.

LSH also lies at the core of \textsf{cBV-HB} \cite{DBLP:conf/edbt/KarapiperisVVC16,DBLP:journals/kais/KarapiperisV16}, which embeds the textual values of selected attributes into a compact binary Hamming space that is efficient, due to the limited size of its embeddings (e.g., 120 bits for 4 attributes), and preserves the original distances in the sense that certain types of errors correspond to specific distance bounds. Special care is taken to support composite matching rules that involve the main logical operators (i.e., AND, OR and NOT).

Similarly, \textit{HARRA} \cite{DBLP:conf/edbt/KimL10} 
uses \textsf{LSH} to hash similar entities into the same buckets.
Inside every bucket, all pairwise comparisons are executed and 
duplicates
are merged into new profiles. The new profiles are hashed into the existing hash tables and the process is repeated until no entities are merged or another stopping criterion is met (e.g., the portion of merged profiles drops below a predetermined threshold). In every iteration, special care is taken to avoid redundant and superfluous comparisons. 

Finally, spatial hybrid methods combine 
spatial joins with Block Building. The core approach is \textit{StringMap}~\cite{DBLP:conf/dasfaa/JinLM03}, which converts schema-aware blocking keys to a similarity-preserving Euclidean space, whose dimensionality $d$ is heuristically derived from a random sample (typically, $d \in [15, 25]$). For each dimension, a linear algorithm initially selects two pivot attribute values that are (ideally) as far apart as possible. Subsequently, the coordinates of all other attribute values are determined through a comparison with the pivot strings. Using an R-tree or a grid-based index 
in combination 
with two weight
thresholds, similar attribute values are clustered together into overlapping blocks.

This approach is enhanced by \textit{Extended StringMap} \cite{DBLP:journals/tkde/Christen12}, which replaces the weight thresholds with cardinality ones, and the \textit{Double embedding scheme}~\cite{DBLP:conf/dmin/Adly09}. The latter initially maps the input entities to the same $d$-dimensional Euclidean space. Next, the embedded attribute values are mapped to another Euclidean space of lower dimensionality $d' < d$. A similarity join is performed in the second Euclidean space using a $k$-d tree index. The resulting candidate matches are then clustered in the first, $d$-dimensional Euclidean space. The experimental study suggests that the $d'$-dimensional space significantly reduces the runtime of StringMap by 30\% to 60\%.

\section{Blocking vs Filtering: Commonalities and Differences}
\label{sec:discussion}

The timeline in Figure \ref{fig:taxonomyTimeline}(b)
summarizes the landmarks in the evolution of the two frameworks 
showing their gradual convergence. 
We observe that Blocking is the oldest discipline, with the first relevant technique, namely \textsf{SB}, presented in 1969 \cite{fellegi1969theory}. For several decades, research focused on schema-based techniques, with the most significant breakthrough taking place in 1995, with the introduction of \textsf{SN} \cite{DBLP:conf/sigmod/HernandezS95}. The first schema-agnostic Block Building technique is \textit{Semantic Graph Blocking} \cite{DBLP:conf/ideas/NinMML07}, introduced in 2007, but it considers only entity links. In 2011, it was followed by 
\textsf{TB} \cite{DBLP:conf/wsdm/PapadakisINF11}, which exclusively applies to textual values. Block Processing was introduced in 2009 by Iterative Blocking \cite{DBLP:conf/sigmod/WhangMKTG09}, followed by the use of Canopy Clustering for Blocking in 2012 \cite{DBLP:journals/tkde/Christen12} and the introduction of Meta-blocking in 2014 \cite{DBLP:journals/tkde/PapadakisKPN14}.
For Filtering, the first similarity join to be used in an RDBMS can be traced back to 2001 \cite{DBLP:conf/vldb/GravanoIJKMS01}, while the techniques for in-memory execution were coined in 2007  \cite{DBLP:conf/www/BayardoMS07}. Attempts to further increase efficiency by allowing approximate results were first presented in 2011~\cite{DBLP:conf/sigmod/ZhaiLG11}. The first works on massive parallelization for Filtering appear in 2010 \cite{DBLP:conf/sigmod/VernicaCL10}, for Blocking in 2012 \cite{DBLP:journals/pvldb/KolbTR12}, and for Block Processing in 2015 \cite{DBLP:conf/bigdataconf/Efthymiou0PSP15}. The convergence of the two frameworks essentially starts in 2011 with MultiBlock \cite{isele2011efficient}, which introduces Join-based Blocking,
whereas
multiple-predicate Filtering for efficient Matching, 
is introduced by \textsf{Smurf} in 2018 \cite{DBLP:journals/pvldb/CADA18}.

Regarding the qualitative comparison of the two frameworks, we observe that they 
have a number of commonalities: (i) Both serve the same purpose: they increase ER efficiency by reducing the number of performed comparisons. To this end, both employ a stage producing candidate matches, which are subsequently examined analytically in order to remove false positives. (ii) Both usually operate either on two clean but overlapping data collections (Record Linkage for Blocking, Cross-table Join for Filtering) or on a single dirty data collection (Deduplication for Blocking, Self-join for Filtering). (iii) Both extract signatures such that the similarity of two entities is reflected in the similarity of their signatures. (iv) Both also apply similar implementation-level optimizations, representing signatures with integer ids, instead of strings, so as to reduce the memory footprint and facilitate in-memory execution. (v) Both include character- and token-based methods. For Blocking, the former methods mainly pertain to schema-aware techniques that apply character-level transformations to blocking keys (e.g., $q$-grams, suffixes etc), while token-based methods primarily pertain to schema-agnostic methods. For Filtering, similarity measures can also be distinguished between character-based (e.g., edit similarity) and token-based ones (e.g., Jaccard), even though many algorithms can be adapted to handle both. (vi) In both cases, textual data have been combined with other types of data, particularly with spatial or spatio-temporal data, including \cite{DBLP:conf/semweb/Ngomo13} for Blocking and \cite{DBLP:reference/db/Gao09, DBLP:journals/tods/JacoxS07, DBLP:journals/pvldb/BourosGM12, DBLP:journals/vldb/BelesiotisSEKP18} for Filtering. (vii) Both 
can be used in real-time applications, where the input comprises a query entity and the goal is to identify the most similar ones in the minimum possible time. This is called Similarity Search in the case of Filtering and Real-time ER in the case of Blocking (see Section \ref{sec:futureDirections} for more details).

Due to these commonalities, several works use the two frameworks interchangeably, considering Filtering as a means for Blocking (e.g., \cite{DBLP:series/synthesis/2015Christophides}).
In reality, though, Blocking and Filtering have several distinguishing characteristics: (i)
By definition, a blocking scheme applies to a single entity, considering all its attribute values (schema-agnostic methods), or combinations of multiple values (schema-aware techniques). In contrast, Filtering usually applies to a pair of values from the same attribute of two entities. (ii) Blocking relies on positive evidence, clustering together similar entities, while Filtering relies on negative evidence, detecting dissimilar entities early on. (iii) Blocking is typically independent of Entity Matching, whereas Filtering is interwoven with it, as its goal is to optimize the execution of a matching rule. (iv) Blocking is an inherently approximate procedure that falls short of perfect recall ($PC$), even when providing probabilistic guarantees (e.g., LSH Blocking in DeepER \cite{DBLP:journals/corr/abs-1710-00597}). In contrast, most Filtering methods provide an exact solution, returning all pairs of values that exceed the predetermined threshold along with false positives. (v) Blocking trades slightly lower recall ($PC$) for much higher precision ($PQ$), while Filtering trades filtering power for filtering cost. (vi) Blocking may be modelled as a learning problem, where the goal is to define supervised blocking schemes that simultaneously optimize $PC$, $PQ$ and $RR$, but Filtering requires no labelled set for learning to mark a comparison as true negative. Instead, it relies on a theoretical analysis based on the given similarity measure and threshold. (vii) Preserving privacy is orthogonal to Filtering, with very few works examining privacy-preserving similarity joins \cite{DBLP:conf/icde/LiC08,DBLP:journals/dke/KantarciogluIJM09,DBLP:journals/tifs/YuanWWYN17}. In contrast, Blocking constitutes an integral part of privacy-preserving ER, with several relevant works (for details, refer to a recent survey~\cite{DBLP:journals/is/VatsalanCV13}). (viii) Blocking constitutes an integral part of pay-as-you-go ER applications, conveying a significant body of relevant works, as described below. This does not apply to Filtering, given that the only relevant technique is TopkJoin~\cite{DBLP:conf/icde/XiaoWLS09}. 

Regarding the quantitative comparison between 
Blocking and Filtering, few works have actually examined their relative performance. The two frameworks are experimentally juxtaposed in \cite{DBLP:conf/semweb/SongH11,DBLP:journals/tkde/SongLH17,DBLP:conf/semweb/Song12} in terms of effectiveness and time efficiency. 
Using a series of real-world datasets, 
\textsf{RDFKeyLearner} is compared against \textsf{AllPairs}, \textsf{PPJoin}(+) and \textsf{EdJoin} in \cite{DBLP:conf/semweb/SongH11,DBLP:conf/semweb/Song12} 
and against \textsf{EdJoin}, \textsf{PPJoin+} and \textsf{FastJoin} in \cite{DBLP:journals/tkde/SongLH17}.
All methods are fine-tuned using a sample of each dataset.
The outcomes indicate no significant difference in effectiveness,
but regarding time efficiency, Filtering is consistently faster in generating candidate matches and consistently slower in executing the corresponding pairwise comparisons, due to their larger number. In \cite{DBLP:journals/tkde/SongLH17}, the relative scalability of \textsf{RDFKeyLearner} and \textsf{EdJoin} is examined over synthetic datasets of 10$^5$, 2$\cdot$10$^5$, ..., 10$^6$ entities. Again, \textsf{EdJoin} produces more candidate matches and, thus, is slower than \textsf{RDFKeyLearner}.

In \cite{DBLP:journals/pvldb/KopckeTR10}, an experimental analysis over 4 real-world datasets investigates the combined effect of Blocking and Filtering on ER efficiency, implementing the workflow in Figure \ref{fig:computationalCostPlusWorkflow}(b). The results suggest that together, the two frameworks reduce the overall ER running time from 33\% to 76\%, with an average of 50\%. However, only one method per framework is considered:
the manually fine-tuned \textsf{SB} and \textsf{PPJoin} in combination with Cosine and Jaccard similarity. Note that, due to its careful, manual fine-tuning, Blocking has no impact on ER effectiveness. 

However, more experimental analyses are required for drawing safe conclusions about the relative performance of Blocking and Filtering. These analyses should include 
a large, representative variety of techniques per framework along with several established benchmark datasets and should examine the benefits of combining the two frameworks in more depth.

\section{Blocking and Filtering in Entity Resolution Systems}
\label{sec:tools}

We now present the main systems that address ER, examining whether they incorporate any of the aforementioned methods to improve the runtime and the scalability of their workflows. We analytically examined the 18 non-commercial and 15 commercial systems listed in the extended version of \cite{konda2016magellan}\footnote{The extended version of \cite{konda2016magellan} is available here: \url{http://pages.cs.wisc.edu/~anhai/papers/magellan-tr.pdf}.} along with the 10 Link Discovery frameworks surveyed in \cite{DBLP:journals/semweb/NentwigHNR17}. 
Table \ref{tab:LinkDiscoveryToolkits} summarizes the characteristics of 12 open-source ER systems that include at least one Blocking~or~Filtering~method. 

Half of the tools offer a graphical user interface and are implemented in Java. Regarding the type of the input data, most systems support structured data. The only exceptions are the three Link Discovery frameworks, which are crafted for semi-structured data. JedAI is the only tool that applies uniformly to both structured and semi-structured data.

We also observe that all systems include Blocking methods, with Standard Blocking (\textsf{SB}) and Sorted Neighborhood (\textsf{SN}) being the most popular ones. The first four systems are Link Discovery frameworks that implement custom approaches: KnoFuss and SERIMI apply Token Blocking only to the literal values of RDF tiples, while Silk and LIMES implement hybrid methods, MultiBlock and LIMES, respectively (see Section \ref{sec:hybrid}).
Febrl and JedAI offer the largest variety of established techniques. The former provides their original, schema-aware implementation, while the latter provides their schema-agnostic adaptations. For this reason, JedAI is the only tool that implements Block Processing techniques, as well.

Note that Block Building is also a core part of the ER workflow in several commercial systems, such as IBM Infosphere and Informatica Data Quality \cite{konda2016magellan}. These systems are generally required to handle diverse types of data, focusing on data exploration and cleaning. They typically provide variations of \textsf{SB}, allowing users to extract blocking keys from specific attributes through a sophisticated GUI that provides statistics and data analysis. As a result, users' expertise and experience with specific domains is critical for the performance of these systems' blocking components.

Surprisingly, only two systems currently include Filtering algorithms for improving the runtime of their matching process: LIMES and Magellan. The latter actually offers the largest variety of established techniques through the \texttt{py\_stringsimjoin} package. Filtering techniques are also provided by FEVER \cite{DBLP:journals/pvldb/KopckeTR09}, which is a closed-source ER tool, as well as by JedAI's forthcoming version 3. Still, a mere minority of ER tools enables users to combine the benefits of Blocking and Filtering, despite the promising potential of their synergy (see below for more details). Most importantly, these tools exclusively consider traditional Filtering algorithms that apply to the values of individual attributes. Hence, they disregard the recent Filtering techniques for Complex Matching (cf. Section \ref{subsec:filtering_advanced}), which are more suitable for Entity Resolution. Therefore, more effort should be devoted on developing ER tools that make the most of the synergy between Blocking and Filtering.

\begin{table*}[tbp]
\centering
\caption{Blocking and Filtering methods in open-source systems for Entity Resolution. 
}
\label{tab:LinkDiscoveryToolkits}
\vspace{-10pt}
\begin{scriptsize}
\begin{tabular}{|p{1.5cm}|p{4cm}|p{1.5cm}|p{0.4cm}|p{1.0cm}|p{3cm}|}
\toprule
\textbf{Tool}	&	\textbf{Blocking}	&	\textbf{Filtering}	&	\textbf{GUI}	&	\textbf{Language}	&	\textbf{Data Formats}	\\

\midrule
KnoFuss  \cite{nikolov2007knofuss}	&	Literal Blocking	&	-	&	No	&	Java	&	RDF, SPARQL	\\	\hline
SERIMI  \cite{DBLP:journals/tkde/AraujoTVS15}	&	Literal Blocking	&	-	&	No	&	Ruby	&	SPARQL	\\	\hline
Silk \cite{volz2009silk}	&	Multiblock	&	-	&	Yes	&	Scala	&	RDF, SPARQL, CSV	\\	\hline
LIMES \cite{DBLP:conf/ijcai/NgomoA11} & custom methods & PPJoin+, EdJoin, 
custom methods, e.g., ORCHID \cite{DBLP:conf/semweb/Ngomo13}
&	Yes	& Java	&	RDF, SPARQL, CSV	\\	\hline
Dedupe \cite{bilenko2003adaptive}	&	SB with learning-based techniques	&	-	&	No	&	Python	&	CSV, SQL	\\	\hline
DuDe \cite{draisbach2010dude}	&	SB, \textsf{SN}, Sorted blocks	&	-	&	No	&	Java	&	CSV, JSON, XML, BibTex, Databases(Oracle, DB2, MySQL and PostgreSQL)	\\	\hline
Febrl \cite{christen2008febrl}	&	SB, \textsf{SN}, Sorted Blocks, 
Suffix Arrays, Extended Q-Grams, Canopy Clustering, StringMap&	-	&	Yes	&	Python	&	CSV, text-based 	\\	\hline
FRIL \cite{jurczyk2008fine}	&	SB, \textsf{SN}	&	-	&	Yes	&	Java	&	CSV, Excel, COL, Database	\\	\hline
OYSTER \cite{nelson2011entity}	&	SB	&	-	&	No	&	Java	&	text-based	\\	\hline
RecordLinkage \cite{sariyar2011controlling}	&	SB (with SOUNDEX)	&	-	&	No	&	R	&	Database	\\	\hline
Magellan \cite{konda2016magellan} &	SB, \textsf{SN}, it also supports user-specified blocking methods	&	Overlap, Length, Prefix, Position, Suffix	&	Yes	&	Python	&	CSV	\\	\hline
JedAI \cite{DBLP:journals/pvldb/PapadakisTTGPK18}	&	SB, \textsf{SN}, Extended \textsf{SN}, Suffix Arrays, Extended Suffix Arrays, LSH, Q-Grams, Extended Q-Grams + Block Processing	&	to be added in the forthcoming version 3	&	Yes	&	Java	&	CSV, RDF, SPARQL, XML, Database	\\	
& & & & & \\
\bottomrule
\end{tabular}
\end{scriptsize}
\vspace{-12pt}
\end{table*}

It is worth noting that Filtering plays an important role in modern systems. 
For example, \textit{Corleone} \cite{DBLP:conf/sigmod/GokhaleDDNRSZ14} introduced a novel filtering approach that leverages machine learning: active learning is used to minimize the number of examples labeled by users, and then, random forests are trained to learn the matching rules that will be used in filtering. The resulting robust model is scaled by \textit{Falcon} \cite{DBLP:conf/sigmod/DasCDNKDARP17}, while \textit{CloudMatcher} \cite{DBLP:journals/pvldb/GovindPNCDPFCCS18} provides an end-to-end implementation based on this approach. These systems have been successfully applied to real-world domains~\cite{DBLP:conf/sigmod/GovindKCMNLSMBZ19}.

\section{Future Directions}
\label{sec:futureDirections}

Various directions seem promising for future work, from entity evolution \cite{DBLP:conf/jcdl/PapadakisGNPN11} to deep learning \cite{DBLP:journals/corr/abs-1710-00597} and summarization algorithms \cite{DBLP:conf/edbt/KarapiperisGV18}, which minimize the memory footprint of blocks, while accelerating their processing. The following are more mature fields, having assembled a critical mass of methods already.

\textbf{Progressive Entity Resolution.} Due to the constant increase of data volumes, new \textit{progressive} or \textit{pay-as-you-go} ER applications have emerged. Their goal is to provide the best possible \textit{partial solution} within a limited budget of temporal or computational resources. In such applications, Blocking lays the ground for \textit{Prioritization}, which schedules the processing of entities, comparisons or blocks according to the likelihood that they involve duplicates. We distinguish the relevant techniques into schema-aware and schema-agnostic ones.

The schema-aware progressive methods require domain knowledge \cite{DBLP:journals/tkde/PapenbrockHN15,DBLP:journals/tkde/WhangMG13}. \textit{Progressive Sorted Neighborhood} (PSN) \cite{DBLP:journals/tkde/WhangMG13} uses schema-based \textsf{SN} to create a sorted list of entities and then applies an incremental window size $w$. Starting from the top of the list, all entities in consecutive positions ($w$=1) are compared; then, all entities at distance $w$=2 are compared and so on and so forth, until reaching the user-defined budget. \textit{Dynamic PSN} \cite{DBLP:journals/tkde/PapenbrockHN15} extends this static approach by adjusting the processing order of comparisons on-the-fly, according to the results of a perfect matcher. It arranges the sorted entities in a two-dimensional array $A$, and if $A(i, j)$ corresponds to a pair of duplicates, the processing moves on to check $A(i+1, j)$ and $A(i, j+1)$, as well. \textit{Progressive Blocking} \cite{DBLP:journals/tkde/PapenbrockHN15} generalizes this principle to \textsf{SB}. \textit{Hierarchy of Record Partitions} \cite{DBLP:journals/tkde/WhangMG13} creates a static hierarchy of blocks, where the matching likelihood of two entities is proportional to the level in which they co-occur for the first time. This hierarchy is then progressively resolved, level by level, from leaves to root. A variation of this approach is adapted to MapReduce in \cite{DBLP:conf/icde/AltowimM17}, while the \textit{Ordered List of Records} \cite{DBLP:journals/tkde/WhangMG13} converts it into a list of entities that are sorted by their likelihood to produce matches. A progressive solution for relational Multi-source ER over different entity types is proposed in \cite{DBLP:journals/pvldb/AltowimKM14}. Finally, \textsf{P-RDS} adapts LSH-based blocking to a progressive functionality by rearranging the processing order of its hash tables according to the number of matching and unnecessary comparisons in their buckets that have been resolved so far.

The schema-agnostic methods, which disregard any domain knowledge, are classified into two
types \cite{simonini2018schema}: (i) The \textit{sort-based methods} order all entities alphabetically, according to their attribute value tokens, leveraging schema-agnostic SN. \textit{Local Schema-agnostic Progressive SN} \cite{simonini2018schema} slides an incremental window over the sorted list of entities and, for each window size, it orders the non-redundant comparisons according to the co-occurrence frequency of their entities and the number of blocking keys per entity. \textit{Global Schema-agnostic Progressive SN} \cite{simonini2018schema} does the same, but for a predetermined range of windows, eliminating all redundant comparisons they contain. (ii) The \textit{hash-based methods} leverage the blocking graph for Prioritization. \textit{Progressive Block Scheduling} \cite{simonini2018schema} orders the blocks in ascending number of comparisons and then prioritizes all comparisons per block in decreasing edge weight. \textit{Progressive Profile Scheduling} \cite{simonini2018schema} orders entities in decreasing average edge weight and then prioritizes all comparisons per entity in decreasing edge weight.

The schema-agnostic methods excel in recall and precision \cite{simonini2018schema}, but exclusively support \textit{static} prioritization, defining an immutable processing order that disregards the detection of duplicates. Hence, more research is needed for developing \textit{dynamic schema-agnostic} progressive methods.

\textbf{Real-time Entity Resolution.} This is the task of matching an entity that is given as query
to the available entity collections in (ideally) sub-second run-time. To meet this goal, several
specialized \textit{dynamic indexing} techniques have been proposed in the literature. An early approach is presented in \cite{christenDI}. The core idea is to pre-calculate similarities between the attribute values of entities co-occurring in the blocks of Standard Blocking, thus avoiding similarity calculations at query time. Three indexes are created for this purpose, containing all the necessary information. This approach is extended by \textit{DySimII} \cite{10.1007/978-3-642-40319-4_5} so that all three indexes are updated as query entities arrive. The experimental results demonstrate that both the average record insertion time and the average query time remain practically stable, even when the index size grows.

Another family of relevant techniques extends SN. \textit{F-DySNI}  \cite{DBLP:conf/cikm/RamadanC14,Ramadan:2015:DSN:2836847.2816821} converts the sorted list of blocking keys into an index that is faster to search: it creates a braided AVL tree \cite{rice2007braided} that combines a height balanced binary tree with a double-linked list, where every node is linked to its alphabetically sorted predecessor node, to its successor node and to the ids of all entities that correspond to its blocking key. There is one tree for each blocking key definition that gets updated whenever a query entity arrives. The window is fixed or adaptive, considering as neighbors the nodes that exceed a specific similarity threshold. F-DySNI is extended in \cite{DBLP:conf/pakdd/RamadanC15} with an automatic approach for selecting blocking keys; the weak training set of \cite{DBLP:conf/icdm/KejriwalM13} is coupled with a scoring function that assesses the coverage of each key along with the distribution of its block sizes.

Another group of methods relies on LSH. MinHash LSH is combined with SN in \cite{DBLP:conf/pakdd/LiangWCG14}: when searching for the nearest neighbors of a query entity, the entities in large LSH blocks are sorted via a custom scoring function and, then, a window of fixed size slides over the sorted list of entities. \textsf{CF-RDS} \cite{DBLP:journals/datamine/KarapiperisGV18} leverages Hamming LSH, ranking the most similar entities to each query without performing any profile comparison. Instead, it merely aggregates the number of occurrences of each candidate match in the buckets associated with the query entity.

On another line of research, \textit{BlockSketch} \cite{DBLP:conf/edbt/KarapiperisGV18} organizes the entities inside every block into sub-blocks according to their similarity. A representative is assigned to each sub-block based on its distance from the corresponding blocking key. In this way, every query suffices to be compared with a constant number of entities in the target block in order to detect its most similar entities. \textit{SBlockSketch} \cite{DBLP:conf/edbt/KarapiperisGV18} adapts this approach to a stream of query entities through an eviction strategy that bounds the number of blocks that need to be maintained in memory.

All these methods are crafted for structured data, assuming a fixed schema of known quality.
New techniques are required, though, for the noisy, heterogeneous entities of semi-structured data.

\textbf{Parameter Configuration.} Except \textsf{TB}, all Blocking methods involve at least one internal parameter that affects their performance to a large extent \cite{DBLP:journals/tkde/Christen12,DBLP:journals/pvldb/0001SGP16}. This affects their relative performance, rendering the selection of the best performing method for the data at hand into a non-trivial task.

To mitigate this issue, parameter fine-tuning is modelled as an optimization problem in \cite{DBLP:journals/tlsdkcs/MaskatPE16}. The large, heterogeneous space of possible configurations is searched through a genetic algorithm, whose fitness function exploits the labels (i.e., \texttt{match} vs \texttt{non-match}) of part of the candidate matches. After applying the typical series of genetic operators, (i.e., mutation, crossover, elite capture and parental selection) is applied for a specific number of generations, the configuration maximizing the fitness function is selected as optimal. However, this approach involves a large number of parameters itself. In another direction, \textit{MatchCatcher} \cite{DBLP:conf/edbt/LiKCDSPKDR18} implements a human-in-the-loop approach combining expert knowledge with labelled instances in order to learn composite blocking schemes. Using string similarity joins, duplicates sharing no block are efficiently detected. To capture them, the expert user adapts the transformation and assignment functions iteratively. Finally, a method’s performance over several labelled datasets is used for fine-tuning its parameters over a given unlabelled dataset in \cite{o2018new}. At its core lies a two-dimensional metric space formed by the overall running time and F-Measure (horizontal and vertical axis, respectively). The closer a method is mapped to the ideal point (0,1), the better is its performance. A graph is then built such that every node corresponds to a different configuration or blocking method, while a directed edge points from node $n_i$ to $n_j$ if $n_j$ is closer to (0,1). The node with no outgoing edges or the largest difference between incoming and outgoing edges corresponds to the best choice. However, this is a rather time-consuming approach, given the large number of computations it requires.

None of the above methods satisfies the requirement for automatic, data-driven, a-priori parameter configuration of Blocking methods, which thus remains an open problem.

\textbf{Filtering for Entity Resolution.} We believe that more opportunities exist for transferring ideas and approaches between Blocking and Filtering. Another interesting direction is to investigate in practical settings to what extent similarity joins suffice for ER, i.e., representing entity profiles by strings or sets and defining a matching function based on a similarity threshold. We expect that techniques supporting relaxed matching criteria and/or lower similarity thresholds will be required to achieve high recall. Yet, as explained in Section \ref{subsec:filtering_advanced}, relatively few Filtering techniques are designed for these cases. Moreover, scalability remains an open challenge for string and set similarity joins, as shown in \cite{DBLP:journals/pvldb/FierABLF18}. Finally, there is a need for extensible, open-source ER tools that incorporate the majority of established Blocking and Filtering methods and apply seamlessly to structured, semi-structured and unstructured data \cite{DBLP:conf/pods/GolshanHMT17}.

\section{Conclusions}
\label{sec:conclusions}

Efficiency techniques are an integral part of Entity Resolution, since its infancy. We organize the 
relevant works
into Blocking, Filtering and hybrid techniques, facilitating their understanding and use. We also provide an in-depth coverage of each category, further classifying its 
works into novel sub-categories. Lately, 
the rise of big semi-structured data
poses challenges 
to the scalability of efficiency techniques and
to their core assumptions: the requirement of Blocking for schema knowledge and of Filtering for high similarity thresholds. The former led to the introduction of schema-agnostic Blocking and of Block Processing techniques, while the latter led to 
more relaxed criteria of similarity. We cover these new fields in detail, putting in context all relevant works. 

\vspace{4pt}
\noindent
\textbf{Acknowledgements.} This work was partially funded by EU H2020 projects ExtremeEarth (825258) and SmartDataLake (825041).

\def\thebibliography#1{
  \section*{References}
    \vspace{-2pt}
	\scriptsize
  \list
    {[\arabic{enumi}]}
    {\settowidth\labelwidth{[#1]}
     \leftmargin\labelwidth
     \parsep 0pt                
     \itemsep 0pt               
     \advance\leftmargin\labelsep
     \usecounter{enumi}
    }
  \def\newblock{\hskip .11em plus .33em minus .07em}
  \sloppy\clubpenalty10000\widowpenalty10000
  \sfcode`\.=1000\relax
}

\balance
\bibliographystyle{abbrv}
\bibliography{bibliographia}

\begin{thebibliography}{100}

\bibitem{DBLP:conf/dmin/Adly09}
N.~Adly.
\newblock Efficient record linkage using a double embedding scheme.
\newblock In {\em {DMIN}}, pages 274--281, 2009.

\bibitem{DBLP:conf/icde/AfratiSMPU12}
F.~Afrati, A.~D. Sarma, D.~Menestrina, A.~Parameswaran, and J.~Ullman.
\newblock Fuzzy joins using mapreduce.
\newblock In {\em {ICDE}}, pages 498--509, 2012.

\bibitem{DBLP:conf/wiri/AizawaO05}
A.~N. Aizawa and K.~Oyama.
\newblock A fast linkage detection scheme for multi-source information
  integration.
\newblock In {\em WIRI}, pages 30--39, 2005.

\bibitem{DBLP:journals/dke/AllamSK18}
A.~Allam, S.~Skiadopoulos, and P.~Kalnis.
\newblock Improved suffix blocking for record linkage and entity resolution.
\newblock {\em DKE}, 117:98--113, 2018.

\bibitem{DBLP:journals/pvldb/AltowimKM14}
Y.~Altowim, D.~V. Kalashnikov, and S.~Mehrotra.
\newblock Progressive approach to relational entity resolution.
\newblock {\em {PVLDB}}, 7(11):999--1010, 2014.

\bibitem{DBLP:conf/icde/AltowimM17}
Y.~Altowim and S.~Mehrotra.
\newblock Parallel progressive approach to entity resolution using mapreduce.
\newblock In {\em {ICDE}}, pages 909--920, 2017.

\bibitem{DBLP:conf/vldb/ArasuGK06}
A.~Arasu, V.~Ganti, and R.~Kaushik.
\newblock Efficient exact set-similarity joins.
\newblock In {\em VLDB}, pages 918--929, 2006.

\bibitem{DBLP:journals/tkde/AraujoTVS15}
S.~Ara{\'{u}}jo, D.~T. Tran, A.~P. de~Vries, and D.~Schwabe.
\newblock {SERIMI:} class-based matching for instance matching across
  heterogeneous datasets.
\newblock {\em {IEEE} TKDE}, 27(5):1397--1410, 2015.

\bibitem{DBLP:conf/iscc/AraujoPN17}
T.~B. Ara{\'{u}}jo, C.~E.~S. Pires, and T.~P. da~N{\'{o}}brega.
\newblock Spark-based streamlined metablocking.
\newblock In {\em {IEEE} {ISCC}}, pages 844--850, 2017.

\bibitem{DBLP:series/synthesis/2013Augsten}
N.~Augsten and M.~H. B{\"{o}}hlen.
\newblock {\em Similarity Joins in Relational Database Systems}.
\newblock Morgan {\&} Claypool Publishers, 2013.

\bibitem{baxter2003comparison}
R.~Baxter, P.~Christen, and T.~Churches.
\newblock A comparison of fast blocking methods for record linkage.
\newblock In {\em KDD Workshops}, 2003.

\bibitem{DBLP:conf/www/BayardoMS07}
R.~J. Bayardo, Y.~Ma, and R.~Srikant.
\newblock Scaling up all pairs similarity search.
\newblock In {\em WWW}, pages 131--140, 2007.

\bibitem{DBLP:journals/vldb/BelesiotisSEKP18}
A.~Belesiotis, D.~Skoutas, C.~Efstathiades, V.~Kaffes, and D.~Pfoser.
\newblock Spatio-textual user matching and clustering based on set similarity
  joins.
\newblock {\em {VLDB} J.}, 27(3):297--320, 2018.

\bibitem{DBLP:journals/vldb/BenjellounGMSWW09}
O.~Benjelloun, H.~Garcia{-}Molina, D.~Menestrina, Q.~Su, S.~E. Whang, and
  J.~Widom.
\newblock Swoosh: a generic approach to entity resolution.
\newblock {\em {VLDB} J.}, 18(1):255--276, 2009.

\bibitem{DBLP:journals/is/BiancoGD18}
G.~D. Bianco, M.~A. Gon{\c{c}}alves, and D.~Duarte.
\newblock {BLOSS:} effective meta-blocking with almost no effort.
\newblock {\em Inf. Syst.}, 75:75--89, 2018.

\bibitem{DBLP:conf/icdm/BilenkoKM06}
M.~Bilenko, B.~Kamath, and R.~J. Mooney.
\newblock Adaptive blocking: Learning to scale up record linkage.
\newblock In {\em ICDM}, pages 87--96, 2006.

\bibitem{bilenko2003adaptive}
M.~Bilenko and R.~J. Mooney.
\newblock Adaptive duplicate detection using learnable string similarity
  measures.
\newblock In {\em KDD}, pages 39--48, 2003.

\bibitem{BoHuSt07}
T.~Bocek, E.~Hunt, and B.~Stiller.
\newblock {Fast Similarity Search in Large Dictionaries}.
\newblock Technical Report ifi-2007.02, Department of Informatics, University
  of Zurich, April 2007.
\newblock http://fastss.csg.uzh.ch/.

\bibitem{DBLP:journals/pvldb/BourosGM12}
P.~Bouros, S.~Ge, and N.~Mamoulis.
\newblock Spatio-textual similarity joins.
\newblock {\em {PVLDB}}, 6(1):1--12, 2012.

\bibitem{DBLP:conf/sequences/Broder97}
A.~Z. Broder.
\newblock On the resemblance and containment of documents.
\newblock In {\em SEQUENCES}, pages 21--29, 1997.

\bibitem{DBLP:conf/ijcai/CaoCZYLY11}
Y.~Cao, Z.~Chen, J.~Zhu, P.~Yue, C.~Lin, and Y.~Yu.
\newblock Leveraging unlabeled data to scale blocking for record linkage.
\newblock In {\em IJCAI}, pages 2211--2217, 2011.

\bibitem{DBLP:conf/icde/ChaudhuriGK06}
S.~Chaudhuri, V.~Ganti, and R.~Kaushik.
\newblock A primitive operator for similarity joins in data cleaning.
\newblock In {\em ICDE}, pages 5--5, 2006.

\bibitem{christen2008febrl}
P.~Christen.
\newblock Febrl-: an open source data cleaning, deduplication and record
  linkage system with a graphical user interface.
\newblock In {\em KDD}, pages 1065--1068, 2008.

\bibitem{DBLP:books/daglib/0030287}
P.~Christen.
\newblock {\em Data Matching - Concepts and Techniques for Record Linkage,
  Entity Resolution, and Duplicate Detection}.
\newblock Springer, 2012.

\bibitem{DBLP:journals/tkde/Christen12}
P.~Christen.
\newblock A survey of indexing techniques for scalable record linkage and
  deduplication.
\newblock {\em {IEEE} TKDE}, 24(9):1537--1555, 2012.

\bibitem{christenDI}
P.~Christen, R.~W. Gayler, and D.~Hawking.
\newblock Similarity-aware indexing for real-time entity resolution.
\newblock In {\em {CIKM}}, pages 1565--1568, 2009.

\bibitem{DBLP:conf/stoc/ChristianiP17}
T.~Christiani and R.~Pagh.
\newblock Set similarity search beyond minhash.
\newblock In {\em {STOC}}, pages 1094--1107, 2017.

\bibitem{DBLP:conf/icde/ChristianiPS18}
T.~Christiani, R.~Pagh, and J.~Sivertsen.
\newblock Scalable and robust set similarity join.
\newblock In {\em {ICDE}}, pages 1240--1243, 2018.

\bibitem{DBLP:series/synthesis/2015Christophides}
V.~Christophides, V.~Efthymiou, and K.~Stefanidis.
\newblock {\em Entity Resolution in the Web of Data}.
\newblock Morgan {\&} Claypool Publishers, 2015.

\bibitem{DBLP:journals/pvldb/ChuIK16}
X.~Chu, I.~F. Ilyas, and P.~Koutris.
\newblock Distributed data deduplication.
\newblock {\em {PVLDB}}, 9(11):864--875, 2016.

\bibitem{clauset2004finding}
A.~Clauset, M.~E. Newman, and C.~Moore.
\newblock Finding community structure in very large networks.
\newblock {\em Physical review E}, 70(6):066111, 2004.

\bibitem{DBLP:conf/sigmod/DasCDNKDARP17}
S.~Das, P.~S.~G. C., A.~Doan, J.~F. Naughton, G.~Krishnan, R.~Deep, E.~Arcaute,
  V.~Raghavendra, and Y.~Park.
\newblock Falcon: Scaling up hands-off crowdsourced entity matching to build
  cloud services.
\newblock In {\em Proceedings of the 2017 {ACM} International Conference on
  Management of Data, {SIGMOD}}, pages 1431--1446, 2017.

\bibitem{DBLP:conf/compgeom/DatarIIM04}
M.~Datar, N.~Immorlica, P.~Indyk, and V.~S. Mirrokni.
\newblock Locality-sensitive hashing scheme based on p-stable distributions.
\newblock In {\em {SOCG}}, pages 253--262, 2004.

\bibitem{DBLP:conf/cikm/VriesKCC09}
T.~de~Vries, H.~Ke, S.~Chawla, and P.~Christen.
\newblock Robust record linkage blocking using suffix arrays.
\newblock In {\em CIKM}, pages 305--314, 2009.

\bibitem{DBLP:journals/tkdd/VriesKCC11}
T.~de~Vries, H.~Ke, S.~Chawla, and P.~Christen.
\newblock Robust record linkage blocking using suffix arrays and bloom filters.
\newblock {\em {TKDD}}, 5(2):9:1--9:27, 2011.

\bibitem{DeanG04}
J.~Dean and S.~Ghemawat.
\newblock Mapreduce: simplified data processing on large clusters.
\newblock {\em Commun. {ACM}}, 51(1):107--113, 2008.

\bibitem{DBLP:journals/pvldb/DengKMS17}
D.~Deng, A.~Kim, S.~Madden, and M.~Stonebraker.
\newblock Silkmoth: An efficient method for finding related sets with maximum
  matching constraints.
\newblock {\em {PVLDB}}, 10(10):1082--1093, 2017.

\bibitem{DBLP:journals/pvldb/DengLWF15}
D.~Deng, G.~Li, H.~Wen, and J.~Feng.
\newblock An efficient partition based method for exact set similarity joins.
\newblock {\em {PVLDB}}, 9(4):360--371, 2015.

\bibitem{DBLP:series/synthesis/2015Dong}
X.~L. Dong and D.~Srivastava.
\newblock {\em Big Data Integration}.
\newblock Morgan {\&} Claypool Publishers, 2015.

\bibitem{draisbach2010dude}
U.~Draisbach and F.~Naumann.
\newblock Dude: The duplicate detection toolkit.
\newblock In {\em QDB}, 2010.

\bibitem{DBLP:conf/nss/DraisbachN11}
U.~Draisbach and F.~Naumann.
\newblock A generalization of blocking and windowing algorithms for duplicate
  detection.
\newblock In {\em ICDKE}, pages 18--24, 2011.

\bibitem{DBLP:conf/icde/DraisbachNSW12}
U.~Draisbach, F.~Naumann, S.~Szott, and O.~Wonneberg.
\newblock Adaptive windows for duplicate detection.
\newblock In {\em ICDE}, pages 1073--1083, 2012.

\bibitem{DBLP:conf/semweb/DuanFHKSW12}
S.~Duan, A.~Fokoue, O.~Hassanzadeh, A.~Kementsietsidis, K.~Srinivas, and M.~J.
  Ward.
\newblock Instance-based matching of large ontologies using locality-sensitive
  hashing.
\newblock In {\em {ISWC}}, pages 49--64, 2012.

\bibitem{DBLP:journals/corr/abs-1710-00597}
M.~Ebraheem, S.~Thirumuruganathan, S.~R. Joty, M.~Ouzzani, and N.~Tang.
\newblock Distributed representations of tuples for entity resolution.
\newblock {\em {PVLDB}}, 11(11):1454--1467, 2018.

\bibitem{DBLP:conf/bigdataconf/Efthymiou0PSP15}
V.~Efthymiou, G.~Papadakis, G.~Papastefanatos, K.~Stefanidis, and T.~Palpanas.
\newblock Parallel meta-blocking: Realizing scalable entity resolution over
  large, heterogeneous data.
\newblock In {\em IEEE Big Data}, pages 411--420, 2015.

\bibitem{DBLP:journals/is/Efthymiou0PSP17}
V.~Efthymiou, G.~Papadakis, G.~Papastefanatos, K.~Stefanidis, and T.~Palpanas.
\newblock Parallel meta-blocking for scaling entity resolution over big
  heterogeneous data.
\newblock {\em Inf. Syst.}, 65:137--157, 2017.

\bibitem{DBLP:conf/bigdataconf/EfthymiouSC15}
V.~Efthymiou, K.~Stefanidis, and V.~Christophides.
\newblock Big data entity resolution: From highly to somehow similar entity
  descriptions in the web.
\newblock In {\em {IEEE} Big Data}, pages 401--410, 2015.

\bibitem{DBLP:journals/tkde/ElmagarmidIV07}
A.~K. Elmagarmid, P.~G. Ipeirotis, and V.~S. Verykios.
\newblock Duplicate record detection: {A} survey.
\newblock {\em {IEEE} TKDE}, 19(1):1--16, 2007.

\bibitem{DBLP:journals/jidm/EvangelistaCSM10}
L.~Evangelista, E.~Cortez, A.~da~Silva, and W.~M. Jr.
\newblock Adaptive and flexible blocking for record linkage tasks.
\newblock {\em {JIDM}}, 1(2):167--182, 2010.

\bibitem{fellegi1969theory}
I.~P. Fellegi and A.~B. Sunter.
\newblock A theory for record linkage.
\newblock {\em Journal of the American Statistical Association},
  64(328):1183--1210, 1969.

\bibitem{DBLP:journals/pvldb/FierABLF18}
F.~Fier, N.~Augsten, P.~Bouros, U.~Leser, and J.~Freytag.
\newblock Set similarity joins on mapreduce: An experimental survey.
\newblock {\em {PVLDB}}, 11(10):1110--1122, 2018.

\bibitem{DBLP:conf/kdd/FisherCWR15}
J.~Fisher, P.~Christen, Q.~Wang, and E.~Rahm.
\newblock A clustering-based framework to control block sizes for entity
  resolution.
\newblock In {\em KDD}, pages 279--288, 2015.

\bibitem{DBLP:reference/db/Gao09}
D.~Gao.
\newblock Temporal joins.
\newblock In {\em Encyclopedia of Database Systems}, pages 2982--2987. 2009.

\bibitem{DBLP:journals/pvldb/GetoorM12}
L.~Getoor and A.~Machanavajjhala.
\newblock Entity resolution: Theory, practice {\&} open challenges.
\newblock {\em {PVLDB}}, 5(12):2018--2019, 2012.

\bibitem{giang2015machine}
P.~Giang.
\newblock A machine learning approach to create blocking criteria for record
  linkage.
\newblock {\em Health care manag. science}, 18(1):93--105, 2015.

\bibitem{DBLP:conf/vldb/GionisIM99}
A.~Gionis, P.~Indyk, and R.~Motwani.
\newblock Similarity search in high dimensions via hashing.
\newblock In {\em VLDB}, pages 518--529, 1999.

\bibitem{DBLP:conf/sigmod/GokhaleDDNRSZ14}
C.~Gokhale, S.~Das, A.~Doan, J.~F. Naughton, N.~Rampalli, J.~W. Shavlik, and
  X.~Zhu.
\newblock Corleone: hands-off crowdsourcing for entity matching.
\newblock In {\em International Conference on Management of Data ({SIGMOD})},
  pages 601--612, 2014.

\bibitem{DBLP:conf/pods/GolshanHMT17}
B.~Golshan, A.~Y. Halevy, G.~A. Mihaila, and W.~Tan.
\newblock Data integration: After the teenage years.
\newblock In {\em ACM PODS}, pages 101--106, 2017.

\bibitem{DBLP:conf/sigmod/GovindKCMNLSMBZ19}
Y.~Govind, P.~Konda, P.~S.~G. C., P.~Martinkus, P.~Nagarajan, H.~Li,
  A.~Soundararajan, S.~Mudgal, J.~R. Ballard, H.~Zhang, A.~Ardalan, S.~Das,
  D.~Paulsen, A.~S. Saini, E.~Paulson, Y.~Park, M.~Carter, M.~Sun, G.~M. Fung,
  and A.~Doan.
\newblock Entity matching meets data science: {A} progress report from the
  magellan project.
\newblock In {\em Proceedings of the 2019 International Conference on
  Management of Data}, pages 389--403, 2019.

\bibitem{DBLP:journals/pvldb/GovindPNCDPFCCS18}
Y.~Govind, E.~Paulson, P.~Nagarajan, P.~S.~G. C., A.~Doan, Y.~Park, G.~Fung,
  D.~Conathan, M.~Carter, and M.~Sun.
\newblock Cloudmatcher: {A} hands-off cloud/crowd service for entity matching.
\newblock {\em Proc. {VLDB} Endow.}, 11(12):2042--2045, 2018.

\bibitem{DBLP:conf/vldb/GravanoIJKMS01}
L.~Gravano, P.~G. Ipeirotis, H.~V. Jagadish, N.~Koudas, S.~Muthukrishnan, and
  D.~Srivastava.
\newblock Approximate string joins in a database (almost) for free.
\newblock In {\em VLDB}, pages 491--500, 2001.

\bibitem{DBLP:journals/corr/abs-1907-08667}
T.~Gschwind, C.~Miksovic, K.~Mirylenka, and P.~Scotton.
\newblock Fast record linkage for company entities.
\newblock {\em CoRR}, abs/1907.08667, 2019.

\bibitem{DBLP:conf/sdm/GuB04}
L.~Gu and R.~A. Baxter.
\newblock Adaptive filtering for efficient record linkage.
\newblock In {\em SIAM}, pages 477--481, 2004.

\bibitem{DBLP:conf/sigmod/HernandezS95}
M.~A. Hern{\'{a}}ndez and S.~J. Stolfo.
\newblock The merge/purge problem for large databases.
\newblock In {\em SIGMOD}, pages 127--138, 1995.

\bibitem{DBLP:journals/datamine/HernandezS98}
M.~A. Hern{\'{a}}ndez and S.~J. Stolfo.
\newblock Real-world data is dirty: Data cleansing and the merge/purge problem.
\newblock {\em Data Min. Knowl. Discov.}, 2(1):9--37, 1998.

\bibitem{isele2011efficient}
R.~Isele, A.~Jentzsch, and C.~Bizer.
\newblock Efficient multidimensional blocking for link discovery without losing
  recall.
\newblock In {\em WebDB}, 2011.

\bibitem{DBLP:journals/tods/JacoxS07}
E.~H. Jacox and H.~Samet.
\newblock Spatial join techniques.
\newblock {\em {ACM} TODS}, 32(1):7, 2007.

\bibitem{DBLP:journals/pvldb/JiangLFL14}
Y.~Jiang, G.~Li, J.~Feng, and W.~Li.
\newblock String similarity joins: An experimental evaluation.
\newblock {\em {PVLDB}}, 7(8):625--636, 2014.

\bibitem{DBLP:conf/dasfaa/JinLM03}
L.~Jin, C.~Li, and S.~Mehrotra.
\newblock Efficient record linkage in large data sets.
\newblock In {\em {DASFAA}}, pages 137--146, 2003.

\bibitem{jurczyk2008fine}
P.~Jurczyk, J.~J. Lu, L.~Xiong, J.~D. Cragan, and A.~Correa.
\newblock Fine-grained record integration and linkage tool.
\newblock {\em Birth Defects Research Part A: Clinical and Molecular
  Teratology}, 82(11):822--829, 2008.

\bibitem{DBLP:journals/dke/KantarciogluIJM09}
M.~Kantarcioglu, A.~Inan, W.~Jiang, and B.~Malin.
\newblock Formal anonymity models for efficient privacy-preserving joins.
\newblock {\em DKE}, 68(11):1206--1223, 2009.

\bibitem{DBLP:journals/datamine/KarapiperisGV18}
D.~Karapiperis, A.~Gkoulalas{-}Divanis, and V.~S. Verykios.
\newblock Fast schemes for online record linkage.
\newblock {\em Data Min. Knowl. Discov.}, 32(5):1229--1250, 2018.

\bibitem{DBLP:conf/edbt/KarapiperisGV18}
D.~Karapiperis, A.~Gkoulalas{-}Divanis, and V.~S. Verykios.
\newblock Summarization algorithms for record linkage.
\newblock In {\em {EDBT}}, pages 73--84, 2018.

\bibitem{DBLP:conf/edbt/KarapiperisVVC16}
D.~Karapiperis, D.~Vatsalan, V.~S. Verykios, and P.~Christen.
\newblock Efficient record linkage using a compact hamming space.
\newblock In {\em {EDBT}}, pages 209--220, 2016.

\bibitem{DBLP:journals/kais/KarapiperisV16}
D.~Karapiperis and V.~S. Verykios.
\newblock A fast and efficient hamming lsh-based scheme for accurate linkage.
\newblock {\em KAIS}, 49(3):861--884, 2016.

\bibitem{DBLP:conf/icdm/KejriwalM13}
M.~Kejriwal and D.~P. Miranker.
\newblock An unsupervised algorithm for learning blocking schemes.
\newblock In {\em ICDM}, pages 340--349, 2013.

\bibitem{DBLP:conf/semweb/KejriwalM14a}
M.~Kejriwal and D.~P. Miranker.
\newblock A two-step blocking scheme learner for scalable link discovery.
\newblock In {\em OM Workshop}, pages 49--60, 2014.

\bibitem{DBLP:journals/corr/KejriwalM15}
M.~Kejriwal and D.~P. Miranker.
\newblock A {DNF} blocking scheme learner for heterogeneous datasets.
\newblock {\em CoRR}, abs/1501.01694, 2015.

\bibitem{DBLP:journals/is/KenigG13}
B.~Kenig and A.~Gal.
\newblock Mfiblocks: An effective blocking algorithm for entity resolution.
\newblock {\em Inf. Syst.}, 38(6):908--926, 2013.

\bibitem{DBLP:conf/edbt/KimL10}
H.~Kim and D.~Lee.
\newblock {HARRA:} fast iterative hashed record linkage for large-scale data
  collections.
\newblock In {\em EDBT}, pages 525--536, 2010.

\bibitem{DBLP:journals/pvldb/KolbTR12}
L.~Kolb, A.~Thor, and E.~Rahm.
\newblock Dedoop: Efficient deduplication with hadoop.
\newblock {\em {PVLDB}}, 5(12):1878--1881, 2012.

\bibitem{DBLP:conf/icde/KolbTR12}
L.~Kolb, A.~Thor, and E.~Rahm.
\newblock Load balancing for mapreduce-based entity resolution.
\newblock In {\em ICDE}, pages 618--629, 2012.

\bibitem{DBLP:journals/ife/KolbTR12}
L.~Kolb, A.~Thor, and E.~Rahm.
\newblock Multi-pass sorted neighborhood blocking with mapreduce.
\newblock {\em Computer Science - R{\&}D}, 27(1):45--63, 2012.

\bibitem{konda2016magellan}
P.~Konda, S.~Das, P.~S.~G. C., A.~Doan, A.~Ardalan, J.~R. Ballard, H.~Li,
  F.~Panahi, H.~Zhang, J.~F. Naughton, S.~Prasad, G.~Krishnan, R.~Deep, and
  V.~Raghavendra.
\newblock Magellan: Toward building entity matching management systems.
\newblock {\em PVLDB}, 9(12):1197--1208, 2016.

\bibitem{DBLP:journals/pvldb/KopckeTR09}
H.~K{\"{o}}pcke, A.~Thor, and E.~Rahm.
\newblock Comparative evaluation of entity resolution approaches with {FEVER}.
\newblock {\em {PVLDB}}, 2(2):1574--1577, 2009.

\bibitem{DBLP:journals/pvldb/KopckeTR10}
H.~K{\"{o}}pcke, A.~Thor, and E.~Rahm.
\newblock Evaluation of entity resolution approaches on real-world match
  problems.
\newblock {\em {PVLDB}}, 3(1):484--493, 2010.

\bibitem{DBLP:conf/icde/LiLL08}
C.~Li, J.~Lu, and Y.~Lu.
\newblock Efficient merging and filtering algorithms for approximate string
  searches.
\newblock In {\em ICDE}, pages 257--266, 2008.

\bibitem{DBLP:journals/pvldb/LiDWF11}
G.~Li, D.~Deng, J.~Wang, and J.~Feng.
\newblock {PASS-JOIN:} {A} partition-based method for similarity joins.
\newblock {\em {PVLDB}}, 5(3):253--264, 2011.

\bibitem{DBLP:conf/sigmod/LiHDL15}
G.~Li, J.~He, D.~Deng, and J.~Li.
\newblock Efficient similarity join and search on multi-attribute data.
\newblock In {\em SIGMOD}, pages 1137--1151, 2015.

\bibitem{DBLP:conf/edbt/LiKCDSPKDR18}
H.~Li, P.~Konda, P.~Suganthan, A.~Doan, B.~Snyder, Y.~Park, G.~Krishnan,
  R.~Deep, and V.~Raghavendra.
\newblock Matchcatcher: {A} debugger for blocking in entity matching.
\newblock In {\em {EDBT}}, pages 193--204, 2018.

\bibitem{DBLP:conf/icde/LiC08}
Y.~Li and M.~Chen.
\newblock Privacy preserving joins.
\newblock In {\em ICDE}, pages 1352--1354, 2008.

\bibitem{DBLP:conf/pakdd/LiangWCG14}
H.~Liang, Y.~Wang, P.~Christen, and R.~W. Gayler.
\newblock Noise-tolerant approximate blocking for dynamic real-time entity
  resolution.
\newblock In {\em {PAKDD}}, pages 449--460, 2014.

\bibitem{DBLP:journals/tkde/LuDHO14}
W.~Lu, X.~Du, M.~Hadjieleftheriou, and B.~C. Ooi.
\newblock Efficiently supporting edit distance based string similarity search
  using {B} {\textdollar}{\^{}}+{\textdollar}-trees.
\newblock {\em {IEEE} TKDE}, 26(12):2983--2996, 2014.

\bibitem{DBLP:conf/vldb/LvJWCL07}
Q.~Lv, W.~Josephson, Z.~Wang, M.~Charikar, and K.~Li.
\newblock Multi-probe {LSH:} efficient indexing for high-dimensional similarity
  search.
\newblock In {\em {VLDB}}, pages 950--961, 2007.

\bibitem{DBLP:journals/cj/MaDY15}
K.~Ma, F.~Dong, and B.~Yang.
\newblock Large-scale schema-free data deduplication approach with adaptive
  sliding window using mapreduce.
\newblock {\em Comput. J.}, 58(11):3187--3201, 2015.

\bibitem{DBLP:conf/wsdm/MaT13}
Y.~Ma and T.~Tran.
\newblock Typimatch: type-specific unsupervised learning of keys and key values
  for heterogeneous web data integration.
\newblock In {\em WSDM}, pages 325--334, 2013.

\bibitem{DBLP:conf/edbt/MalhotraAS14}
P.~Malhotra, P.~Agarwal, and G.~Shroff.
\newblock Graph-parallel entity resolution using {LSH} {\&} {IMM}.
\newblock In {\em {EDBT Workshops}}, pages 41--49, 2014.

\bibitem{DBLP:conf/gvd/MannA14}
W.~Mann and N.~Augsten.
\newblock {PEL:} position-enhanced length filter for set similarity joins.
\newblock In {\em Grundlagen Datenbanken}, pages 89--94, 2014.

\bibitem{DBLP:journals/pvldb/MannAB16}
W.~Mann, N.~Augsten, and P.~Bouros.
\newblock An empirical evaluation of set similarity join techniques.
\newblock {\em {PVLDB}}, 9(9):636--647, 2016.

\bibitem{DBLP:journals/tlsdkcs/MaskatPE16}
R.~Maskat, N.~W. Paton, and S.~M. Embury.
\newblock Pay-as-you-go configuration of entity resolution.
\newblock {\em T-LSD-KCS}, 29:40--65, 2016.

\bibitem{DBLP:conf/kdd/McCallumNU00}
A.~McCallum, K.~Nigam, and L.~H. Ungar.
\newblock Efficient clustering of high-dimensional data sets with application
  to reference matching.
\newblock In {\em KDD}, pages 169--178, 2000.

\bibitem{mcneill2012dynamic}
W.~McNeill, H.~Kardes, and A.~Borthwick.
\newblock Dynamic record blocking: efficient linking of massive databases in
  mapreduce.
\newblock In {\em {QDB}}, 2012.

\bibitem{DBLP:conf/sac/MestrePN15}
D.~G. Mestre, C.~E.~S. Pires, and D.~C. Nascimento.
\newblock Adaptive sorted neighborhood blocking for entity matching with
  mapreduce.
\newblock In {\em SAC}, pages 981--987, 2015.

\bibitem{DBLP:conf/aaai/MichelsonK06}
M.~Michelson and C.~A. Knoblock.
\newblock Learning blocking schemes for record linkage.
\newblock In {\em AAAI}, pages 440--445, 2006.

\bibitem{DBLP:conf/nips/MikolovSCCD13}
T.~Mikolov, I.~Sutskever, K.~Chen, G.~S. Corrado, and J.~Dean.
\newblock Distributed representations of words and phrases and their
  compositionality.
\newblock In {\em NIPS}, pages 3111--3119, 2013.

\bibitem{nascimento2019exploiting}
D.~C. Nascimento, C.~E.~S. Pires, and D.~G. Mestre.
\newblock Exploiting block co-occurrence to control block sizes for entity
  resolution.
\newblock {\em Knowledge and Information Systems}, pages 1--42, 2019.

\bibitem{DBLP:conf/cikm/NegahbanRG12}
S.~Negahban, B.~I.~P. Rubinstein, and J.~Gemmell.
\newblock Scaling multiple-source entity resolution using statistically
  efficient transfer learning.
\newblock In {\em CIKM}, pages 2224--2228, 2012.

\bibitem{nelson2011entity}
E.~Nelson and J.~Talburt.
\newblock Entity resolution for longitudinal studies in education using oyster.
\newblock In {\em IKE}, 2011.

\bibitem{DBLP:journals/semweb/NentwigHNR17}
M.~Nentwig, M.~Hartung, A.~Ngomo, and E.~Rahm.
\newblock A survey of current link discovery frameworks.
\newblock {\em Semantic Web}, 8(3):419--436, 2017.

\bibitem{DBLP:conf/semweb/Ngomo13}
A.~Ngomo.
\newblock {ORCHID} - reduction-ratio-optimal computation of geo-spatial
  distances for link discovery.
\newblock In {\em ISWC}, pages 395--410, 2013.

\bibitem{DBLP:conf/ijcai/NgomoA11}
A.~Ngomo and S.~Auer.
\newblock {LIMES} - {A} time-efficient approach for large-scale link discovery
  on the web of data.
\newblock In {\em IJCAI}, pages 2312--2317, 2011.

\bibitem{nikolov2007knofuss}
A.~Nikolov, V.~Uren, and E.~Motta.
\newblock Knofuss: a comprehensive architecture for knowledge fusion.
\newblock In {\em K-CAP}, pages 185--186, 2007.

\bibitem{DBLP:conf/ideas/NinMML07}
J.~Nin, V.~Munt{\'{e}}s{-}Mulero, N.~Mart{\'{\i}}nez{-}Bazan, and
  J.~Larriba{-}Pey.
\newblock On the use of semantic blocking techniques for data cleansing and
  integration.
\newblock In {\em IDEAS}, pages 190--198, 2007.

\bibitem{o2018new}
K.~O'Hare, A.~Jurek, and C.~de~Campos.
\newblock A new technique of selecting an optimal blocking method for better
  record linkage.
\newblock {\em Inf. Syst.}, 77:151--166, 2018.

\bibitem{o2019review}
K.~O'Hare, A.~Jurek-Loughrey, and C.~de~Campos.
\newblock A review of unsupervised and semi-supervised blocking methods for
  record linkage.
\newblock In {\em Linking and Mining Heterogeneous and Multi-view Data}, pages
  79--105. 2019.

\bibitem{DBLP:journals/pvldb/0001APK15}
G.~Papadakis, G.~Alexiou, G.~Papastefanatos, and G.~Koutrika.
\newblock Schema-agnostic vs schema-based configurations for blocking methods
  on homogeneous data.
\newblock {\em {PVLDB}}, 9(4):312--323, 2015.

\bibitem{DBLP:conf/i-semantics/0001BPK17}
G.~Papadakis, K.~Bereta, T.~Palpanas, and M.~Koubarakis.
\newblock Multi-core meta-blocking for big linked data.
\newblock In {\em SEMANTICS}, pages 33--40, 2017.

\bibitem{DBLP:conf/iiwas/PapadakisDFK10}
G.~Papadakis, G.~Demartini, P.~Fankhauser, and P.~K{\"{a}}rger.
\newblock The missing links: discovering hidden same-as links among a billion
  of triples.
\newblock In {\em {iiWAS}}, pages 453--460, 2010.

\bibitem{DBLP:conf/jcdl/PapadakisGNPN11}
G.~Papadakis, G.~Giannakopoulos, C.~Nieder{\'{e}}e, T.~Palpanas, and W.~Nejdl.
\newblock Detecting and exploiting stability in evolving heterogeneous
  information spaces.
\newblock In {\em JCDL}, pages 95--104, 2011.

\bibitem{DBLP:conf/wsdm/PapadakisINF11}
G.~Papadakis, E.~Ioannou, C.~Nieder{\'{e}}e, and P.~Fankhauser.
\newblock Efficient entity resolution for large heterogeneous information
  spaces.
\newblock In {\em WSDM}, pages 535--544, 2011.

\bibitem{DBLP:conf/jcdl/PapadakisINPN11}
G.~Papadakis, E.~Ioannou, C.~Nieder{\'{e}}e, T.~Palpanas, and W.~Nejdl.
\newblock Eliminating the redundancy in blocking-based entity resolution
  methods.
\newblock In {\em JCDL}, pages 85--94, 2011.

\bibitem{DBLP:conf/sigmod/PapadakisINPN11}
G.~Papadakis, E.~Ioannou, C.~Nieder{\'{e}}e, T.~Palpanas, and W.~Nejdl.
\newblock To compare or not to compare: making entity resolution more
  efficient.
\newblock In {\em {SWIM}}, page~3, 2011.

\bibitem{DBLP:conf/wsdm/PapadakisINPN12}
G.~Papadakis, E.~Ioannou, C.~Nieder{\'{e}}e, T.~Palpanas, and W.~Nejdl.
\newblock Beyond 100 million entities: large-scale blocking-based resolution
  for heterogeneous data.
\newblock In {\em WSDM}, pages 53--62, 2012.

\bibitem{DBLP:journals/tkde/PapadakisIPNN13}
G.~Papadakis, E.~Ioannou, T.~Palpanas, C.~Nieder{\'{e}}e, and W.~Nejdl.
\newblock A blocking framework for entity resolution in highly heterogeneous
  information spaces.
\newblock {\em {IEEE} TKDE}, 25(12):2665--2682, 2013.

\bibitem{DBLP:journals/tkde/PapadakisKPN14}
G.~Papadakis, G.~Koutrika, T.~Palpanas, and W.~Nejdl.
\newblock Meta-blocking: Taking entity resolution to the next level.
\newblock {\em {IEEE} TKDE}, 26(8):1946--1960, 2014.

\bibitem{DBLP:conf/icde/PapadakisN11}
G.~Papadakis and W.~Nejdl.
\newblock Efficient entity resolution methods for heterogeneous information
  spaces.
\newblock In {\em ICDE Workshops}, pages 304--307, 2011.

\bibitem{7498364}
G.~Papadakis and T.~Palpanas.
\newblock Blocking for large-scale entity resolution: Challenges, algorithms,
  and practical examples.
\newblock In {\em {ICDE}}, pages 1436--1439, 2016.

\bibitem{PapadakisTutorialWww18}
G.~Papadakis and T.~Palpanas.
\newblock Web-scale, schema-agnostic, end-to-end entity resolution.
\newblock In {\em WWW Companion Volume}, 2018.

\bibitem{DBLP:journals/pvldb/0001PK14}
G.~Papadakis, G.~Papastefanatos, and G.~Koutrika.
\newblock Supervised meta-blocking.
\newblock {\em {PVLDB}}, 7(14):1929--1940, 2014.

\bibitem{DBLP:journals/bdr/PapadakisPPK16}
G.~Papadakis, G.~Papastefanatos, T.~Palpanas, and M.~Koubarakis.
\newblock Boosting the efficiency of large-scale entity resolution with
  enhanced meta-blocking.
\newblock {\em Big Data Research}, 6:43--63, 2016.

\bibitem{DBLP:conf/edbt/0001PPK16}
G.~Papadakis, G.~Papastefanatos, T.~Palpanas, and M.~Koubarakis.
\newblock Scaling entity resolution to large, heterogeneous data with enhanced
  meta-blocking.
\newblock In {\em EDBT}, pages 221--232, 2016.

\bibitem{DBLP:journals/pvldb/0001SGP16}
G.~Papadakis, J.~Svirsky, A.~Gal, and T.~Palpanas.
\newblock Comparative analysis of approximate blocking techniques for entity
  resolution.
\newblock {\em {PVLDB}}, 9(9):684--695, 2016.

\bibitem{DBLP:journals/pvldb/PapadakisTTGPK18}
G.~Papadakis, L.~Tsekouras, E.~Thanos, G.~Giannakopoulos, T.~Palpanas, and
  M.~Koubarakis.
\newblock The return of jedai: End-to-end entity resolution for structured and
  semi-structured data.
\newblock {\em {PVLDB}}, 11(12):1950--1953, 2018.

\bibitem{DBLP:journals/tkde/PapenbrockHN15}
T.~Papenbrock, A.~Heise, and F.~Naumann.
\newblock Progressive duplicate detection.
\newblock {\em {IEEE} TKDE.}, 27(5):1316--1329, 2015.

\bibitem{DBLP:conf/emnlp/PenningtonSM14}
J.~Pennington, R.~Socher, and C.~D. Manning.
\newblock Glove: Global vectors for word representation.
\newblock In {\em {EMNLP}}, pages 1532--1543, 2014.

\bibitem{DBLP:conf/edbt/PuhlmannWN06}
S.~Puhlmann, M.~Weis, and F.~Naumann.
\newblock {XML} duplicate detection using sorted neighborhoods.
\newblock In {\em {EDBT}}, pages 773--791, 2006.

\bibitem{DBLP:conf/sigmod/QinWLXL11}
J.~Qin, W.~Wang, Y.~Lu, C.~Xiao, and X.~Lin.
\newblock Efficient exact edit similarity query processing with the asymmetric
  signature scheme.
\newblock In {\em {SIGMOD}}, pages 1033--1044, 2011.

\bibitem{DBLP:journals/pvldb/QinX18}
J.~Qin and C.~Xiao.
\newblock Pigeonring: {A} principle for faster thresholded similarity search.
\newblock {\em {PVLDB}}, 12(1):28--42, 2018.

\bibitem{DBLP:conf/cikm/RamadanC14}
B.~Ramadan and P.~Christen.
\newblock Forest-based dynamic sorted neighborhood indexing for real-time
  entity resolution.
\newblock In {\em CIKM}, pages 1787--1790, 2014.

\bibitem{DBLP:conf/pakdd/RamadanC15}
B.~Ramadan and P.~Christen.
\newblock Unsupervised blocking key selection for real-time entity resolution.
\newblock In {\em {PAKDD}}, pages 574--585, 2015.

\bibitem{Ramadan:2015:DSN:2836847.2816821}
B.~Ramadan, P.~Christen, H.~Liang, and R.~W. Gayler.
\newblock Dynamic sorted neighborhood indexing for real-time entity resolution.
\newblock {\em J. Data and Information Quality}, 6(4):15:1--15:29, 2015.

\bibitem{10.1007/978-3-642-40319-4_5}
B.~Ramadan, P.~Christen, H.~Liang, R.~W. Gayler, and D.~Hawking.
\newblock Dynamic similarity-aware inverted indexing for real-time entity
  resolution.
\newblock In {\em PAKDD Workshops}, pages 47--58, 2013.

\bibitem{DBLP:conf/icdm/RanbadugeVC16}
T.~Ranbaduge, D.~Vatsalan, and P.~Christen.
\newblock Scalable block scheduling for efficient multi-database record
  linkage.
\newblock In {\em ICDM}, pages 1161--1166, 2016.

\bibitem{DBLP:journals/is/RibeiroH11}
L.~A. Ribeiro and T.~H{\"{a}}rder.
\newblock Generalizing prefix filtering to improve set similarity joins.
\newblock {\em Inf. Syst.}, 36(1):62--78, 2011.

\bibitem{rice2007braided}
S.~V. Rice.
\newblock Braided avl trees for efficient event sets and ranked sets in the
  simscript iii simulation programming language.
\newblock In {\em Western MultiConference on Computer Simulation}, pages
  150--155, 2007.

\bibitem{DBLP:conf/icde/RongLSWLD17}
C.~Rong, C.~Lin, Y.~N. Silva, J.~Wang, W.~Lu, and X.~Du.
\newblock Fast and scalable distributed set similarity joins for big data
  analytics.
\newblock In {\em {ICDE}}, pages 1059--1070, 2017.

\bibitem{DBLP:journals/tkde/RongLWDCT13}
C.~Rong, W.~Lu, X.~Wang, X.~Du, Y.~Chen, and A.~K.~H. Tung.
\newblock Efficient and scalable processing of string similarity join.
\newblock {\em {IEEE} TKDE}, 25(10):2217--2230, 2013.

\bibitem{DBLP:conf/sigmod/Sarawagi04}
S.~Sarawagi and A.~Kirpal.
\newblock Efficient set joins on similarity predicates.
\newblock In {\em SIGMOD}, pages 743--754, 2004.

\bibitem{sariyar2011controlling}
M.~Sariyar, A.~Borg, and K.~Pommerening.
\newblock Controlling false match rates in record linkage using extreme value
  theory.
\newblock {\em Journal of biomedical informatics}, 44(4):648--654, 2011.

\bibitem{DBLP:conf/cikm/SarmaJMB12}
A.~D. Sarma, A.~Jain, A.~Machanavajjhala, and P.~Bohannon.
\newblock An automatic blocking mechanism for large-scale de-duplication tasks.
\newblock In {\em CIKM}, pages 1055--1064, 2012.

\bibitem{DBLP:journals/pvldb/SatuluriP12}
V.~Satuluri and S.~Parthasarathy.
\newblock Bayesian locality sensitive hashing for fast similarity search.
\newblock {\em {PVLDB}}, 5(5):430--441, 2012.

\bibitem{DBLP:journals/tkde/ShenWH15}
W.~Shen, J.~Wang, and J.~Han.
\newblock Entity linking with a knowledge base: Issues, techniques, and
  solutions.
\newblock {\em {IEEE} TKDE}, 27(2):443--460, 2015.

\bibitem{DBLP:conf/icde/ShuCXM11}
L.~Shu, A.~Chen, M.~Xiong, and W.~Meng.
\newblock Efficient spectral neighborhood blocking for entity resolution.
\newblock In {\em ICDE}, pages 1067--1078, 2011.

\bibitem{DBLP:journals/pvldb/SimoniniBJ16}
G.~Simonini, S.~Bergamaschi, and H.~V. Jagadish.
\newblock {BLAST:} a loosely schema-aware meta-blocking approach for entity
  resolution.
\newblock {\em {PVLDB}}, 9(12):1173--1184, 2016.

\bibitem{DBLP:journals/is/SimoniniGBJ19}
G.~Simonini, L.~Gagliardelli, S.~Bergamaschi, and H.~V. Jagadish.
\newblock Scaling entity resolution: {A} loosely schema-aware approach.
\newblock {\em Inf. Syst.}, 83:145--165, 2019.

\bibitem{simonini2018schema}
G.~Simonini, G.~Papadakis, T.~Palpanas, and S.~Bergamaschi.
\newblock Schema-agnostic progressive entity resolution.
\newblock {\em IEEE TKDE}, 31(6):1208--1221, 2019.

\bibitem{DBLP:conf/semweb/Song12}
D.~Song.
\newblock Scalable and domain-independent entity coreference: Establishing high
  quality data linkages across heterogeneous data sources.
\newblock In {\em ISWC}, pages 424--432, 2012.

\bibitem{DBLP:conf/semweb/SongH11}
D.~Song and J.~Heflin.
\newblock Automatically generating data linkages using a domain-independent
  candidate selection approach.
\newblock In {\em ISWC}, pages 649--664, 2011.

\bibitem{DBLP:journals/tkde/SongLH17}
D.~Song, Y.~Luo, and J.~Heflin.
\newblock Linking heterogeneous data in the semantic web using scalable and
  domain-independent candidate selection.
\newblock {\em {IEEE} TKDE}, 29(1):143--156, 2017.

\bibitem{DBLP:conf/icde/StefanidisCE17}
K.~Stefanidis, V.~Christophides, and V.~Efthymiou.
\newblock Web-scale blocking, iterative and progressive entity resolution.
\newblock In {\em {ICDE}}, pages 1459--1462, 2017.

\bibitem{DBLP:conf/www/StefanidisEHC14}
K.~Stefanidis, V.~Efthymiou, M.~Herschel, and V.~Christophides.
\newblock Entity resolution in the web of data.
\newblock In {\em {WWW} Companion Volume}, pages 203--204, 2014.

\bibitem{DBLP:conf/psd/SteortsVSF14}
R.~C. Steorts, S.~L. Ventura, M.~Sadinle, and S.~E. Fienberg.
\newblock A comparison of blocking methods for record linkage.
\newblock In {\em Privacy in Statistical Databases}, pages 253--268, 2014.

\bibitem{DBLP:journals/pvldb/CADA18}
P.~Suganthan, A.~Ardalan, A.~Doan, and A.~Akella.
\newblock Smurf: Self-service string matching using random forests.
\newblock {\em {PVLDB}}, 12(3):278--291, 2018.

\bibitem{DBLP:journals/pvldb/SunS0BD19}
J.~Sun, Z.~Shang, G.~Li, Z.~Bao, and D.~Deng.
\newblock Balance-aware distributed string similarity-based query processing
  system.
\newblock {\em {PVLDB}}, 12(9):961--974, 2019.

\bibitem{DBLP:journals/pvldb/SunSLDB17}
J.~Sun, Z.~Shang, G.~Li, D.~Deng, and Z.~Bao.
\newblock Dima: {A} distributed in-memory similarity-based query processing
  system.
\newblock {\em {PVLDB}}, 10(12):1925--1928, 2017.

\bibitem{DBLP:journals/pvldb/TaoDS17}
W.~Tao, D.~Deng, and M.~Stonebraker.
\newblock Approximate string joins with abbreviations.
\newblock {\em {PVLDB}}, 11(1):53--65, 2017.

\bibitem{DBLP:conf/sigmod/TaoYSK09}
Y.~Tao, K.~Yi, C.~Sheng, and P.~Kalnis.
\newblock Quality and efficiency in high dimensional nearest neighbor search.
\newblock In {\em {SIGMOD}}, pages 563--576, 2009.

\bibitem{DBLP:journals/corr/abs-1809-11084}
S.~Thirumuruganathan, S.~A.~P. Parambath, M.~Ouzzani, N.~Tang, and S.~R. Joty.
\newblock Reuse and adaptation for entity resolution through transfer learning.
\newblock {\em CoRR}, abs/1809.11084, 2018.

\bibitem{DBLP:journals/is/VatsalanCV13}
D.~Vatsalan, P.~Christen, and V.~S. Verykios.
\newblock A taxonomy of privacy-preserving record linkage techniques.
\newblock {\em Inf. Syst.}, 38(6):946--969, 2013.

\bibitem{DBLP:conf/sigmod/VernicaCL10}
R.~Vernica, M.~J. Carey, and C.~Li.
\newblock Efficient parallel set-similarity joins using mapreduce.
\newblock In {\em {SIGMOD}}, pages 495--506, 2010.

\bibitem{volz2009silk}
J.~Volz, C.~Bizer, M.~Gaedke, and G.~Kobilarov.
\newblock Silk-a link discovery framework for the web of data.
\newblock {\em LDOW}, 538, 2009.

\bibitem{DBLP:journals/pvldb/WangLF10}
J.~Wang, G.~Li, and J.~Feng.
\newblock Trie-join: Efficient trie-based string similarity joins with
  edit-distance constraints.
\newblock {\em {PVLDB}}, 3(1):1219--1230, 2010.

\bibitem{DBLP:conf/sigmod/WangLF12}
J.~Wang, G.~Li, and J.~Feng.
\newblock Can we beat the prefix filtering?: an adaptive framework for
  similarity join and search.
\newblock In {\em SIGMOD}, pages 85--96, 2012.

\bibitem{DBLP:journals/tods/WangLF14}
J.~Wang, G.~Li, and J.~Feng.
\newblock Extending string similarity join to tolerant fuzzy token matching.
\newblock {\em {ACM} TODS}, 39(1):7:1--7:45, 2014.

\bibitem{DBLP:conf/icde/WangLZ19}
J.~Wang, C.~Lin, and C.~Zaniolo.
\newblock Mf-join: Efficient fuzzy string similarity join with multi-level
  filtering.
\newblock In {\em {ICDE}}, pages 386--397, 2019.

\bibitem{DBLP:journals/corr/WangSSJ14}
J.~Wang, H.~T. Shen, J.~Song, and J.~Ji.
\newblock Hashing for similarity search: {A} survey.
\newblock {\em CoRR}, abs/1408.2927, 2014.

\bibitem{DBLP:journals/tkde/WangYWL17}
J.~Wang, X.~Yang, B.~Wang, and C.~Liu.
\newblock Ls-join: Local similarity join on string collections.
\newblock {\em {IEEE} TKDE}, 29(9):1928--1942, 2017.

\bibitem{DBLP:conf/sigmod/WangXQWZI16}
P.~Wang, C.~Xiao, J.~Qin, W.~Wang, X.~Zhang, and Y.~Ishikawa.
\newblock Local similarity search for unstructured text.
\newblock In {\em {SIGMOD}}, pages 1991--2005, 2016.

\bibitem{DBLP:journals/tkde/WangCL16}
Q.~Wang, M.~Cui, and H.~Liang.
\newblock Semantic-aware blocking for entity resolution.
\newblock {\em {IEEE} TKDE}, 28(1):166--180, 2016.

\bibitem{DBLP:journals/tkde/WangQXLS13}
W.~Wang, J.~Qin, C.~Xiao, X.~Lin, and H.~T. Shen.
\newblock Vchunkjoin: An efficient algorithm for edit similarity joins.
\newblock {\em {IEEE} TKDE}, 25(8):1916--1929, 2013.

\bibitem{DBLP:journals/pvldb/WangQLZC17}
X.~Wang, L.~Qin, X.~Lin, Y.~Zhang, and L.~Chang.
\newblock Leveraging set relations in exact set similarity join.
\newblock {\em {PVLDB}}, 10(9):925--936, 2017.

\bibitem{DBLP:journals/tkde/WhangMG13}
S.~E. Whang, D.~Marmaros, and H.~Garcia{-}Molina.
\newblock Pay-as-you-go entity resolution.
\newblock {\em {IEEE} TKDE.}, 25(5):1111--1124, 2013.

\bibitem{DBLP:conf/sigmod/WhangMKTG09}
S.~E. Whang, D.~Menestrina, G.~Koutrika, M.~Theobald, and H.~Garcia{-}Molina.
\newblock Entity resolution with iterative blocking.
\newblock In {\em SIGMOD}, pages 219--232, 2009.

\bibitem{DBLP:journals/pvldb/XiaoWL08}
C.~Xiao, W.~Wang, and X.~Lin.
\newblock Ed-join: an efficient algorithm for similarity joins with edit
  distance constraints.
\newblock {\em {PVLDB}}, 1(1):933--944, 2008.

\bibitem{DBLP:conf/icde/XiaoWLS09}
C.~Xiao, W.~Wang, X.~Lin, and H.~Shang.
\newblock Top-k set similarity joins.
\newblock In {\em {ICDE}}, pages 916--927, 2009.

\bibitem{DBLP:conf/www/XiaoWLY08}
C.~Xiao, W.~Wang, X.~Lin, and J.~X. Yu.
\newblock Efficient similarity joins for near duplicate detection.
\newblock In {\em WWW}, pages 131--140, 2008.

\bibitem{DBLP:journals/tods/XiaoWLYW11}
C.~Xiao, W.~Wang, X.~Lin, J.~X. Yu, and G.~Wang.
\newblock Efficient similarity joins for near-duplicate detection.
\newblock {\em {ACM} TODS}, 36(3):15:1--15:41, 2011.

\bibitem{DBLP:journals/pvldb/XuL19}
P.~Xu and J.~Lu.
\newblock Towards a unified framework for string similarity joins.
\newblock {\em {PVLDB}}, 12(11):1289--1302, 2019.

\bibitem{DBLP:conf/jcdl/YanLKG07}
S.~Yan, D.~Lee, M.~Kan, and C.~L. Giles.
\newblock Adaptive sorted neighborhood methods for efficient record linkage.
\newblock In {\em {JCDL}}, pages 185--194, 2007.

\bibitem{DBLP:conf/ipccc/YanXM13}
W.~Yan, Y.~Xue, and B.~Malin.
\newblock Scalable load balancing for mapreduce-based record linkage.
\newblock In {\em {IPCCC}}, pages 1--10, 2013.

\bibitem{DBLP:journals/fcsc/YuLDF16}
M.~Yu, G.~Li, D.~Deng, and J.~Feng.
\newblock String similarity search and join: a survey.
\newblock {\em Frontiers Comput. Sci.}, 10(3):399--417, 2016.

\bibitem{DBLP:journals/vldb/YuWLZDF17}
M.~Yu, J.~Wang, G.~Li, Y.~Zhang, D.~Deng, and J.~Feng.
\newblock A unified framework for string similarity search with edit-distance
  constraint.
\newblock {\em {VLDB} J.}, 26(2):249--274, 2017.

\bibitem{DBLP:journals/tifs/YuanWWYN17}
X.~Yuan, X.~Wang, C.~Wang, C.~Yu, and S.~Nutanong.
\newblock Privacy-preserving similarity joins over encrypted data.
\newblock {\em {IEEE TIFS}}, 12(11):2763--2775, 2017.

\bibitem{DBLP:conf/sigmod/ZhaiLG11}
J.~Zhai, Y.~Lou, and J.~Gehrke.
\newblock {ATLAS:} a probabilistic algorithm for high dimensional similarity
  search.
\newblock In {\em {SIGMOD}}, pages 997--1008, 2011.

\bibitem{zhang2017pruning}
F.~Zhang, Z.~Gao, and K.~Niu.
\newblock A pruning algorithm for meta-blocking based on cumulative weight.
\newblock In {\em Journal of Physics: Conference Series}, volume 887, 2017.

\bibitem{DBLP:conf/icde/ZhangLWZXY17}
Y.~Zhang, X.~Li, J.~Wang, Y.~Zhang, C.~Xing, and X.~Yuan.
\newblock An efficient framework for exact set similarity search using tree
  structure indexes.
\newblock In {\em {ICDE}}, pages 759--770, 2017.

\bibitem{zhang2018transformation}
Y.~Zhang, J.~Wu, J.~Wang, and C.~Xing.
\newblock A transformation-based framework for knn set similarity search.
\newblock {\em IEEE TKDE}, 2018.

\bibitem{DBLP:conf/sigmod/ZhangHOS10}
Z.~Zhang, M.~Hadjieleftheriou, B.~C. Ooi, and D.~Srivastava.
\newblock Bed-tree: an all-purpose index structure for string similarity search
  based on edit distance.
\newblock In {\em {SIGMOD}}, pages 915--926, 2010.

\bibitem{DBLP:conf/sigmod/ZhuDNM19}
E.~Zhu, D.~Deng, F.~Nargesian, and R.~J. Miller.
\newblock {JOSIE:} overlap set similarity search for finding joinable tables in
  data lakes.
\newblock In {\em {SIGMOD}}, pages 847--864, 2019.

\end{thebibliography}

\end{document}